\documentclass[11pt,a4paper]{article}
\pdfoutput=1
\usepackage{jcappub}
\usepackage{lineno}
\usepackage{epstopdf}
\usepackage{rotating}
\newcommand\ion[2]{#1$\;${\small \uppercase\expandafter{\romannumeral #2}}}%
\newcommand\ionalt[2]{#1$\;${\scriptsize \uppercase\expandafter{\romannumeral #2}}}%
\newcommand{\lya}{Lyman-$\alpha$\ }
\newcommand{\lyb}{Lyman-$\beta$\ }
\newcommand{\lyaf}{Lyman-$\alpha$ forest\ }
\newcommand{\mpch}{h^{-1}{\rm Mpc}}

\newcommand{\mpc}{\, {\rm Mpc}}
\newcommand{\kpc}{\, {\rm kpc}}
\newcommand{\hmpc}{\, h^{-1} \mpc}

\newcommand{\hkpc}{\, h^{-1} \kpc}
\newcommand{\vxperp}{\mathbf{x_\perp}}

\newcommand{\cm}{{\rm cm}}

\title{The Lyman-$\alpha$ forest in three dimensions: measurements of
  large scale flux correlations from BOSS 1st-year data}

\author[a]{An\v{z}e Slosar,}
\author[b]{Andreu Font-Ribera,}
\author[c,d]{Matthew M. Pieri,}
\author[e]{James Rich,}
\author[e]{Jean-Marc Le Goff,}
\author[f,e]{\'{E}ric Aubourg,}
\author[g]{Jon Brinkmann,}
\author[f]{Nicolas Busca,}
\author[h]{Bill Carithers,}
\author[e]{Romain Charlassier,}
\author[h]{Marina Cort\^{e}s,}
\author[i]{Rupert Croft,}
\author[j]{Kyle S. Dawson,}
\author[k]{Daniel Eisenstein,}
\author[f]{Jean-Christophe Hamilton,}
\author[h]{Shirley Ho,}
\author[l]{Khee-Gan Lee,}
\author[l]{Robert Lupton,}
\author[h,a]{Patrick McDonald,}
\author[m]{Bumbarija Medolin,}
\author[n,o]{Jordi Miralda-Escud\'{e},}
\author[p]{Demitri Muna,}
\author[q,r]{Adam D. Myers,}
\author[s]{Robert C. Nichol,}
\author[e]{Nathalie Palanque-Delabrouille,}
\author[t]{Isabelle P\^aris,}
\author[t]{Patrick Petitjean,}
\author[l]{Yodovina Pi\v{s}kur,}
\author[t]{Emmanuel Rollinde,}
\author[h]{Nicholas P. Ross,}
\author[h]{David J. Schlegel,}
\author[u]{Donald P. Schneider,}
\author[a]{Erin Sheldon,}
\author[n]{Benjamin A. Weaver,}
\author[d]{David H. Weinberg,}
\author[e]{Christophe Yeche,}
\author[v,w]{Donald G. York}

\affiliation[a]{Brookhaven National Laboratory,
  Blgd 510, Upton NY 11375, USA}
\affiliation[b]{Institut de Ci\`{e}ncies de l'Espai (CSIC-IEEC), 
  Campus UAB, Fac. Ci\`{e}ncies, torre C5 parell 2, Bellaterra, Catalonia}
\affiliation[c]{Center for Astrophysics and Space Astronomy, University
of Colorado, 389 UCB, Boulder, Colorado 80309, USA}
\affiliation[d]{Department of Astronomy, Ohio State University, 140 
West 18th Avenue, Columbus, OH 43210, USA}
\affiliation[e]{CEA, Centre de Saclay, IRFU, 91191 Gif-sur-Yvette,
  France}
\affiliation[f]{APC, Universit\'{e} Paris Diderot-Paris 7, CNRS/IN2P3, CEA, Observatoire de Paris, 10, rue
            A. Domon \& L. Duquet,  Paris, France}
\affiliation[g]{Apache Point Observatory, P.O. Box 59, Sunspot, NM 88349,USA}
\affiliation[h]{Lawrence Berkeley National Laboratory, 1 Cyclotron Road, Berkeley, CA 94720, USA.}
\affiliation[i]{Bruce and Astrid McWilliams Center for Cosmology,
  Carnegie Mellon University, Pittsburgh, PA 15213, USA}
\affiliation[j]{University of Utah, Dept. of Physics \& Astronomy, 115 S 1400 E, Salt Lake City, UT 84112, USA}
\affiliation[k]{Harvard College Observatory, 60 Garden St., Cambridge
  MA 02138, USA}
\affiliation[l]{Department of Astrophysical Sciences, Princeton  University,   Princeton, New Jersey 08544, USA}
\affiliation[m]{104-20 Queens Blvd \#17A, Forest Hills, NY 11375, USA}
\affiliation[n]{Instituci\'{o} Catalana de Recerca i Estudis  Avan\c{c}ats, Barcelona, Catalonia}
\affiliation[o]{Institut de Ci\`{e}ncies del Cosmos, Universitat de Barcelona/IEEC, Barcelona 08028, Catalonia}
\affiliation[p]{Center for Cosmology and Particle Physics, New York University, New York, NY 10003 USA}
\affiliation[q]{Department of Astronomy, MC-221, University of Illinois, 1002 West Green Street, Urbana, IL
61801, USA}
\affiliation[r]{Department of Physics and Astronomy, University of
  Wyoming, Laramie, WY 82071, USA}
\affiliation[s]{Institute of Cosmology and Gravitation (ICG), Dennis
  Sciama Building, Burnaby Road, Univ. of Portsmouth, Portsmouth, PO1
  3FX, UK.}
\affiliation[t]{Universit\'e Paris 6 et CNRS, Institut d'Astrophysique
  de Paris, 98bis blvd. Arago, 75014 Paris, France}
\affiliation[u]{Department of Astronomy and Astrophysics, 
  The Pennsylvania State University, 525 Davey Lab, University Park,
  PA 16802, USA}
\affiliation[v]{Department of Astronomy and Astrophysics, University of
Chicago, 5640 South Ellis Avenue, Chicago, IL 60637, USA}
\affiliation[w]{Enrico Fermi Institute, University of Chicago, 5640 South
Ellis Avenue, Chicago, IL 60637, USA}

\emailAdd{anze@bnl.gov}

\abstract{Using a sample of approximately 14,000 $z>2.1$ quasars
  observed in the first year of the Baryon Oscillation Spectroscopic
  Survey (BOSS), we measure the three-dimensional correlation function
  of absorption in the Lyman-$\alpha$ forest. The angle-averaged
  correlation function of transmitted flux ($F = e^{-\tau}$) is
  securely detected out to comoving separations of $60\hmpc$, the
  first detection of flux correlations across widely separated
  sightlines.  A quadrupole distortion of the redshift-space
  correlation function by peculiar velocities, the signature of the
  gravitational instability origin of structure in the Lyman-$\alpha$
  forest, is also detected at high significance. We obtain a good fit
  to the data assuming linear theory redshift-space distortion and
  linear bias of the transmitted flux, relative to the matter
  fluctuations of a standard $\Lambda$CDM cosmological model
  (inflationary cold dark matter with a cosmological constant).  At
  95\% confidence, we find a linear bias parameter $0.16 < b < 0.24$
  and redshift-distortion parameter $0.44 < \beta < 1.20$, at central
  redshift $z=2.25$, with a well constrained combination
  $b~(1+\beta)=0.336 \pm 0.012$. The errors on $\beta$ are asymmetric,
  with $\beta=0$ excluded at over $5\sigma$ confidence level. The
  value of $\beta$ is somewhat low compared to theoretical
  predictions, and our tests on synthetic data suggest that it is
  depressed (relative to expectations for the Lyman-$\alpha$ forest
  alone) by the presence of high column density systems and metal line
  absorption.  These results set the stage for cosmological parameter
  determinations from three-dimensional structure in the
  Lyman-$\alpha$ forest, including anticipated constraints on dark
  energy from baryon acoustic oscillations.}  


\begin{document}
\maketitle

\section{Introduction}

Early spectra of high-redshift ($z>2$) quasars showed ubiquitous
absorption lines blueward of their \lya emission, which were identified
as arising primarily from \lya absorption by intervening concentrations
of neutral hydrogen \cite{1971ApJ...164L..73L}. While
early models described the absorbers as discrete clouds analogous to
those in the interstellar medium, a combination of theoretical and
observational advances in the mid-1990s led to a revised view of the
forest as a continuous phenomenon, analogous to Gunn-Peterson
\cite{1965ApJ...142.1633G} absorption but tracing an inhomogeneous,
fluctuating intergalactic medium (see, e.g., the review by
\cite{1998ARAA..36..267R}).  This revision of theoretical
understanding also transformed the promise of the \lya forest as a
tool for cosmology, by showing that the forest traces the distribution
of dark matter in the high-redshift universe in a relatively simple
way.  Adopting this ``continuous medium'' view, several groups
\cite{1999ApJ...520....1C,2000ApJ...543....1M,2002ApJ...581...20C,
  2004MNRAS.354..684V,2005ApJ...635..761M,2006ApJS..163...80M}
measured the power spectrum of \lya forest flux in successively larger
samples of quasar spectra and used it to constrain the power spectrum
of the underlying dark matter distribution.  The most recent of these
measurements \cite{2006ApJS..163...80M}, using a large quasar sample from the 
Sloan Digital Sky Survey (SDSS;
\cite{2000AJ....120.1579Y,2002AJ....123..485S,2002AJ....123.2121S,2003AJ....125.1559P,2008ApJ...674.1217P,2006AJ....131.2332G,1998AJ....116.3040G,1996AJ....111.1748F}),
have provided some of the strongest constraints on inflation, neutrino
masses, and the ``coldness'' of dark matter, especially when combined
with data sets that probe other redshifts and/or physical scales
(e.g., \cite{2005PhRvD..71j3515S,2006PhRvL..97s1303S,
  2006JCAP...10..014S,2006MNRAS.365..231V,2008PhRvL.100d1304V,
  2010JCAP...06..015V}).  However, while the underlying density field
is three-dimensional, all of these cosmological analyses have treated
the forest as a collection of independent 1-dimensional maps.  

This paper presents measurements of \lya forest flux
correlations across parallel lines of sight with the widest separation
reached so far, taking advantage
of the large sample of high-redshift quasars observed during the first
year of BOSS, the Baryon Oscillation Spectroscopic Survey of SDSS-III
\cite{2011arXiv1101.1529E}.

Measurements of correlated absorption towards gravitational lenses and
closely separated quasar pairs \cite{1992ApJ...389...39S,
  1994ApJ...437L..83B, 1994ApJ...437L..87D, 1995Natur.373..223D}
provided some of the critical observational evidence for the revised
understanding of the \lya forest.  These transverse correlation
measurements implied a coherence scale of several hundred $\hkpc$
(comoving) for individual absorbing structures.  The large sizes and
low neutral column densities in turn implied that the absorbing gas has
low densities and is highly photoionized by the intergalactic UV
background, making the total baryon content of the forest a large
fraction of the baryons allowed by Big Bang nucleosynthesis
\cite{1995MNRAS.275L..76R}.  Hydrodynamic cosmological simulations
naturally explained these large coherence scales and many other
statistical properties of the forest \cite{1994ApJ...437L...9C,
  1995ApJ...453L..57Z, 1996ApJ...457L..51H, 1996ApJ...471..582M}, with
typical absorption features arising in filamentary structures of
moderate overdensity, $\delta \equiv \rho/\bar{\rho}-1 \sim
20$.  Detailed investigation of these simulations revealed an
approximate description of the forest that is both surprisingly simple
and surprisingly accurate
\cite{1998ASPC..148...21W,1998ApJ...495...44C}: the \lya forest arises
in gas that traces the underlying dark matter distribution, except for
small scale pressure support, with the neutral hydrogen fraction
determined by photoionization equilibrium, and with a power-law
temperature-density relation $T \propto (\rho/\bar{\rho})^{\gamma-1}$.
This relation is established by the competition of
photoionization heating and adiabatic cooling
\cite{1997MNRAS.292...27H}.  In this approximation, the transmitted
flux fraction $F$ is related to the dark matter overdensity $\delta$
by
\begin{equation}
\label{eqn:fgpa}
F = e^{-\tau} = \exp\left[-A(1+\delta)^{2-0.7(\gamma-1)}\right]~,
\end{equation}
where the temperature-density slope $(\gamma-1)$ depends on the
inter-galactic medium (IGM) reionization history, and the constant $A$
depends on redshift and on a variety of physical parameters (see
\cite{2010MNRAS.404.1281P} for a recent discussion).  On large scales,
the three-dimensional power spectrum of the field $\delta_F \equiv
F/\bar{F}-1$ should have the same shape as the power spectrum of
$\delta = \rho/\bar{\rho}-1$
\cite{1998ApJ...495...44C,1999ApJ...520....1C,2000ApJ...543....1M}.
Thermal motions of atoms smooth spectra on small scales, producing a
turnover in the one-dimensional flux power spectrum at high $k$.
(In practice, peculiar velocities, which we discuss next, produce a
turnover at larger scales than thermal motions.)

The most significant correction to equation~(\ref{eqn:fgpa}) is from
peculiar velocities, which shift the apparent locations of the
absorbing neutral hydrogen in the radial direction.  On large scales,
the power spectrum should approximately follow the linear theory model
of redshift-space distortions,
\begin{equation}
\label{eqn:kaiser}
P_F(k,\mu_k) = b^2 P_L(k)(1+\beta\mu_k^2)^2 ~,
\end{equation}
where $P_L(k)$ is the real-space linear power spectrum and $\mu_k$ is
the cosine of the angle between the wavevector ${\bf k}$ and the line
of sight.   The bias factor $b$ of the forest is the bias factor of
the contrast of the flux fluctuations and not the bias factor of the
neutral hydrogen. It is typically low because
the full range of density variations is mapped into the transmitted
flux range $0 < F < 1$ ($b$ is actually technically negative, because
overdensities in mass produce smaller $F$, but this is irrelevant to our paper
so we will just quote $|b|$). 
While the functional form in Eq. \ref{eqn:kaiser}
is that of Kaiser \cite{1987MNRAS.227....1K}, the parameter $\beta$ does not
have the same interpretation, coming from a more general linear theory 
calculation of redshift-space distortions in the case where the directly 
distorted field, in this case optical depth, undergoes a subsequent non-linear
transformation, in this case $F=\exp(-\tau)$ 
\cite{2000ApJ...543....1M,2003ApJ...585...34M}.
For a galaxy survey, $\beta \approx
[\Omega_m(z)]^{0.55}/b$, but for the \lya forest $\beta$ can vary
independently of $\Omega_m$ and $b$, i.e., it is generally a free parameter.  

The simulations of \cite{2003ApJ...585...34M} suggest an approximate value
of $\beta \approx 1.47$ at $z \approx 2.25$ (interpolating over the parameter
dependences in Table 1 of \cite{2003ApJ...585...34M}, since the central model
there is antiquated).  Lower resolution
calculations of \cite{Slosar:2009iv} ($500 h^{-1} {\rm kpc}$ mean
particle spacing and particle-mesh (PM) cell size) found a lower value
$\beta\sim 1$. Unpublished work by Martin White demonstrates that the value of 
$\beta$ predicted by PM simulations with different smoothing applied to
the mass density field decreases with increasing smoothing length (based  
simulations with $73~ h^{-1}{\rm kpc}$
resolution, and in agreement with the $\sim 180~ h^{-1}{\rm kpc}$ resolution 
simulation in \cite{2010ApJ...713..383W}). Reference \cite{2003ApJ...585...34M}
concurred that
low resolution simulations produce lower $\beta$, and agree with White on the
value at similar smoothing, so the fundamental 
outstanding question for predicting $\beta$ is apparently how smooth is the 
gas in the 
IGM (\cite{2003ApJ...585...34M} quantifies other parameter dependences, but 
none of them make so much difference, given the relatively small uncertainty in
those parameters, for vanilla models at least). The smoothing scale of the IGM 
is determined mostly by the  
pressure, i.e., Jeans smoothing, which is determined by the temperature history
of the gas \cite{1998MNRAS.296...44G,2001ApJ...562...52M}. A linear theory 
calculation of the 
smoothing scale for reasonable thermal history, following 
\cite{1998MNRAS.296...44G}, predicts $\sim 100 \hkpc$ or 
slightly more (this is the rms smoothing length for a Gaussian kernel applied 
to the density field).   
\cite{2003ApJ...585...34M} used the hydro-PM (HPM) approximation of 
\cite{1998MNRAS.296...44G} to include pressure in the non-linear evolution of
otherwise PM simulations and found results consistent with smoothing PM by a 
smaller amount, $\sim 40 \hkpc$.
We know of no reason to doubt the qualitative accuracy of the HPM simulations, 
but ultimately fully hydrodynamic simulations will
be required to decisively compute the expected value of $\beta$ for
a given model. 
\cite{2010arXiv1010.5250M} plotted the 3D power spectrum from a hydro 
simulation, and suggested $\beta$ could be $\sim 1$, but their single 
$25 \hmpc$ simulation box had too few modes in the comfortably linear regime
to assess the accuracy of this result.
Finally, even if we were sure of our calculation for a given thermal history,
there is some uncertainty due to uncertainty about the thermal history, 
especially the redshift of reionization (we have not quantified how large this
uncertainty is).

\begin{figure}
  \begin{center}
    \includegraphics[width=0.7\linewidth]{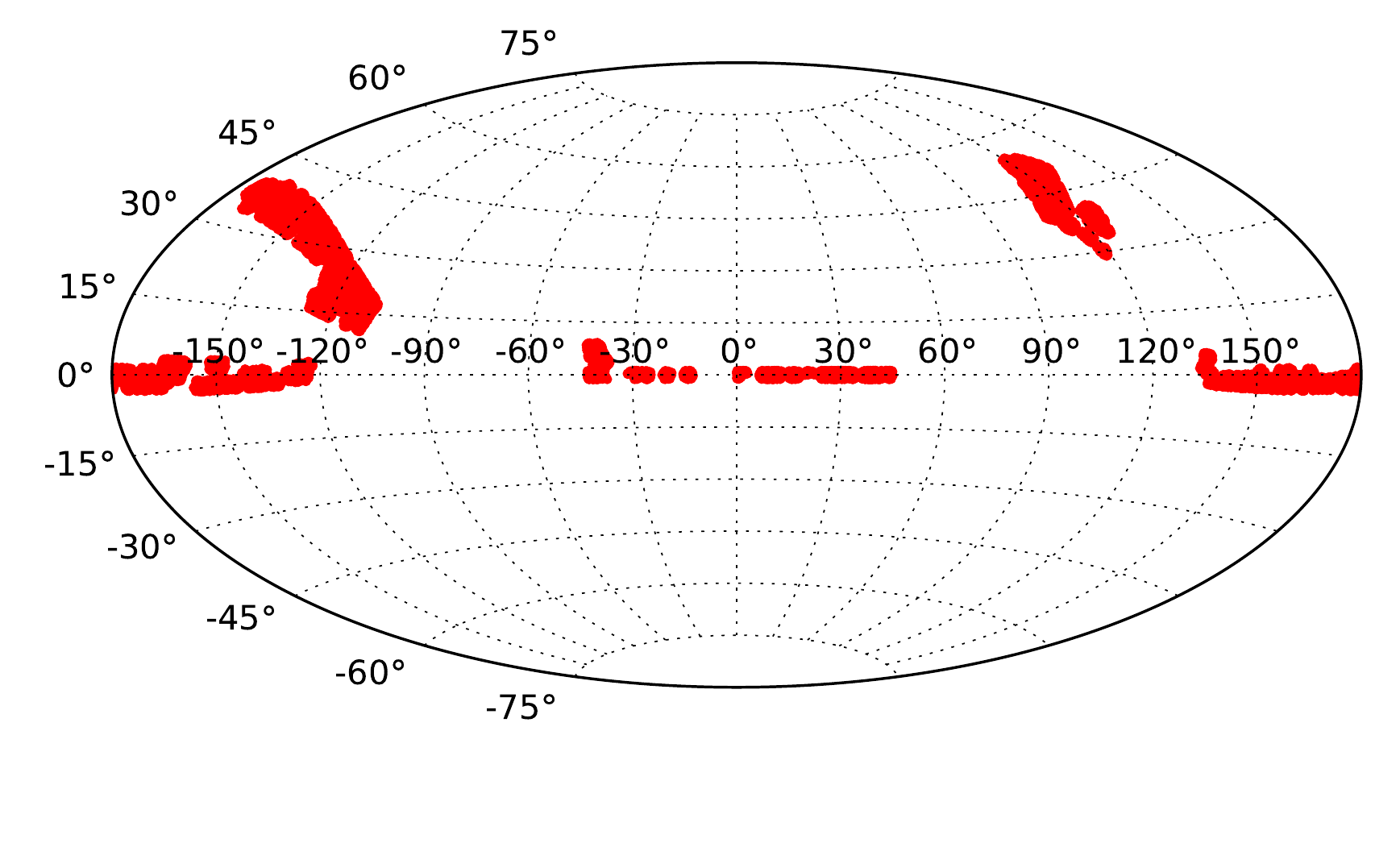}
  \end{center}
  \caption{\small\label{fig:proj} Survey area of the quasars used
in this paper in equatorial coordinates, in the Aitoff projection. }
\end{figure}

\begin{figure}
  \begin{center}
    \includegraphics[width=1.0\linewidth]{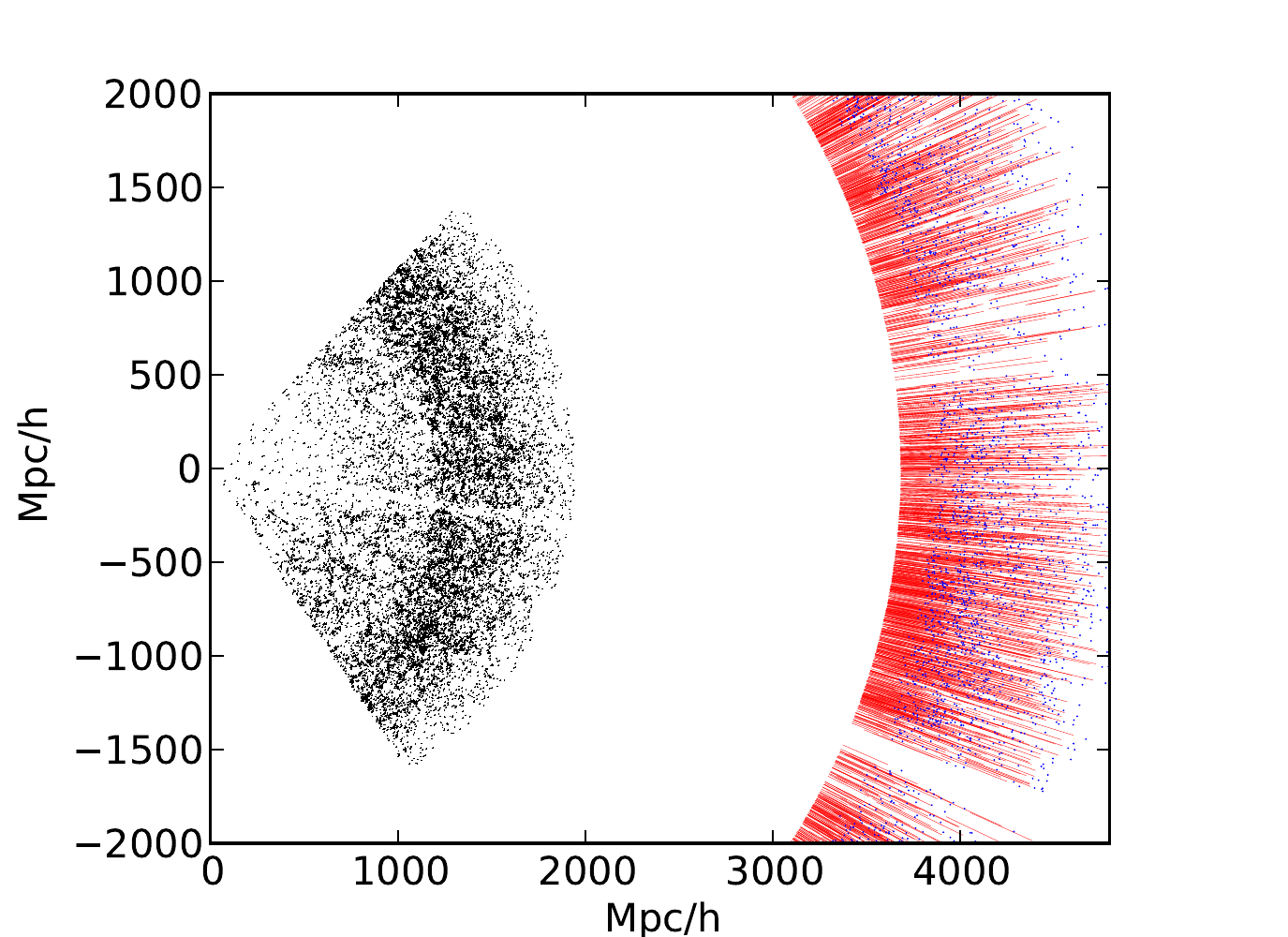}
  \end{center}
  \caption{\small\label{fig:wplot2} The geometry of the BOSS survey
    for a thin slice in the equatorial plane. Our galaxy is at the
    origin. The dark dots are the galaxies measured in the BOSS survey
    and the blue markers show the positions of quasars whose \lya
    forests we use. The actual \lya forest regions are shown in as red lines. 
    Apparent differences between geometry of quasar and galaxy
    distribution arise from small differences in slice thickness and 
    time-span. }
\end{figure}

\begin{figure}
  \begin{center}
    \begin{tabular}{cc}
    \includegraphics[width=0.45\linewidth]{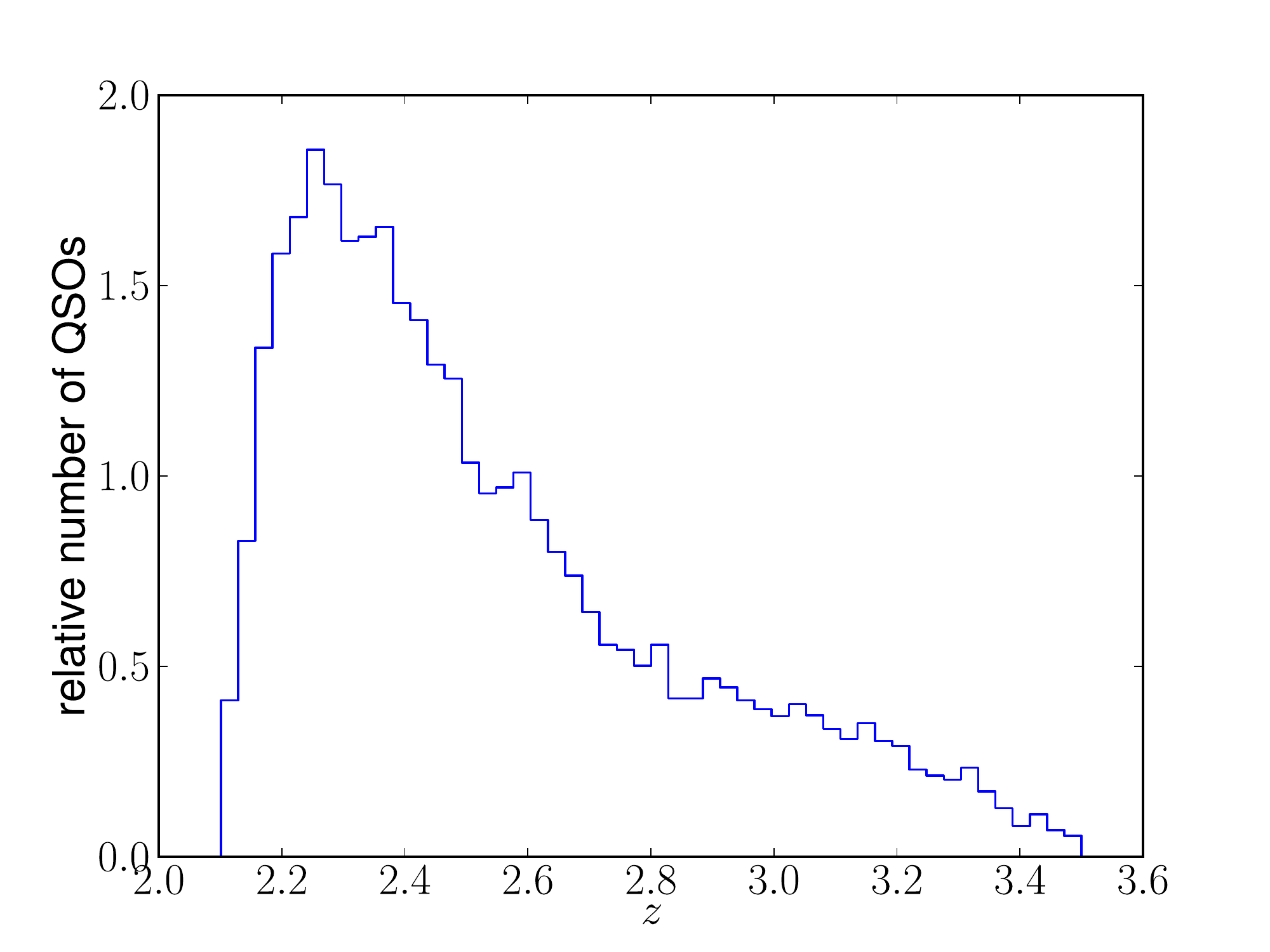} &
    \includegraphics[width=0.45\linewidth]{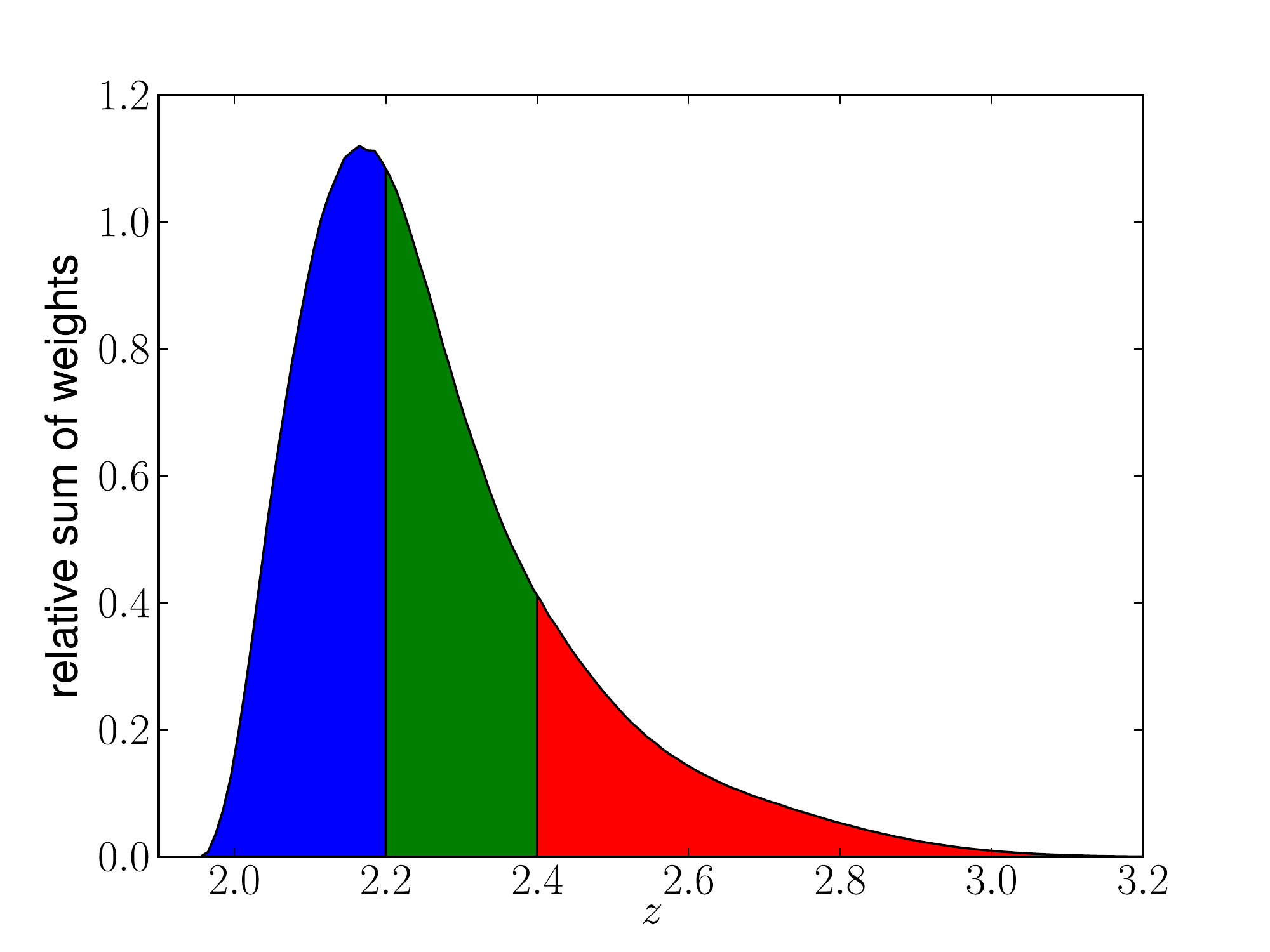} \\
    \end{tabular}
  \end{center}
  \caption{\small \label{fig:Nofz} Redshift distribution of quasars
    for the sample used in this analysis (left panel), and weighted
    distribution of \lyaf pixel redshifts for the three redshift bins
    considered in this paper (right panel). The quasar redshifts are
    cut to be between $2.1<z<3.5$. The pixel weights are limited by
    the UV coverage of the spectroscope at low redshift end and by the
    redshift of quasars at high redshift.  We show the sample of
    quasars without DLA flag; however, the two plots are virtually
    identical for the sample in which DLA flagged quasars are
    included.}
\end{figure}

At the percent level needed to interpret 1D power spectrum measurements, the 
value of the bias parameter $b$ is well known to depend on the 
mean absorption level (shown directly in \cite{2003ApJ...585...34M}), which is 
difficult to determine accurately; however, at the level of this paper it is 
actually quite precisely predicted, with \cite{2003ApJ...585...34M} and 
\cite{2010ApJ...713..383W} (verified by Martin White, private communication) 
agreeing on a value $\sim 0.13$ to $\sim 10$\% with
less sensitivity to smoothing than for $\beta$ 
(\cite{Slosar:2009iv} did find a higher value $\sim 0.18$ for their much
lower $500 h^{-1} {\rm kpc}$ resolution simulation).

In addition to quasar pair measurements, there have been some studies
of closely spaced groupings of quasars that provide hints of
three-dimensional structure in the \lya forest
\cite{2000ApJ...532...77W,Rollinde:2003tf, 2004ApJ...613...61B,
  2006MNRAS.372.1333D,2006MNRAS.370.1804C,
  2008AJ....136..181C,2008ApJS..175...29M,2009MNRAS.392.1539T,
  2010MNRAS.407.1290C}.  However, large scale flux correlations are
weak, so detecting them with high significance across widely separated
sightlines requires a large and dense quasar sample.  By design, the
BOSS quasar survey provides precisely such a sample, probing a large
comoving volume with $\sim 15$ sightlines per deg$^2$.

The ultimate goal of the BOSS survey is to measure the baryon acoustic
oscillation (BAO) feature to high precision in the {\it galaxy}
redshift survey at $z<0.7$ and \lya forest at $z\approx 2.5$
\cite{2011arXiv1101.1529E}
(the possibility of measuring the BAO feature using the \lyaf\ was
suggested in \cite{2003ApJ...585...34M} and some preliminary
calculations were done by \cite{white03}, but the potential of the
measurement was not really quantified until \cite{2007PhRvD..76f3009M}
(see also \cite{Slosar:2009iv,2010ApJ...713..383W,2011arXiv1102.1752M}).  
These measurements will be used as a
standard ruler to measure the angular diameter distance $D_A(z)$ and
Hubble parameter $H(z)$. The first proof-of-concept paper for
the galaxy part of the survey is \cite{2011ApJ...728..126W}, and this
paper attempts to do the same for the \lyaf part of the survey.  The
sample of 14,000 quasars analyzed in this paper is too small to yield
a confident detection of the BAO feature, but it does allow the first
precise measurements of three-dimensional structure in the \lya
forest.  The basic statistic that we use is the \textit{flux
  correlation function}
\begin{equation}
\label{eqn:xidef}
\xi_F(r,\mu) = \langle \delta_F({\bf x})\delta_F({\bf x}+{\bf r})\rangle ~,
\end{equation}
which is the Fourier transform of the power spectrum~(\ref{eqn:kaiser}).
Our measurements provide novel tests of the basic cosmological
understanding of high-redshift structure and the nature of the
\lya forest.

The next section describes our data sample, based on quasars observed
by BOSS between December 2009 and July 2010.  During these first few
months of BOSS, which included commissioning of new hardware,
software, and observing procedures, data taking was relatively
inefficient.  Our sample of 14,598 $z>2.1$ quasars over $\sim 880$ square
degrees is less than 10\% of the anticipated final sample of 150,000
quasars selected over 10,000 deg$^2$, but it is already comparable to
the total number of $z>2.1$ quasars found by SDSS-I and II
\cite{2010AJ....139.2360S}.  Section 3 describes our methods for
creating realistic synthetic data sets, which are crucial to testing
our measurement and error estimation procedures and which we use as a
source of theoretical predictions for comparison to our measurements.
Section 4 describes our analysis procedures including removal of
continuum, estimation of the flux correlation function and its
statistical errors, and model fitting, presenting detailed tests of
these procedures on the synthetic data.  Section 5 presents the main
results of the BOSS analysis.  We describe several internal tests for
systematic errors in \S 6, and we summarize our results  in \S 7 with
final remarks given \S 8.

\section{Data sample}
\label{sec:data-sample}

In this section we give a brief overview of the BOSS Quasar data
sample that was used in our analysis, and the French Participation
Group (FPG) visual inspections of the BOSS spectra. The quasar target
selection for the first year of BOSS observations is described in
detail in \cite{rossprep}, which draws together the various methods
presented in \cite{2009ApJS..180...67R, 2010A&A...523A..14Y} and
\cite{kirkprep}. In summary, selecting $z>2.1$, and in particular
$2.1<z<3.5$ objects, has always been challenging due to the
colours of quasars at these redshifts crossing over the stellar locus
in, e.g., SDSS {\it ugriz} optical colour-space
\cite{2002AJ....123.2945R}. Therefore, the ``UV Excess'' method, generally
used for $z<2$ objects, fails. As such, the $2.2<z<3.5$
redshift range was sparsely sampled in the original SDSS,
and the number density of $z>3.5$ objects was very low
($\sim$ few deg$^{-2}$).

In BOSS, selecting $2.2<z<3.5$ is key to meeting our science
requirements. We have the advantage that when one selects the
quasars only as backlights for the detection of neutral hydrogen,
one is at liberty to use a combination of methods to select 
quasars without having to understand the completeness of the sample in
any variable relevant for studying quasar properties or clustering. In
total 54,909 spectra were taken by BOSS that were 
targeted as high, $z>2.2$, quasar targets between Modified Julian
Dates (MJDs) 55176 (11 Dec 2009) and 55383 (6 July 2010). From this
sample, 52,238 are unique objects, and have SDSS PSF magnitudes in the
range $18.0<r<22.0$.  Of these 52,238 objects, 13,580 (14,810) quasars have
$z\geq 2.20\ (2.10)$ and a secure redshift as given by the BOSS
spectro-pipeline.

Although the BOSS spectro-pipeline is secure at the 90-95\% level, a
non-negligible number of objects are classified as QSOs when they are
stars and vice-versa.  Because of this, the nearly 55,000 objects were
also visually inspected by one, or more, of the BOSS team members. These
inspections are generally referred to as the FPG inspections.
The manual inspections of the BOSS spectra give a further level of
confidence in the data, and result in a net gain of real high-$z$
quasars that the pipeline does not select, at the level of $\sim 1-3$
objects per 7 deg$^2$ plate.  Moreover, the visual inspection has the option to
manually label high-$z$ quasars which have either Damped
Lyman-$\alpha$ (DLA) or Broad Absorption Line (BAL)
features. Currently, this option is binary in the sense that the
visual inspector has the option of setting the BAL and DLA flags but no
attempt is made to measure $N_{\rm H I}$ column density or the
position of the absorber. We have checked that the flagged DLAs have
log~$N$(H~{\sc i})(cm$^{-2}$)~$> \sim 20.3$. The list is not complete
however, since DLAs are difficult to detect along lines of sight of
SNR~$<$~4 \cite{2009A&A...505.1087N}. We have also checked that
the so-called balnicity
index is larger than 2000~km~s$^{-1}$ (see
e.g. \cite{2006ApJS..165....1T}). The removal of DLA systems in our
data hence still depends on a human factor and signal-to-noise ratio
and is currently not a very well quantified process. This will become
better defined in future work.

For the analysis presented here, those objects labeled by the FPG
visual inspections as ``BAL'' are not used, and we don't include objects
labeled as ``DLA'' unless mentioned specifically.

In our sample we restrict the analysis to quasars from the FPG sample
with $2.1<z<3.5$. Compared to the targeted science sample described
above, this sample has no magnitude cut, expands the lower redshift to
$z=2.1$ and uses some quasars that are on plates nominally marked as
``bad'' by the pipeline, but contain useful data based on manual
inspections.  After removing approximately 1300 quasars marked as
BALs, our complete sample contains \hbox{14,598} quasars, of which
\hbox{13,743} are not labeled as having a DLA.  Several quasars in the
first year sample had repeated observations - we do not use these
repeats and limit ourselves to the best quality observation (as
determined by FPG). The FPG visual inspections are continuously being
updated and revised.  We froze the data used on 1$^{\rm st}$ October
2010.

Further details about the target selection and details about the
instrument and the observational procedure can be found in 
\cite{rossprep}.

Figure~\ref{fig:proj} shows the position of the BOSS Year One quasars
on the sky in equatorial projections in the Aitoff
projection. Figure~\ref{fig:wplot2} shows the geometry of BOSS probes
of large scale structure in comoving
coordinates. Figure~\ref{fig:Nofz} gives the redshift distribution of
the $z>2.1$ quasars and the weighted histogram of pixel redshifts
contributing to our correlation measurement (explained and discussed
later), as a function of redshift.

\section{Synthetic data}

To test our data reduction pipeline and to examine the impact of
various systematics, we  created extensive synthetic
datasets. These synthetic data shared the geometry with the observed set
of quasars, i.e. the positions and redshifts of quasars were identical, 
and we used the observed SDSS $g$ magnitudes to normalize the spectra.

The statistical properties of synthetic data matched our initial estimate
of the properties of the underlying flux field, including the non-linear 
corrections to the linear redshift-space distortions 
\cite{2003ApJ...585...34M}. 

The \lya absorption spectra of each quasar were generated using a
method described below to obtain the variable $\delta_F = F/\bar F - 1$,
where $F$ is the fraction of transmitted flux, 
at each
spectral pixel from a realization of the underlying cosmological density
fluctuations, including linear redshift-space distortions and non-linear
corrections to the flux power spectrum model that is used as input.
A total of 30 different realizations of this underlying cosmological
field were generated. For each realization, the following sets of
synthetic data were created, with increasing level of realism to
evaluate the impact of various observational effects and of our data
reduction procedure:

\begin{itemize}
\item Noiseless measurements of $\delta_F$

\item Noiseless measurements of \lya forest with continua

\item Noisy measurements of \lya forest with continua

\item Noisy measurements of \lya forest with continua and
  high-absorption systems

\item Noisy measurements of \lya forest with continua and forest metal
  contamination

\item Noisy measurements of \lya forest with continua and forest metal
  contamination and high-absorption systems

\end{itemize}

Note that continua here means variable
continua using randomly generated PCA eigenmode amplitudes
\cite{2005ApJ...618..592S}, instead of the same
mean continuum for every QSO.

In the following subsections 
we briefly describe how these synthetic data are
actually generated.  The generation of \lya absorption and
associated structure will be discussed in more detail in \cite{fontprep}.

\subsection{Absorption field}

\subsubsection{Generation of a Gaussian field}

A \lyaf survey samples the intrinsically three-dimensional flux
transmission fraction field along a set of infinitesimally thin lines
of sight to quasars.  The brute force method for generating a survey
with correct correlations would be to generate the full 3D field with
sufficient resolution (tens of kpc), and then read off the density
along lines of sight. For a survey as large as ours, this
requires too much memory.  Fortunately, when generating a Gaussian
field we can greatly simplify the problem, maintaining exact
high-resolution statistics while only creating the field for the
 limited fraction of the volume where it is required.

To generate a Gaussian random field for a general set of
pixels with covariance matrix $\mathbf{C}$ we can Cholesky decompose
the covariance matrix, i.e. first find $\mathbf{L}$ such that
\begin{equation}
\mathbf{L}\cdot \mathbf{L}^T=\mathbf{C}.
\end{equation}
If we then generate a unit variance 
white noise field $w_j$ in the pixels and multiply it by
$L_{ij}$, the resulting field will have the desired correlation function, i.e. 
\begin{equation}
\left<\delta_i \delta_j\right>=\left< L_{ik} w_k L_{jl} w_l \right> =
L_{ik}L^T_{kj}=C_{ij}.
\end{equation}

Assuming the lines of sight are parallel allows another significant
simplification.  Suppose the field $\delta\left(x_\parallel,
  \vxperp\right)$ has power spectrum
$P\left(k_\parallel,k_\perp\right)$.  If we Fourier transform this
field in the radial direction only, the resulting modes have
correlation
\begin{equation}
\left<\delta\left(k_\parallel,\vxperp\right)
\delta\left(k_\parallel^\prime,\vxperp^\prime\right)\right>
=2 \pi~\delta^D\left(k_\parallel+k_\parallel^\prime\right)
P_\times\left(k_\parallel,\left|\vxperp-\vxperp^\prime\right|\right) ~,
\label{eq_mocktrick}
\end{equation}
where
\begin{equation}
P_\times\left(k_\parallel,r_\perp\right) =
\frac{1}{2 \pi} \int_{k_\parallel}^\infty k~dk ~
J_0\left(k_\perp r_\perp\right) ~P\left(k,\mu_k\right) ~,
\end{equation} 
where $k_\perp=\sqrt{k^2-k_\parallel^2}$ and $\mu_k =k_\parallel/k$.
The key point is that modes with different values of $k_\parallel$ 
are uncorrelated. We can generate the field efficiently by following the 
above general procedure for generating a correlated Gaussian random field,
except now independently for every value of $k_\parallel$ required for the 
radial Fourier transform. We never need to manipulate a matrix larger than 
a manageable $N_q \times N_q$, where $N_q$ is the number of
quasars. However, we do take into account the fact that lines of sight
are not fully parallel as discussed in Section \ref{sec:redsh-evol-non}.

We use this procedure to generate any desired Gaussian field $\delta$
at each pixel of each quasar spectrum of the synthetic data sets, once
we have a model for the power spectrum $P(k,\mu_k)$ of this variable.

\subsubsection{Generation of the transmitted flux fraction, $F$}

In order to obtain realistic synthetic data including noise as in the
observed spectra, it is necessary to generate the flux transmission
fraction $F$ with a realistic distribution in the range $0$ to $1$,
and add the noise to this flux variable. The solution we have adopted
is to convert the Gaussian fields $\delta$ to a new field $F(\delta)$
with any desired distribution. In this paper we use a log-normal
distribution in the optical depth, $\tau= -\log F$ (where log denotes
natural logarithm), which implies the following probability
distribution for $F$:

\begin{equation}
 p(F)\, dF = { \exp \left[ - (\log\tau - \log\tau_0)^2/(2\sigma_\tau^2)
 \right] \over \tau F \sqrt{2\pi} \sigma_\tau } \, dF ~,
\end{equation}
with the two parameters $\tau_0$ and $\sigma_\tau$ that determine the
mean and dispersion of $F$.  
We find the function $F(\delta)$ that
results in this distribution for $F$ when $\delta$ is a Gaussian
variable. The correlation function of the variables $\delta$ and $F$
are then related by the following equation:

\begin{align}
   \xi_F(r_{12}) = & \left< F_1 F_2 \right> \nonumber \\
               = & \int_0^1 dF_1  \int_0^1 dF_2 p(F_1,F_2) F_1 F_2
               \nonumber \\
               = & \int_{-\infty}^\infty d\delta_1  \int_{-\infty}^\infty d\delta_2 p(\delta_1,\delta_2) F_1 F_2 \nonumber\\
               = & \int_{-\infty}^\infty d\delta_1  \int_{-\infty}^\infty d\delta_2 \frac{ e^{-\frac{\delta_1^2 
               + \delta_2^2-2\delta_1\delta_2\xi^2(r_{12})}{2(1-\xi^2(r_{12}))}} }{2 \pi \sqrt{1-\xi^2(r_{12})}} F(\delta_1) F(\delta_2) ~ .
\end{align}

We note that this equation relates $\xi_F$ to $\xi$, independently
of the variables $r$, $\mu$ and $z$. Therefore, we simply
tabulate $\xi_F(\xi)$, and invert the relation to figure out the
correlation function $\xi$ that is required in order to obtain any
desired correlation function $\xi_F$.

  We use a model for the flux power spectrum in redshift space,
$P_F(k,\mu_k)$, which was fitted to the results of 
simulations in \cite{2003ApJ...585...34M}. We use the value
of $\beta$ for the central model in that paper, $\beta=1.58$, and a bias 
inspired by the 
central model but roughly adjusted for cosmological model (not
using the parameter dependence Table in \cite{2003ApJ...585...34M}), 
$b=0.145$, at $z=2.25$. When 
generating the synthetic data, we keep the value of $\beta$ constant as
a function of redshift, and vary the amplitude of the power spectrum
according to a power-law, $P_F \propto (1+z)^\alpha$. We compute the
correlation function $\xi_F$ from the Fourier transform, then we
calculate $\xi$ for the Gaussian field, and we Fourier transform back
to obtain the desired power spectrum for $\delta$. We generate this
Gaussian field in all the quasar spectra with the method described
above, and then transform this to the variable $\delta_F$. In this way
we obtain a set of spectra with the desired flux distribution and
redshift-space power spectrum.

We use mean transmission fraction approximately matching the observations
\cite{2005ApJ...635..761M}: 
$\ln \left<F\right>(z)=\ln(0.8) \left[\left(1+z\right)/3.25\right]^{3.2}$.

\subsubsection{Redshift evolution and non-parallel lines of sight} 
\label{sec:redsh-evol-non}

  The field $\delta_F$ is generated as described at a fixed redshift,
and assuming parallel lines of sight. This is done at four different
values of the redshift, by generating the realizations of the field
with fixed random numbers for all the Fourier modes, and changing the
amplitudes according to the different power spectra at every redshift.
The power spectrum varies both because of the intrinsic evolution, and
because the angular diameter distance used to convert angular to
physical separation between the quasars varies with redshift. In this
way, the fact that the lines of sight are non-parallel is incorporated
into the calculation in the same way as the redshift evolution of the
power spectrum. The final field $\delta_F$ is obtained by interpolating
between the fields that have been generated at different redshifts, to
the exact redshift of each point on the line of sight. This field then
has the desired local power spectrum at any redshift, evolving as it
is prescribed according to the linear interpolation.

The redshift evolution of amplitude assumed in this work is described
by a power law $(1+z)^\alpha$, with $\alpha=3.9$ as suggested by the
measured evolution of the 1D flux power spectrum
\cite{2006ApJS..163...80M}.

\subsection{Adding absorption systems: Lyman Limit Systems (LLS) and Damped 
Lyman-$\alpha$ (DLA) systems}

The highest column density absorption systems produce broad damped absorption
wings which can have a strong impact on the correlation function of the
transmitted flux. These systems are traditionally known as Damped Lyman
Alpha absorbers (DLA) and have \ion{H}{1} column densities above 
$10^{20.3}\cm^{-2}$, however, the damping wings can affect the line profile
at lower column densities as well.
Systems with column densities above $10^{17.2}\cm^{-2}$ 
are known as Lyman Limit Systems (LLS) since they are self-shielded
\cite{2002ApJ...568L..71Z}. At $10^{17.2}\cm^{-2}$, the effect of damping wings
on the profile is small, but it becomes significant well before 
$10^{20.3}\cm^{-2}$ \cite{2005MNRAS.360.1471M}. 

The impact of
these absorption systems is two-fold. First, they add noise to the measurement
of the correlation function. Second, the systems trace the underlying
mass density field, and therefore they are correlated with themselves and
with the \lya absorption arising from the intergalactic medium. This
systematically modifies the overall \lya transmission correlation function.

To simulate the effect of these systems in the synthetic data, we introduce
lines with $N_{\rm HI}>10^{17.2}\cm^2$ with an abundance consistent with 
observational constraints \cite{2004PASP..116..622P,2003MNRAS.345..480P,
2003MNRAS.346.1103P}, using the formula of \cite{2005MNRAS.360.1471M}. 
The decrease in $\left<F\right>(z=2.25)$ due to these systems is 0.014.
We also introduce a
correlation between these systems and the rest of the \lya absorption by placing
them only in regions where the optical depth is above a critical value $\tau_0$,
such that the probability of $\tau>\tau_0$ is 1\%. This is performed
to explore the way that the damped absorbers may bias the measured
correlation function, but their detailed impact depends on the specifics of
this correlation. We leave for a future analysis a better
modeling of the cross-correlation of damped absorbers and the \lya forest.

In the rest of the paper we refer to these contaminants as LLS/DLA.

\subsection{Adding forest metal contamination}
\label{subsec:formet}

Absorption by intergalactic metals imprints correlated signal in quasar spectra 
on characteristic scales. These scales are 
 set by the wavelength ratios of metal line
transitions with \lya and with each other. As a
result, this correlated signal is a potential contaminant of
large-scale structure measurements in the forest.  In order to add forest
metal absorption to the synthetic data we measure the strength of
metals. We do this in a self-consistent manner by measuring metal line absorption
in the continuum normalized BOSS Year One spectra (excluding spectra
with known DLAs). We use a modified version of the method set out by
 \cite{2010ApJ...724L..69P} and measure these lines by stacking absorption 
 lines between 1041\AA--1185\AA\ in the quasar rest-frame. We measure
 the signal associated with metal lines correlated with \lya or \ion{Si}{3} in
 the forest. This is described in Appendix~\ref{sec:appMet}, including a
 look-up table of flux decrements measured in the composite spectra.
 
We introduce these metal absorption features to the full suite of
mock data with no noise and no LLS/DLAs, on a pixel-by-pixel
basis; we walk
through the spectra and lower the transmitted flux by values from
interpolation of this table, and then add Gaussian noise. 
As a result, we lower the mean transmitted flux by 0.003 in the mocks.
This approach assumes that metal lines
trace \lya structure monotonically, and as such these 1D
correlations will add metal structure to our 3D analysis. Full line profiles
are recovered by virtue of our easing of the local minimum requirement
(metal absorption is added at the wings of \lya profiles as well as
the center). 

This technique does not provide a full description of metal
absorption and, in particular, neglects the impact of scatter in
metallicity and metal complexes. 
We test these mocks by stacking them in the same manner used for
the BOSS Year One spectra. The metal correlations imprinted in the
noise-free mocks are reasonably well recovered in these composite spectra.
Composite spectra of the mocks with noise added (after metals have been
introduced)
show metal correlations that, where measurable, are up to 10\% weaker 
than those seen in 
the observed data or added to the mocks in all cases except the \ion{Si}{3}
line seen in the strongest bin which is 30\% weaker.
This is caused
by a combination of Gaussian noise and the probability distribution
function of the flux. The noise distribution is symmetrical but, in the
relevant regime, there are always
far more pixels in the higher flux bin, which have weaker associated metal
absorption.
We conclude that metals added are probably an underestimate of the average 
metal 
absorption associated with \lya lines, but these results are sufficient for
an exploration of the approximate impact of forest metals.
It seems
unlikely that LLS/DLA interlopers are able to produce the observed metal signal for 
the reasons given in \cite{2010ApJ...724L..69P}, however, 
we shall explore this issue further in future publications.
We also combine forest metals with LLS/DLA corrected mocks by
introducing these high column lines
after forest metals have been added. Hence only metals associated 
with the \lya forest are included.

\subsection{Generating the spectra}

Once we have created an absorption field for every line of sight, we
proceed to generate the actual spectrum for each quasar, multiplying
it by the ``continuum" of the quasar, i.e. the unabsorbed spectrum.

We generate each quasar continuum shape taking into account the quasar redshift
and using a mean rest-frame continuum and random PCA components derived from 
the low-redshift Hubble data \cite{2005ApJ...618..592S}.
The continuum is  then
normalized using the $g$ magnitude of the quasar (taking into
account the \lyaf absorption).  

Since our data are sampled on precisely the same grid as the
observed data, we can also introduce noise by using the actual values
of noise from the observed data. We assume noise to be Gaussian with the
absolute flux variance given by the pipeline, i.e., we do not
correct the signal-to-noise ratio for differences between our randomly
generated continuum level and the data level.

Because the mocks were generated before the data analysis procedure was
finalized, we have mocks only for quasars with redshift $>2.2$,
while the data analysis uses $z_q>2.1$.

\section{Data analysis}
\label{sec:data-analysis}

In this section we describe the analysis applied to both real and
synthetic data. Briefly, the steps involved in this analysis start
with co-adding the multiple spectra of each quasar. We then fit a
model for the mean quasar continuum and mean absorption from the whole
set of spectra. This is used to determine the fluctuation in the
fraction of transmitted flux, $\delta_F$, from the observed flux, in
each individual spectrum over the \lya forest range. The correlation
function is then measured from these flux fluctuations.  The
information on the distribution of datapoint pairs and the measured
correlations feed into the code that estimates the errors on our
correlation function.  With the estimated correlation function and
error-covariance matrix, we finally proceed to estimate the parameters
in our model of the correlation function, in particular the two bias
parameters. The next subsections explain in detail each of these
steps.

\subsection{Preparing the data}


All spectra in this analysis were obtained with the BOSS spectrograph
on the Sloan 2.5-meter telescope at Apache Point Observatory
\cite{2006AJ....131.2332G}.  The BOSS instrument follows the design of
the spectrograph previously used in SDSS, but with improved
throughput, wider wavelength coverage, reduced sky background from
smaller fibers and a larger number of fibers.  A full description of the
instrument will be reported in a future paper \cite{smeeprep}.

Observations are performed using 15-minute exposures that are
processed immediately with a simple data reduction package.  These
quick reductions provide estimates of data quality that are evaluated
by a pair of observers at APO to insure roughly uniform depth over the
full survey.  The observers cease exposures when estimates of the
accumulated SNR exceed a minimum threshold, leading to an average
exposure time of approximately 75 minutes per plate.  Within 24 hours
of completing a night of observations, the images are fully reduced
into wavelength-calibrated, flux-calibrated, one dimensional spectra.
Adapted from the SDSS-I/II spectroscopic reduction pipeline, version
\texttt{v5\_4\_14} of the BOSS pipeline is used to generate the spectra in this
analysis.  An updated version of the BOSS pipeline \cite{pipeprep} is
used for the first two years of spectra that will be publicly released
in July, 2012.

The modeling of the noise associated with the flux measured at each
wavelength is particularly important for measurements of correlated
Lyman-$\alpha$ absorption.  Noise is characterized for each pixel in
the raw images early in the Spectro-2D stage of the BOSS data
reduction pipeline.  After bias removal, the flux in each pixel is
converted from analog digital units into photon counts using CCD
amplifier gains that were determined independently during the
commissioning of the instrument.  We generate a corresponding image to
represent the variance by taking the sum of square of the readnoise
(measured in the overscan region of each quadrant) and the
gain-corrected flux measured at each pixel for which the flux is
positive.  Pixels that fluctuate to negative flux values are assigned
a variance from the readnoise only.  The relative pixel-to-pixel
response in the data is corrected with a flat-field image and the
variance for each pixel is rescaled accordingly.  For each pixel, we
then invert the variance into units of inverse variance, and set the
inverse variance equal to zero for all pixels that are contaminated by
cosmic rays or known cosmetic defects.

We transform each two-dimensional image into a set of one-dimensional
spectra with an optimal extraction \cite{1986PASP...98..609H}.  The
flat-field exposure is used to determine the shape of the profile and
the fiber-to-fiber variations in throughput.  The arc lamp image is
used to determine the wavelength solution, with linear corrections
made in each exposure to match sky line locations.  The profile is fit
to each spectrum in the two-dimensional image weighted by the inverse
variance.  Each extracted spectrum and corresponding inverse variance
vector are corrected to ensure a uniform response curve as determined
from the fiber-to-fiber variations in throughput mentioned above.  All
corrections to the inverse variance and flux vector preserve the SNR
for each pixel.  Sky subtraction is performed on each individual
exposure using a model for background that varies with fiber position.
The model is determined from the one-dimensional spectra extracted
from fibers that were intentionally set to regions where no objects
had been detected in the SDSS imaging program.  The inverse variance
of all fibers is rescaled by a factor to force the error distribution
of all sky fibers at each wavelength to have a normal distribution
with unit variance.  In the Lyman-alpha region of the spectra, this
typically scales the errors by a factor of 1.1, and this scaling is
never allowed to decrease the errors.  This effectively corrects for
systematic uncertainties in the sky modeling and subtraction. We also
mask spectra around  bright atmospheric emission lines, such as
\ion{Hg}{0} (5460\AA) and \ion{O}{1}  (5577\AA).

At the final step in the Spectro-2D stage of the BOSS reduction
pipeline, the spectra derived from individual exposures are combined
into a coadded spectrum for each fiber.  The Spectro-1D pipeline
spectroscopically classifies and assigns redshifts to each object
based upon a minimum chi-squared fit to a library of PCA templates
derived from BOSS data.  

In this analysis, we use the spectra output from the Spectro-2D
pipeline before the coaddition.  Instead of using the resampled coadds
from Spectro-2D, we co-add the data using the same fixed wavelength
grid as the pipeline, but combining the flux values of the closest
pixels in wavelength from the individual exposures. This ensures, in
the simplest possible way, that the noise of the co-added data is
independent from pixel to pixel, at the expense of a poor treatment of
the small scale fluctuations in the data. Since we are interested in
large-scale correlations, we are not concerned about these effects. In
each pixel the data are coadded using inverse variance weighting. We
apply a small correction to the final variance in each pixel, which
typically increases it by less than 10\%, to ensure that the
inter-exposure variance in pixels is consistent with the noise
predicted by the pipeline for individual exposures.

\begin{figure}
  \begin{center}
    \includegraphics[width=0.7\linewidth]{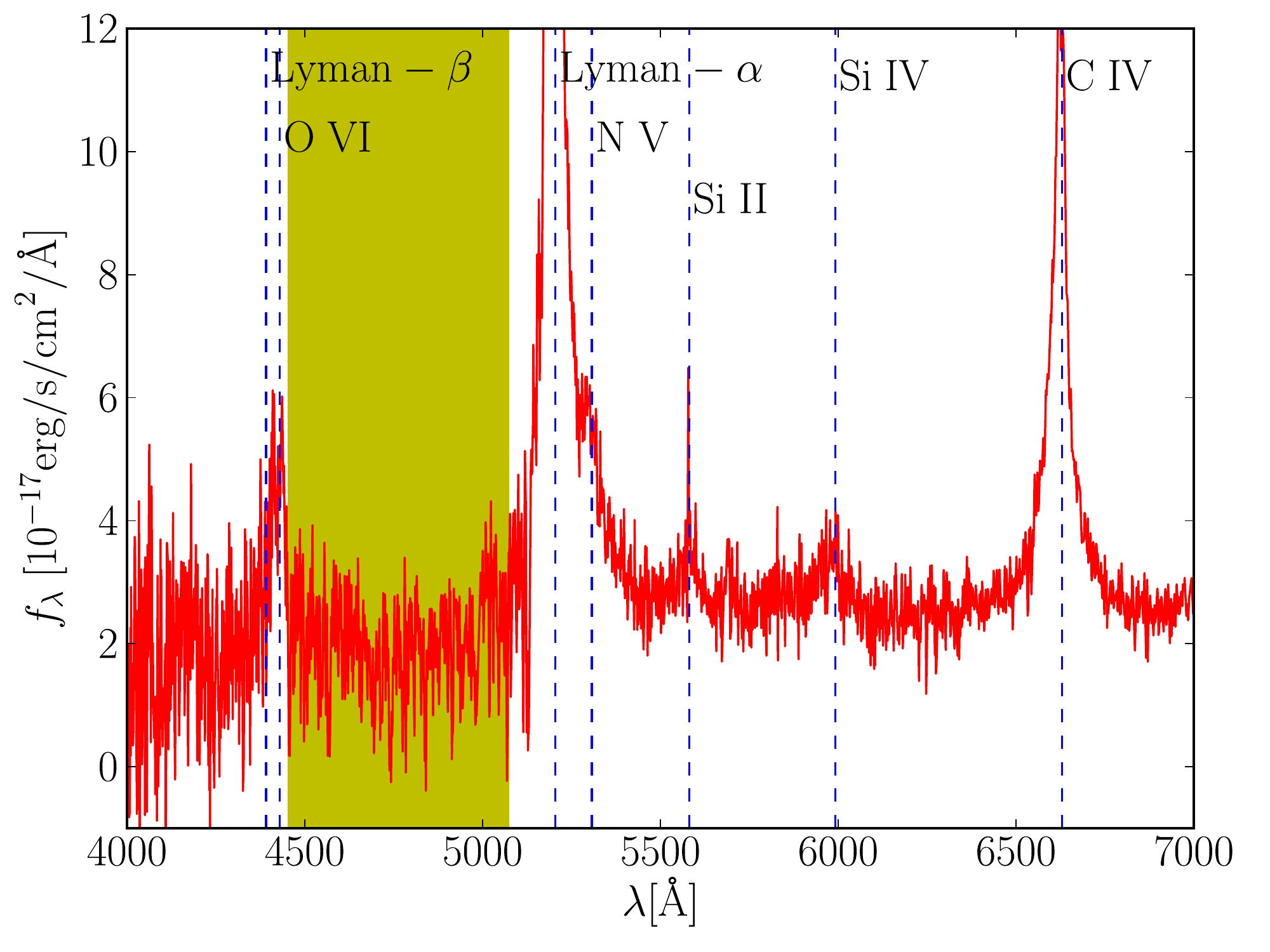}
  \end{center}
  \caption{\small\label{fig:spec}A quasar spectrum
    at redshift $z=3.276$ from the first year of BOSS data. The \lyaf is
    the shaded region between the \lya and the \lyb emission lines. Other
    strong emission lines are also indicated.}
\end{figure}

Typical values of observational parameters for BOSS quasars used in
this work can be found in the Table \ref{tab:typ}.  An example of a
BOSS spectrum after this reduction is shown in Figure
\ref{fig:spec}. This spectrum has a somewhat higher than typical SNR
of quasars in our sample.

\begin{table}
  \centering
  \begin{tabular}{ccc}
    Parameter & Value\\
\hline
Total number of QSOs & 14,598 ($\sim$ 160,000 in the complete survey) \\
QSO sky density & 10-25 deg$^{-2}$ (typically $\sim$15 deg$^{-2}$)\\
QSO redshift range & 2.2 - 3.5 \\
Spectral resolution & $\sim 2.5 {\rm \AA} \sim 2.5 \mpch$\\
Pixel scale & $\sim 1.0 {\rm \AA} \sim 1.0 \mpch$\\
Signal to noise ratio per pixel & 0.5 - 10 (typically $\sim$ 1)\\
Exposure times &  Until SNR thresholds is  reached, typically 5$\times$15 minutes  \\
g-band magnitude  & 18-22 (typically $\sim 21$) \\
  \end{tabular}
  \caption{Typical observational parameters for BOSS quasars}
  \label{tab:typ}
\end{table}

\subsection{Continuum fitting}
\label{sec:cont-fit}

The next step in analyzing the data is to fit the continuum. Our 
model for the flux measured at wavelength $\lambda$ in quasar $i$ at
redshift $z_i$ is as follows:
\begin{equation}
  f(\lambda,i) =  a_i\, [\lambda_r/(1185{\, \rm \AA})]^{b_i}\,
  C(\lambda_r) \bar{F}(\lambda) 
  \left[1+\delta_F (\lambda,i)\right] ~,
\end{equation}
where $\lambda_r=\lambda/(1+z_i)$ is the rest-frame wavelength.
The term $C(\lambda_r)$
denotes the mean rest-frame quasar spectrum, which is multiplied by a
power law $a_i [\lambda/(1185\,{\rm \AA})]^{b_i}$, where $a_i$ and $b_i$
are two parameters determined for each individual quasar. The power-law
is included to allow for large-scale spectro-photometric errors
that are present due to imperfect sky subtraction and calibration, as
well as for any intrinsic variation in the quasar spectra. The
term $\bar{F}(z)$ describes the mean absorption in the forest as a
function of redshift.  The entire product $a_i \lambda^{b_i}
C(\lambda/(1+z_q)) \bar{F}(z)$ is a well constrained quantity, but
individual terms are correlated. For example, we can multiply $a_i$
and $\bar{F}(z)$ by a certain factor and absorb it into
$C(\lambda/(1+z_q))$; this is also true for the power-law parameter
$b_i$. Consequently, some quantities ($\bar{F}$, $a_i$, etc.) are
determined only up to an overall normalization constant.

For the purpose of determining our fit to the mean quasar continuum
and mean transmission, and the parameters $a_i$ and $b_i$,
we calculate an overall likelihood of all the data without taking into
account the correlations of the forest and the continuum fluctuations
in different spectral pixels.
This simplification allows us to write this simple
expression for the likelihood function, 
\begin{equation}
  \log \mathcal{L} = \sum_i \sum_{\lambda} \left[ -
     \frac{ \left[ f(\lambda,i) - a_i [\lambda_r/(1185\, {\rm \AA})]^{b_i}\,
     C(\lambda_r) \bar{F}(z)\right]^2 }{2\sigma^2(i,\lambda)} -
    \log \sigma(i,\lambda) \right] + {\rm const.}.
\end{equation}
Here, $\sigma(i,\lambda)$ is the sum of the variances due to
measurement noise and the \emph{intrinsic variance}, i.e., it contains both
variance due to continuum errors as well a variance due to small scale
flux fluctuations. This quantity is required to be a function of the
observed wavelength only in the region of the \lya forest (or
equivalently, the \lya forest redshift), while outside the forest we
force it to depend only on rest-frame wavelength, which implicitly
assumes that the dominant source of variation in that region
is continuum errors.
We do not necessarily believe his assumption is true, but this 
allows in one sense optimal weighting of different parts of
the quasar spectrum outside the forest, i.e., prevents certain
high-variance parts of the rest-frame spectrum from excessively
affecting the solution. In the long term, this assumption will have
to be relaxed.

We restrict the rest-frame wavelength range used for this fit between
$1041\, $\AA\ and $1600\, $\AA , and we also discard the data at an
observed wavelength bluer than $3600\, $\AA.  We parameterize
$C(\lambda_r)$ using 20 equally-spaced spline nodes inside the forest
($1041\, $\AA\ to $1185\, $\AA), and 20 equally-spaced spline nodes
outside the forest ($1185\, $\AA\ to $1600\, $\AA), with one point
fixed to the center of the \lya emission line. We also use 8 spline
nodes in redshift for the mean transmission fraction $\bar F$, another
8 spline nodes to describe the scatter in the forest as a function of
redshift, and another 20 spline nodes to describe the rms 
variations with rest-frame wavelength outside the
forest. The full model used in continuum fitting is therefore described by 76 
global
parameters and 2 parameters for each quasar, $a_i$ and $b_i$.

Since we are completely ignoring the correlations between pixels, the
numerical values of $\sigma^2(i,\lambda)$ and $\chi^2 = -2 \log
\mathcal{L}$ have no physical interpretation. They should be thought
of as merely assisting the fit to naturally incorporate the variances
in the data. Therefore, it is not crucial that $\sigma^2$ is
parameterized by redshift inside the forest and rest-wavelength
outside.
A possible criticism of our fitting procedure is that redward of the
Lyman-alpha emission, 20 nodes of the spline fit are not enough to
resolve the full structure of the mean continuum at the resolution of
the BOSS spectrograph. However, these points are used only to assist
with determining the values of $a_i$ and $b_i$ (the normalization and power
law slope, relative to a mean continuum), and hence our
determination of continuum must be only good enough to broadly
describe the shape in that region.

We determine all the parameters using a specially crafted Markov Chain
Monte Carlo (MCMC, see e.g. \cite{2002PhRvD..66j3511L}) procedure to
find the maximum of $\mathcal{L}$.  The global parameters are first
varied, while keeping $a_i$ and $b_i$ fixed, taking several hundred
accepted samples (steps in the chain) of the MCMC. Next, the global
parameters are fixed and we vary just $a_i$ and $b_i$ for one quasar
at a time, looping over all quasars.  At this step, only the model for
the particular quasar under consideration matters, and hence
likelihood evaluations are extremely fast. We then go back to varying
the global parameters and repeat the process for several accepted
samples. This is iterated about 50 times (we have tested that using
just 20 iterations negligibly affects the final results), and we
finally take the sample with the optimal fit. The proposal function of
the MCMC process learns the degeneracy directions by using the
covariance of previously accepted samples, separately for the global
parameters and the quasar-specific parameters.  This MCMC process is
simply used here as a functional maximizer, i.e., we take just the
most likely sample.  The process takes about 36 hours on an 8-core
workstation.

\subsection{Determination of $\delta_F$}
\label{sec:determining-delta_F}

  The next step is determining the flux variation $\delta_F$, reducing
the resolution of our data and determining the weighting factors that
are used for calculating the correlation function.
First, the continuum is predicted at each pixel of each quasar
spectrum. Then the observed flux and continuum are averaged over four
consecutive pixels, weighted with the inverse variance. This provides
coarser pixels and an estimate of the noise on the value of flux in each
rebinned pixel, with a reduced resolution.
The typical length of these rebinned pixels is $\sim 3\mpch$.
We assume that the total variance in a coarse pixel is the sum of the
noise plus the intrinsic
variance in the flux field, which is a function of redshift.
We then iteratively determine:
\begin{itemize}
\item The intrinsic variance in the rebinned pixels, determined as a
  function of redshift by splining over 8 points in redshift.
\item A factor by which we multiply our estimate of the continuum in
  each quasar, which ensures that the flux averaged over the entire
  \lya forest of the quasar equals the mean continuum averaged over the
  same interval. This average is calculated by weighting with the
  inverse total variance in each pixel.
\end{itemize}

We find that by forcing the average mean flux over the quasar forest
to match that of the average continuum, we decrease the variance in
the correlation function estimate, although we also erase all modes
along the line of sight of wavelength larger than that of the size of
the forest. The way that this elimination of the mean flux variation
in each quasar alters the measured correlation function can be easily
modeled, as discussed in Appendix \ref{sec:appDC}. The upshot is that
we can correct the predicted correlation function for this effect
using a simple theory with one parameter $\Delta r$, which is the
effective \lyaf length.  In the limit of $\Delta r \rightarrow
\infty$, the correction disappears, as expected. The value of $\Delta
r$ can be predicted from first principles, but we let it be a
nuisance parameter as explained later.

Finally, we find the average value of the fluctuation $\delta_F$ at
each fixed observed wavelength by averaging over all quasars, and we
then subtract this average from all the values of $\delta_F$ in each
individual quasar. The average is calculated in wavelength intervals
$\Delta \lambda / \lambda_{Ly\alpha\ } = 0.001$.  This procedure,
which removes purely radial modes, helps remove some systematics of
imperfect sky subtraction or Galactic CaII absorption
\cite{2008A&A...487..583B} that occur at a fixed wavelength. The
analysis of synthetic data demonstrates that this negligibly impacts
the measured \lya correlation properties.  Finally, we make a cut on
pixels in $\delta_F$ that are over $6-\sigma$ away from zero in terms
of the total variance. A handful of pixels were removed that way, but
this cut makes negligible change in the resulting correlation
function.

\subsection{Estimation of the correlation function and its errors}
\label{sec:estim-corr-funct}

We use the trivial sub-optimal estimator of the correlation function
as simply the weighted average over pixel pairs,

\begin{equation}
  \bar{\xi_F} (r,\mu)  = \frac{\sum_{{\rm pairs}\ i,j} w_i w_j\,
 \delta_{Fi} \delta_{Fj}}{\sum_{{\rm pairs}\ i,j} w_i w_j},
\end{equation}
where the weights $w_i$ are the total inverse variance (from both the
measurement noise as well as the intrinsic variance due to small scale
fluctuations in the forest).  This estimator effectively assumes that
the covariance matrix of the $\delta_{Fi}$ values can be approximated
as diagonal. By construction it is an unbiased estimator, but it is
not an optimal one. To avoid introducing contamination from the
correlated residuals from continuum fitting errors, we include only
pairs of points from \emph{different quasars} when measuring the
three-dimensional correlation function.  We also measure the
one-dimensional correlation function along each quasar, which is,
however, only  used for the error estimation.

Throughout we neglect signal-noise correlations that arise due to the
fact that the Poisson noise associated with counting of the photons
hitting the CCD is measured from the \emph{observed} photon count
rather than the underlying (and unknown) true expectation value of
photon count. In \cite{2006ApJS..163...80M} it was found that this
results in 1-2\% change in the amplitude of the continuum fit, which
induces a corresponding 2-4\% change in the amplitude of the power
spectrum, but they find no effect beyond this change in amplitude. The
effect is likely to be smaller in BOSS data, compared to SDSS data
since objects in BOSS are in average fainter and hence the amount of
(not signal-correlated) Poisson noise coming from sky-subtraction is
larger. 

Another potentially important source of spurious correlations is
correlated noise on different quasars observed on the same plate from
imperfect sky-subtraction.  We have found that including pairs of
pixels from the same observed-frame wavelength produces strong
contamination, which we think comes from residuals of the
sky-subtraction. We therefore eliminate all pairs of pixels with an
observed wavelength that are within $1.5\, $\AA\ of each other.  The
resulting correlation function is well described by the
three-parameter linear cosmological model and thus we believe that we
have managed to remove majority of the contaminating signal.
By applying this cut and ignoring pairs in the same quasar with thus
discard information in purely transverse ($\mu=0$) and purely radial
($\mu=1$) modes.

We measure the correlation function $\xi_F(r,\mu,z)$ in 12 radial bins
up to $r=100\mpch$, 10 angular bins equally spaced in $\mu=\cos
\theta$, where $\theta$ is the angle from the line of sight, and 3
redshift bins ($z<2.2$, $2.2<z<2.4$, $z>2.4$). The redshift bins
correspond to the redshift of the absorbing gas rather that the
redshift of the background quasars backlighting the gas. Together with
estimating the correlation function in individual bins, we have also
calculated the weighted averages of $r$, $\mu$ and $z$ for each bin (i.e., 
weighted in the same way as the $\xi$ measurement in that bin),
which are used in the subsequent analyses. These averages correspond
to the true bin position, although we find the difference with respect
to the nominal bin centre to be small.  
When estimating the correlation function in limited redshift bins, we
take into consideration all pairs whose mean redshift falls in the
redshift bin under consideration.

It can be shown that the error covariance matrix of this estimator is
given by (see Appendix \ref{sec:appErr}):
\begin{equation}
   C_{AB}  = \frac{ \sum_{{\rm pairs}\ i,j \in A, {\rm
       pairs}\ k,l \in B } w_iw_j w_k w_l (\xi_{Fik} \xi_{Fjl}+ \xi_{Fil} \xi_{Fjk})}{ \sum_{{\rm pairs}\ i,j \in A} w_i w_j \sum_{{\rm
         pairs}\ k,l \in B} w_k w_l } ~,
  \label{eq:err}
\end{equation}
where $A$ and $B$ represent two bins in $r$, $\mu$ and $z$ of the
correlation function measurement and $\xi_{ij,obs}$ denotes the actual
observed covariance between the data points $i$ and $j$.
We stress that $\xi_{ij,obs}$ is obtained from the data, so the contribution
from noise and continuum fitting errors, metal absorption or any other
possible systematic effects is automatically included (note that we do use
overall measurements of the correlation function, including even the noise in 
an averaged sense, not pixel-by-pixel noise, however, the covariance is not
dominated by pixel self-products so this detail probably is not important).

Strictly speaking, this estimator for
errors is true only at the level of 2-point contribution to the error
covariance matrix; however, we show using synthetic data that it
accurately reproduces the errors (although our mocks do not contain as much
non-Gaussianity as we expect from the real data).

Evaluating the sum in the numerator of equation \ref{eq:err} is a
computationally daunting task: one would need to make a sum over all
possible pairs of pairs, which is an $O(N^4)$ task for $N$ datapoints
compared to the $O(N^2)$ for the estimation of the correlation
function itself. Since in our case $N\sim10^5$, the extra effort is
$O(10^{10})$ computationally more expensive when compared to
estimation of the correlation function. However, only a small fraction
of all possible 4-point configurations add significantly to the sum,
namely those configurations of points $(i,j,k,l)$ for which $(i,k)$
and $(j,l)$ are close together (so that the corresponding values of
$\xi_F$ are large), and for which the distance between the pairs
$(i,j)$ and $(k,l)$ is in the relevant range over which we want to
estimate the correlation function. We therefore use the following
Monte-Carlo procedure:
\begin{enumerate}
\item For each quasar, identify the list of quasars that are closer
  than the largest distance scale on which we are attempting to measure the
  correlation function ($100 \mpch$). We denote such quasar pairs
  \emph{neighbors}. We additionally identify a subset of neighbours
  which are at distances closer than $r_{\rm cn}=30\mpch$. Such quasar
  pairs are denoted \emph{close neighbors}.

\item Select a random two pixels in the dataset, corresponding to
  \emph{neighbouring} quasars A and B\footnote{Since quasars have
    varying number of pixels, this is not identical to selecting
    random neighbouring quasars A and B and then random pixels in these 
    quasars.}.
  These two points constitute a pair $(i,j)$, which is held fixed
  while we loop over all possible pairs of points $(k,l)$. Pairs
  $(k,l)$ are chosen from all possible pixels in close neighbors of
  $A$ and close neighbours of $B$.  For each such quadruplet and for
  the two possible pairing of points, determine which covariance
  matrix element they belong to and add to the corresponding element
  according to the Equation (\ref{eq:err}).

\item Repeat the step 2 for $N_{\rm MC}$ times and then multiply the
  final sum  with the ratio of all possible pairs $(i,j)$ to the
  actual number of pairs considered $N_{\rm MC}$.

\item Divide this sum by the appropriate sum of weights for each
  covariance matrix bin.
\end{enumerate}

This process converges for about $N_{\rm MC} > 10^{7}$, in the sense
that the resulting $\chi^2$ values change by less than unity.  The
reason why this Monte-Carlo procedure works is fairly intuitive:
quasars are distributed randomly and hence there are no special points
that would anomalously contribute to the variance of our estimator. We
must sample the geometry of the distribution of our points well enough
to cover all typical configurations, and by that time the error
estimation converges. It is also important to note that our error
estimation is complete in the sense that it is based on the measured
two-point statistics in the field. For example, the continuum errors
add a large scale contaminating signal along each quasar. These errors
manifest themselves as larger correlations in the auto-correlation
function of the quasar and ultimately result in larger errors on our
measurements of the three-dimensional correlation function.

Finally, the speed of this error estimation is drastically affected by
the distance chose to denote close neighbors $r_{\rm cn}$. The method
approximates that the contribution to the error covariance from
correlations beyond this distance is negligible. The speed for
converges grows approximately quadratically with $r_{\rm cn}$. We have
tested the procedure with $r_{\rm cn}=10,20,30,50 \mpch$ and noted
that the results converge at $r_{\rm cn}\sim 20\mpch$. The convergence
has been established by inspecting the best fit $\chi^2$ when fitting
bias/beta parameters (see section \ref{sec:fitt-bias-param}) and
demanding that it changes by less than unity. We have therefore chosen
$r_{\rm cn}=30\mpch$.  

\begin{figure}
  \begin{center}
    \includegraphics[width=1.0\linewidth]{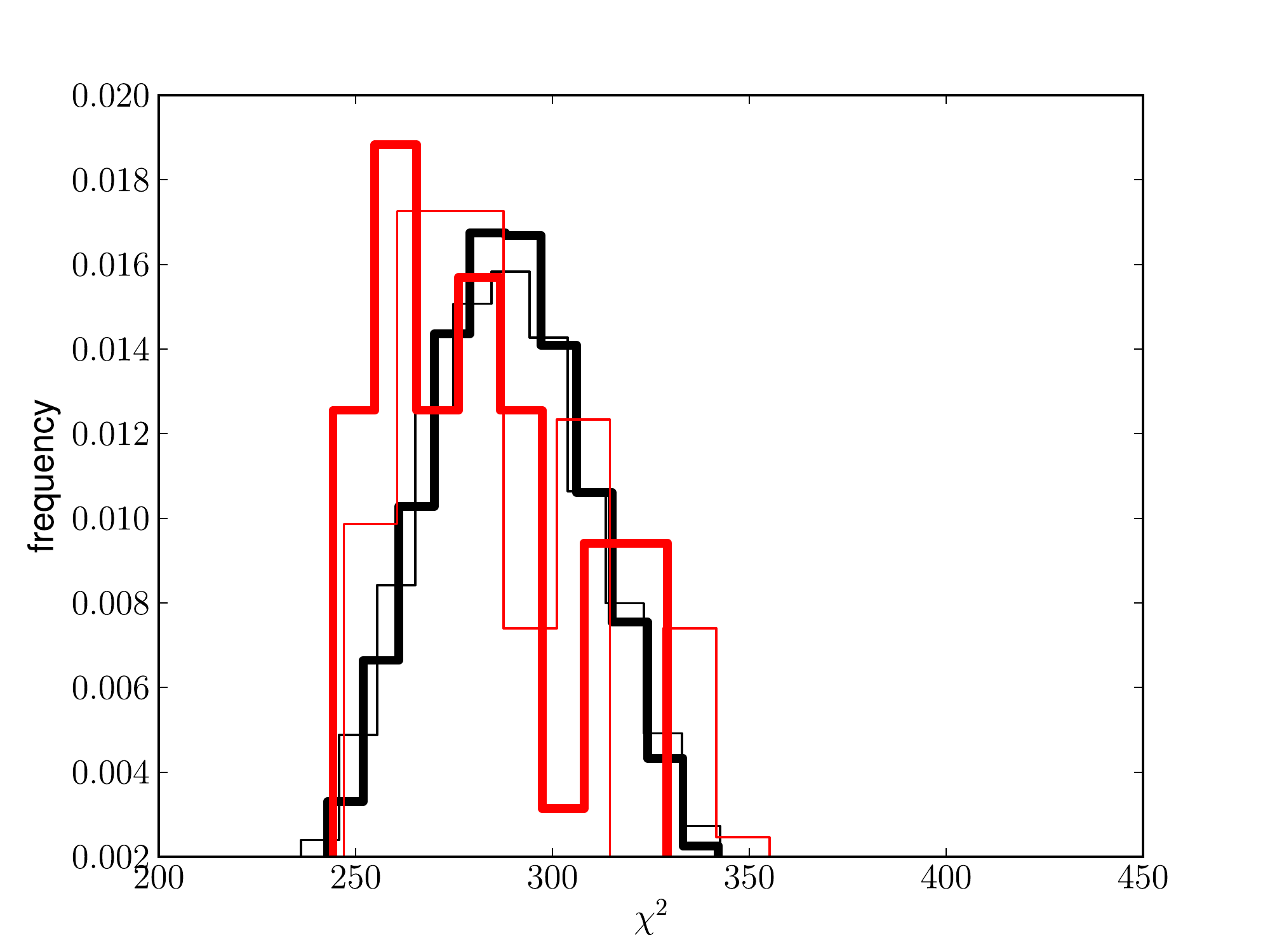}
  \end{center}
  \caption{\label{fig:errtest} The result of the $\chi^2$ distribution
    exercise for radial bins at $r>10\mpch$, which is used in our
    cosmological fitting. The thick black histogram correspond to
    Monte-Carlos with realizations, while the thick red histogram is the
    distribution for the actual 30 realizations of synthetic data. The
    thin histograms of the same colors shows the effect of ignoring
    off-diagonal elements of the covariance matrix. The number of bins
    here is 300 ($10\times 10 \times 3$). See text for discussion.  }
\end{figure}

We tested this procedure using synthetic data as follows. We averaged
30 synthetic realizations of the full dataset (with noise and
continuum) into one single mean measurement, which was assumed to be
the true theory. For each realization we calculated the $\chi^2$,
resulting in 30 $\chi^2$ values. Since we did not input the true
theory, but the mean of thirty synthetic datasets,
the mean $\chi^2$, for $N$ correlation function bins, is expected to be 
$\langle \chi^2 \rangle = \left(1-1/M \right) N$, where
$M=30$ is the number of datasets being averaged over. 
In practice, we Monte-Carlo this
distribution by drawing random vectors from the same covariance matrix
in a set of 30.  The purpose of performing error estimation test this
way is that it allows us to disentangle the accuracy of
error-estimation from systematic effects that might affect the theory
predictions (for example, when continuum fitting can add small
systematic shifts in the mean predicted signal).  Figure
\ref{fig:errtest} presents the results of this exercise on the synthetic
data with continuum and noise and shows that the error
estimation is robust.

\begin{figure}
  \begin{center}
    \includegraphics[width=1.0\linewidth]{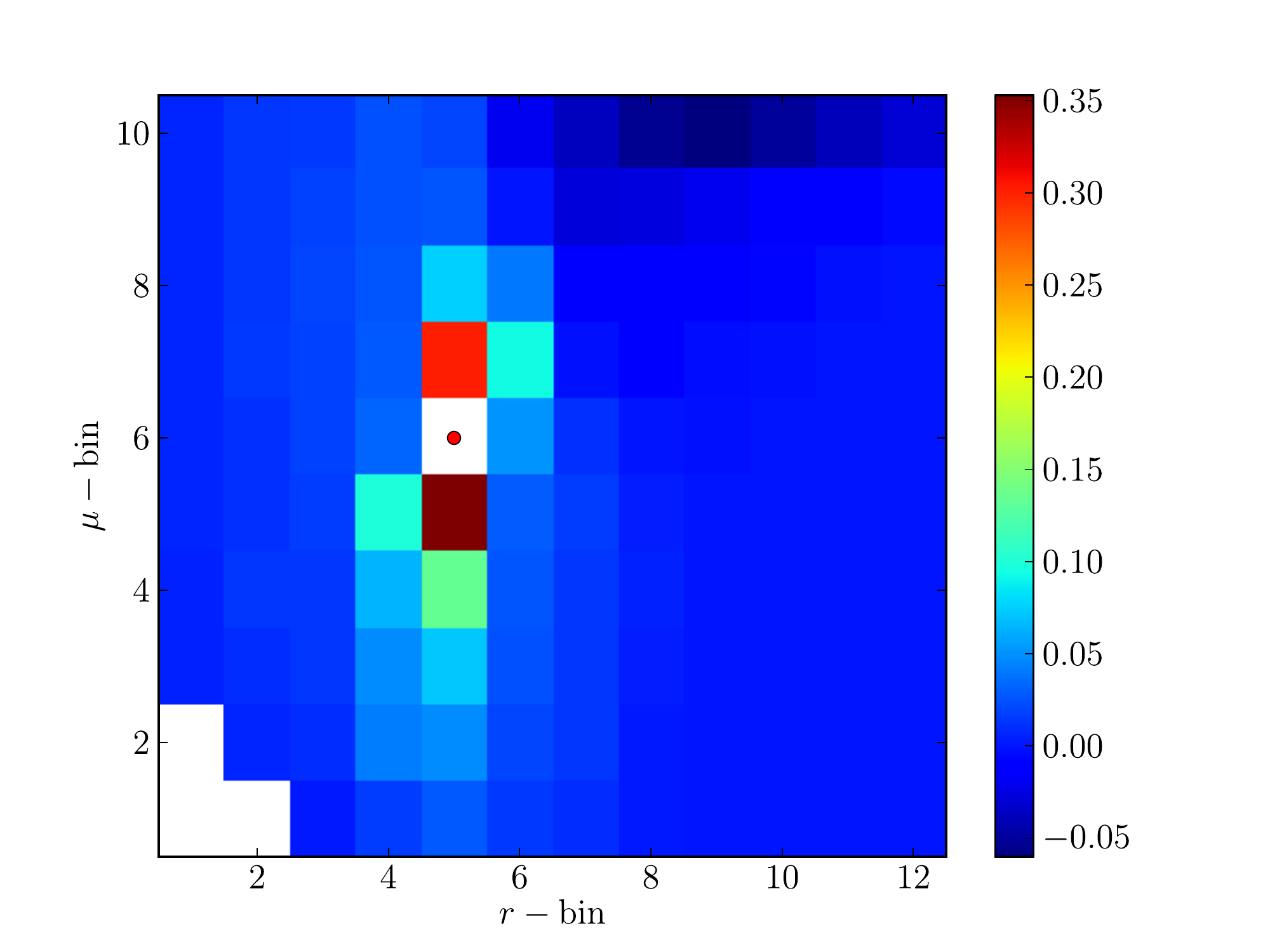}
  \end{center}
  \caption{\label{fig:cove} The structure of the
    error-covariance. Off-diagonal terms for the correlation matrix
    are plotted for one particular bin (denoted with a red point) on
    the plane of $r$ and $\mu$ bins at the middle redshift bin. The covariance
    at the bin itself is unity, but we do not plot it to avoid
    unnecessarily stretching the color scale.  The lower left corner
    is not present in the data due to a cut on observed-frame $\Delta
    \lambda$. Values of $\mu$ bins are linearly spaced between 0.05
    (first bin) and 0.95 (tenth bin). The first four radial bins
    correspond to distances of $3\mpch$, $8\mpch$, $12 \mpch$ and
    $17\mpch$. The remaining 8 radial bins are uniformly spaced between
    $25\mpch$ and $95\mpch$. The reference point is $25\mpch$, 
    $\mu=0.55$. }
\end{figure}

We plot the structure of the error covariance in Figure
\ref{fig:cove} which shows that the neighboring $\mu$ bins
are somewhat correlated, but the neighboring $r$ bins much less so.
The neighboring redshift bins (not plotted) are even less correlated
(at around 1\%). Finally, we note the negative correlation at
high-$\mu$, large $r$ arising due to removal of the mean in
each quasar.

\begin{figure}
  \begin{center}
    \includegraphics[width=1.0\linewidth]{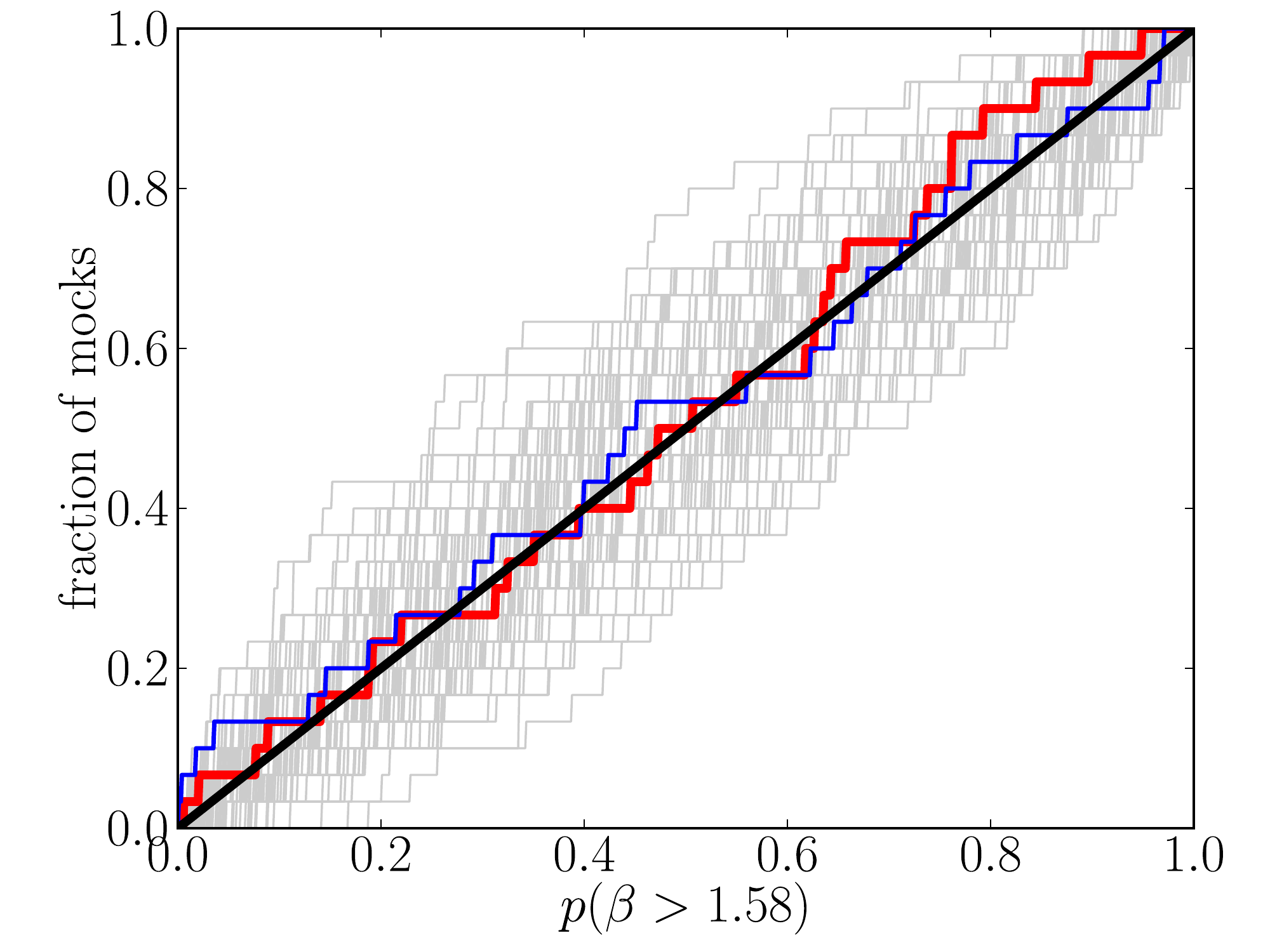}
  \end{center}
  \caption{\label{fig:kare} Cumulative distribution of the fraction of
    samples with value of $\beta$ above the fiducial value when
    fitting synthetic data. For correctly estimated errors, ignoring
    the upper prior on $\beta$, this distribution should be flat
    between zero and unity, giving a straight line for cumulative
    distribution, plotted as thick red.  Thin blue line is the same when
    off-diagonal elements in the covariance matrix are ignored.  Faint grey 
    lines are 100 realizations of 30 points drawn from uniform.}
\end{figure}

The $\chi^2$ distribution test in Figure \ref{fig:errtest} shows that the error
calculation is broadly correct, but is not very sensitive to inaccuracy in an 
important subset of the error matrix, and does not test for systematic
inaccuracy in the measurement.
To address these points we also show the result of another test 
of the error matrix.
In the next section we discuss how we fit the bias parameters,
and in the results section we will quote the tail probability that
$\beta$ is larger than a certain number based on MCMC chains.  We test
this procedure on our 30 mocks by precisely the same test, namely
fitting the bias parameters and then measuring the probability that
the value of $\beta$ is larger than the fiducial value used in
creating the synthetic datasets. If our process is unbiased and
correct, then we expect this probability to be uniformly
distributed between zero and one. We plot the results of this test in
Figure \ref{fig:kare}. Results of this exercise are quite
encouraging - despite evidence for small systematic shifts in the
inferred parameters (Table \ref{tab:resmock}), 
we seem be able to successfully recover limits in
the $\beta$ parameter for confidence limits used in this paper.

Note that Figure \ref{fig:kare} is a test
both of the errors (whether the covariance matrix is correct, and
whether the likelihood is fully described by a covariance matrix) and
any systematic offset in the measurement. There are a few things that are
known to be imperfect about our analysis, although not expected to be
important: We make no attempt to
include non-Gaussianity of the measured field in the covariance matrix
estimation, i.e., to include the connected part of the 4-point
function. The mocks are in any case only a weak test of this
assumption, as they are not designed to have the correct 4-point
function. On the other hand, \cite{2006ApJS..163...80M} found that the
errors in the 1D power spectrum were within $\sim 10$\% of the
Gaussian expectation, and we are probing larger scales here, where we
expect less non-Gaussianity.  There are technical issues that could
add up to biases in the pipeline. For example, the measured
correlation function, which is used to estimate the errors, is assumed
to be constant within radial bins, which is a poor approximation,
especially for the most important bins at small separations. It is
also true that the process of adding noise and fitting and dividing by
continuum could potentially ``renormalize'' the large-scale bias
parameters. Careful exploration of these effects will be left for the
future work. This will inevitably require many more mocks in order to
clearly differentiate between biased error estimation and realization
variance. In any case, Figure \ref{fig:kare} shows that many possible forms of
error in the analysis are unimportant.

\subsection{Fitting bias parameters}
\label{sec:fitt-bias-param}

We have fitted the bias parameters of the real and synthetic data
assuming a fiducial flat cosmology with $\Omega_m=0.27$,
$\sigma_8=0.8$, $n_s=0.96$, $H_0=71~ {\rm km~s^{-1}}\mpc^{-1} = 100~ h~ 
{\rm km~s^{-1}}\mpc^{-1}$, $\Omega_b=0.04$, (where symbols have their standard
meaning). We assume that the correlation function of $\delta_F$ is
linearly biased with respect to the underlying dark-matter correlation
function, and that it follows the linear theory redshift-space
distortions.  We do not use the measured correlation at $r<10\mpch$
for the fits where linear theory 
becomes a poorer approximation. After Fourier
transforming Equation (\ref{eqn:kaiser}), one derives the following
expression for the redshift-space correlation function:

\begin{equation}
  \xi_F (r,\mu) = \sum_{\ell=0,2,4} b^2 C_\ell(\beta) \xi_{F\ell}(r) P_\ell(\mu) ~,
\end{equation}
where $P_\ell$ are Legendre polynomials,
\begin{eqnarray}
  C_0 = 1+2/3 \beta + 1/5 \beta^2 ~,\\
  C_2 = 4/3\beta + 4/7\beta^2 ~,\\
  C_4 = 8/35 \beta^2 ~,
\end{eqnarray}
and
\begin{equation}
  \xi_{F\ell} (r) = (2\pi)^{-3} \int P(k) k_\ell(kr) d^3k ~,
\end{equation}
with the kernels $k_\ell(x)$ given by
\begin{eqnarray}
k_0 (x) & = & \sin(x)/x ~,\\
k_2 (x) & = & (\sin(x) x^2-3\sin(x)+3 \cos(x) x)/x^3 ~, \\
k_4 (x) & = & (x^4 \sin(x)-45 x^2 \sin(x) +105\sin(x)+10 x^3 \cos(x)-105 x \cos(x))/x^5 ~,
\end{eqnarray}
and $P(k)$ is the linear real-space power spectrum.

 We model the redshift evolution of the power spectrum as a simple
power-law,
\begin{equation}
P_F(k,z) = b^2 P (k,z=2.25) \left(\frac{1+z}{1+2.25}\right)^\alpha.
\end{equation}
Therefore, the growth factor never enters our analysis - the redshift
evolution is parametrised purely as power-law deviation from the power
at $z=2.25$. The only exception to this rule is when we fit the
actual bias parameters in the three redshift bins independently (see Figure
\ref{fig:koot}), where bias parameters are measured in individual
redshift bins with respect to the matter power at that redshift
(but are assumed constant across a redshift bin).

Whenever we fit for these parameters, we also include three $\Delta r$
parameters describing the effect of removing the mean quasar component
(as derived in Appendix \ref{sec:appDC}) at three redshifts, as free
nuisance parameters. Although these parameters can be determined
\mbox{\emph{a-priori}}, we have found that fitting for them is easier,
accurate enough (as tested on mocks), and worsens the
errorbars on the inferred bias parameters by a very small amount when
we marginalize over them. The resulting chains constrained
values of $\Delta r$ very weakly to be in the range between $\sim
200\mpch$ and $500\mpch$.

We fit using an MCMC procedure. We put a flat prior on $0<\beta<5$ and
unconstrained flat prior on $b(1+\beta)$, which are well-defined
non-degenerate parameters. 
(Note that $\beta$ can in general be less than zero,
however, it is difficult to imagine a scenario in which this would happen.)
This implies a non-flat prior on $b$. We also assume a flat prior on
$\alpha$.

When we fit the data, we always evaluate the theory at each bin's mean
redshift, $\mu$ value and radius. This approach allows for any linear variations
of the underlying quantity across the bin. This is not a good
approximation at the lowest values of $r$, but, as we show later, it is
a good approximation for the bins that we actually use when fitting
the bias parameters.

\subsection{Pipeline tests and synthetic data}

Our code has been extensively tested on the synthetic data. These
tests have served two goals. First, they helped remove 
coding errors in both the data-reduction code and the synthetic
data making codes. For this purpose, the two codes were written
completely separately by two people. Second, these tests established the
amount of systematic effects introduced by the data-reduction
process. Our guiding principle in this exploratory work was that we
should only correct for the errors that affect our results at the
level of the current statistical errors in the experiment. Namely, as
long as we are able to reproduce the input parameters from our
synthetic data with the highest level of realism and noise at
acceptable $\chi^2$ values, we pronounce the data reduction code
sufficient and apply the same method to the observed data. More data might
require better codes.

While not always strictly enforced, we did attempt to follow the
principles of blind analysis: after our basic reduction code was in
place and gave no visually surprising results on the observed data, we
worked exclusively on synthetic data in order to understand the
reduction procedure and returned to observed data only when we were able
to reproduce the input parameters from the synthetic data. (The analysis 
procedure was not, however, frozen at this point, as our mocks were not 
realistic enough to expect every problem to show up in them.)

We begin by discussing the noiseless synthetic measurements of
$\delta_F$. While these data are noiseless and there is no error
associated with the continuum fitting, the final result still
contains variance associated with sample variance.

\begin{figure}
  \begin{center}
    \includegraphics[width=1.0\linewidth]{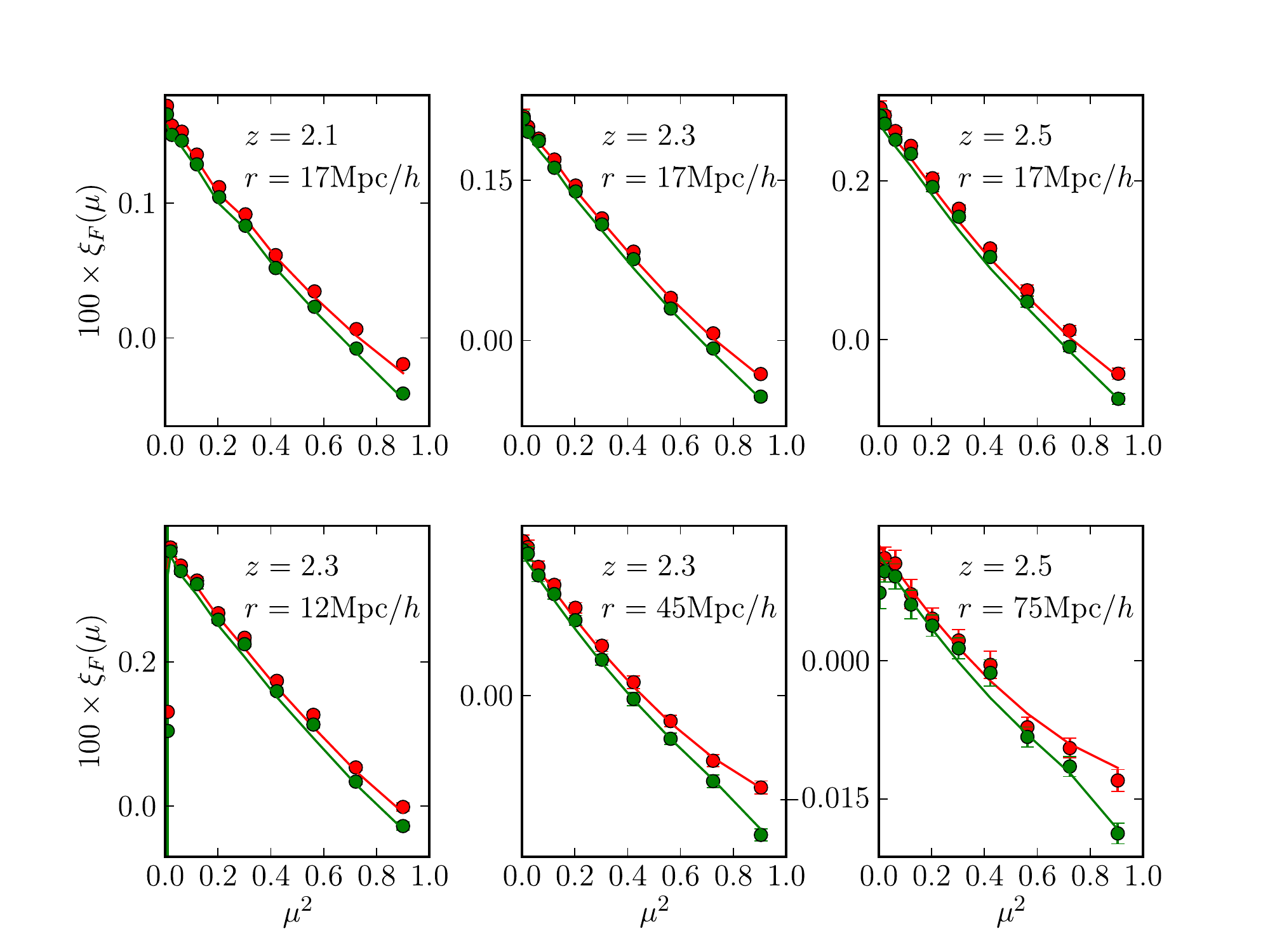}
  \end{center}
  \caption{\label{fig:pure2}The correlation function plotted as a
    function of $\mu^2$ for selected radial and redshift bins.
    Points are averages of 30 noiseless synthetic datasets with
    perfectly known continuum without any processing (red) and with mean
    and radial model removal (green). Red lines are the theory used to
    produce synthetic data, while green lines is the same after we
    corrected for the effect of mean removal using equations in Appendix
    \ref{sec:appDC} with $\Delta r$ of 250, 300 and 350 $\mpch$ for
    the three redshift bins.}
\end{figure}

We have analyzed these datasets using three methods: i) by
calculating the correlation function, ii) by first removing the mean
component of each spectrum and then calculating the correlation
function and iii) by also removing the purely radial modes as
described in Section \ref{sec:determining-delta_F}. The result of ii) and
iii) are practically identical.  We show the results of this exercise
in Figure \ref{fig:pure2}, where we have averaged the individual
results of all 30 realizations in order to make this an essentially
statistical error-free measurement, at least when compared to the 
statistical error in the real
data. The purpose of this figure is essentially two-fold: to
illustrate that, at the level of noiseless synthetic data, we can reproduce the
theory that is being put into the simulations, and, more importantly,
that the simple model in Appendix \ref{sec:appDC} appears to produce fairly 
good results (we will quantify these later). 

In order to proceed we need to understand the effect of adding various
complexities to the data and how they affect the results. It is very
difficult to judge the size of various effects by visually assessing
how the points move on the plots, so we adopted the following
procedure.  We run the parameter estimation of the synthetic data with
the covariance matrix corresponding to one mock noisy dataset, but
with the data vector corresponding to mean over 30 mock
datasets. During the fitting procedure this yields $\chi^2$ values
which are too small, but it allows one to see how central values move
with respect to the size of the error bars. Note that our 3-parameter
model (the bias $b$, the redshift-space distortion $\beta$,
and the redshift-evolution index $\alpha$)
is essentially the same model that has been used to create
mocks, but with the important difference that the latter contains
corrections to the linear theory redshift-space formula arising from matter
power spectrum
non-linearities, scale-dependence of the effective bias and 
fingers-of-God effect.

\begin{figure}
  \begin{center}
    \begin{tabular}{cc}
    \includegraphics[width=0.5\linewidth]{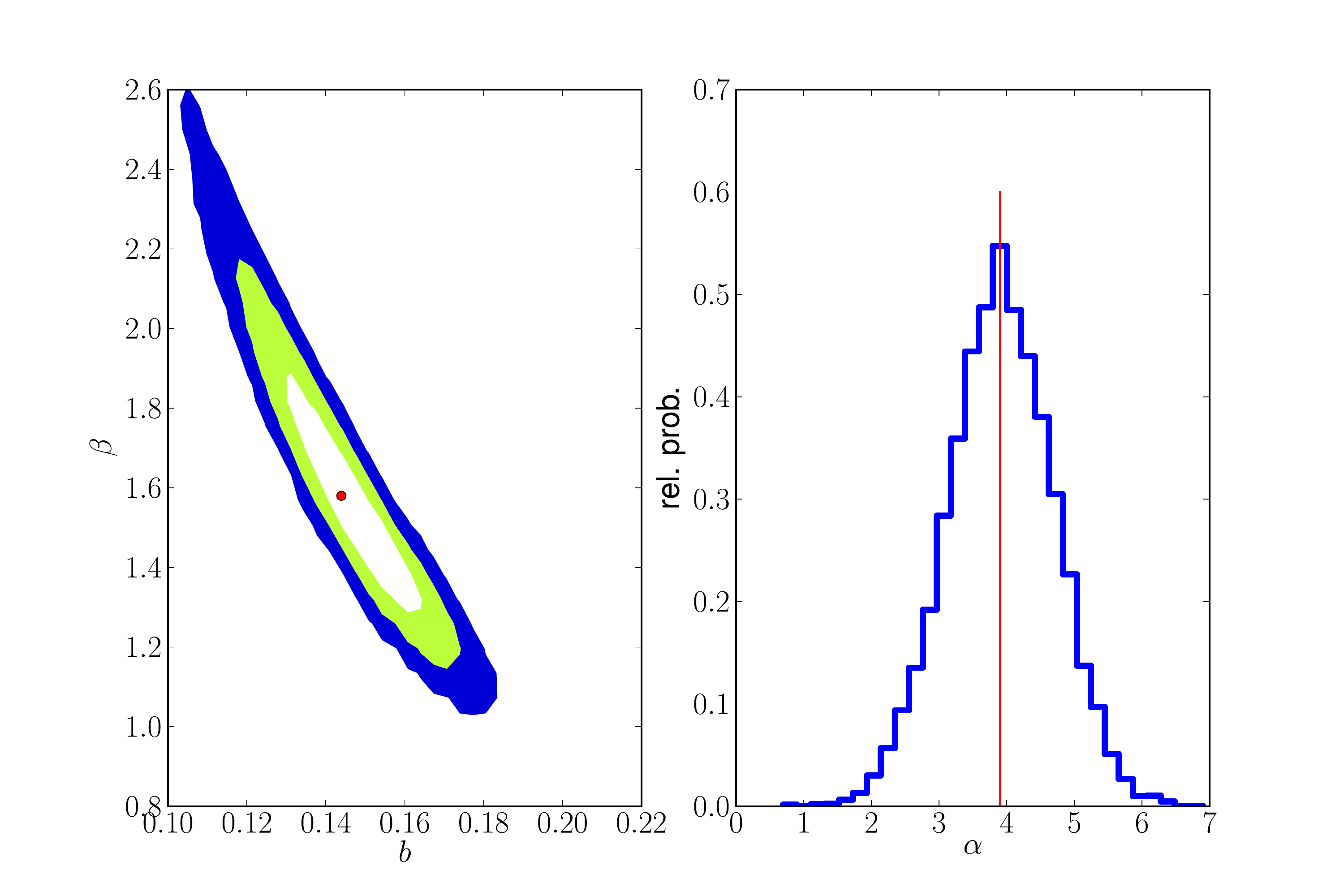}  & 
    \includegraphics[width=0.5\linewidth]{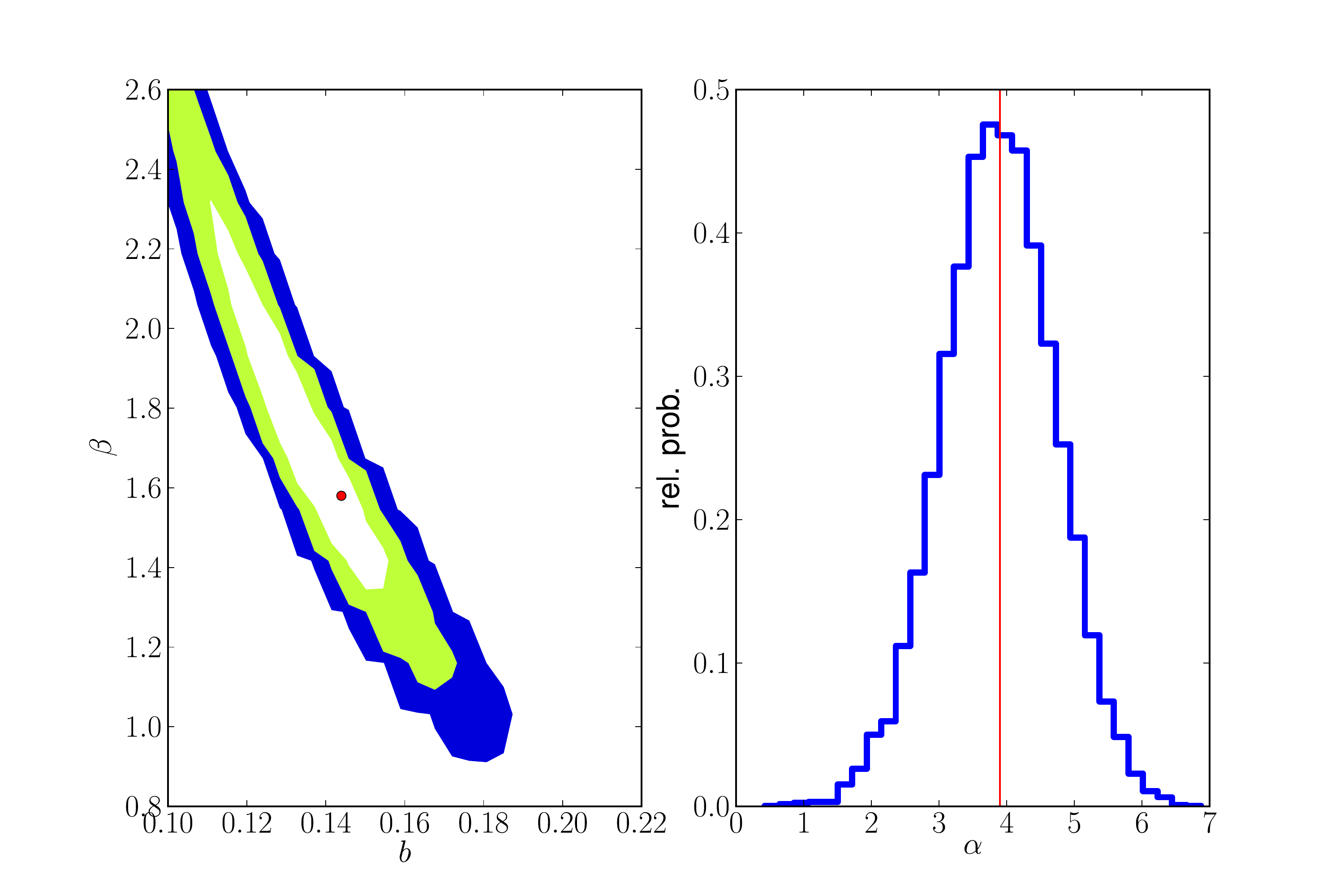} \\
    \includegraphics[width=0.5\linewidth]{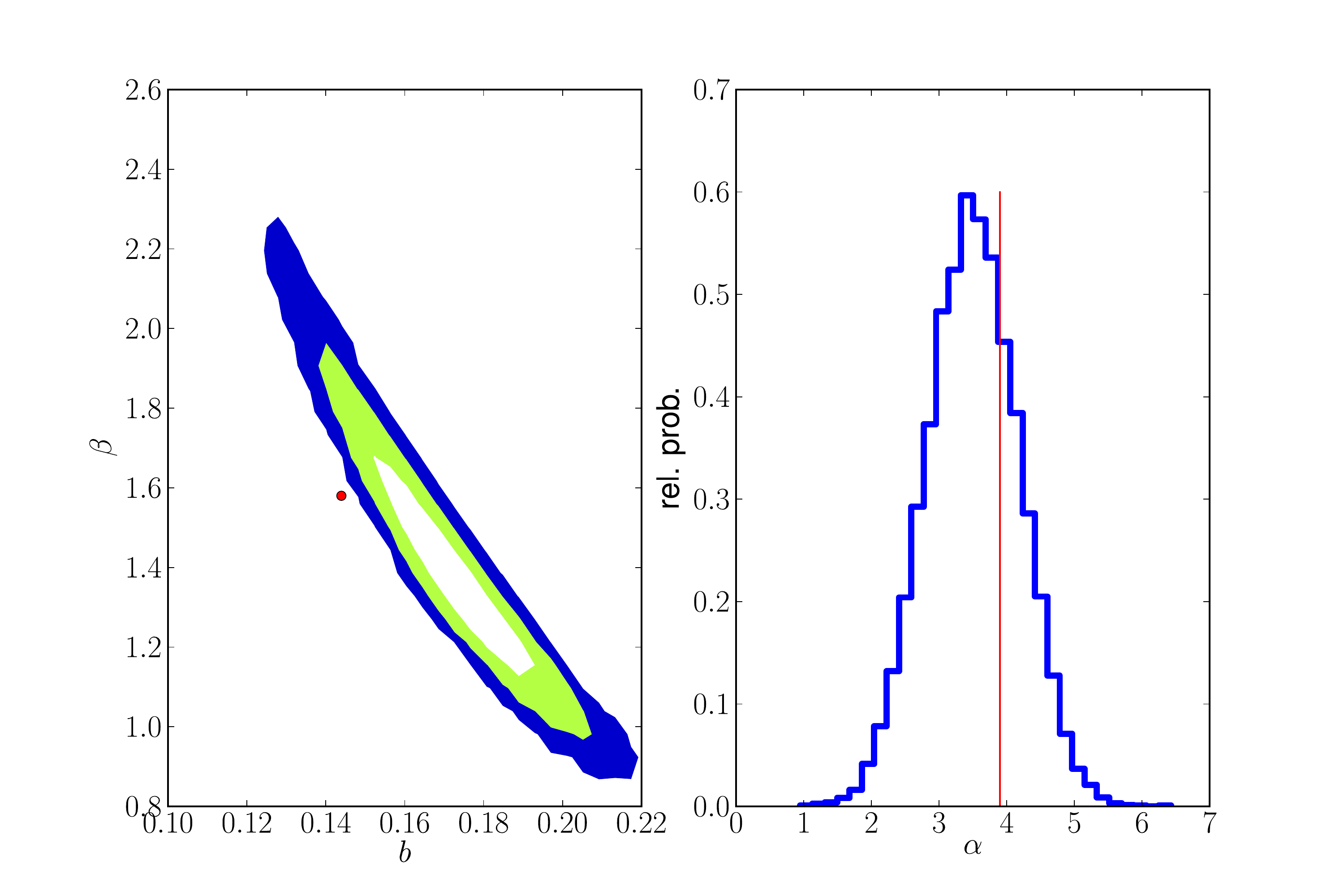} &
    \includegraphics[width=0.5\linewidth]{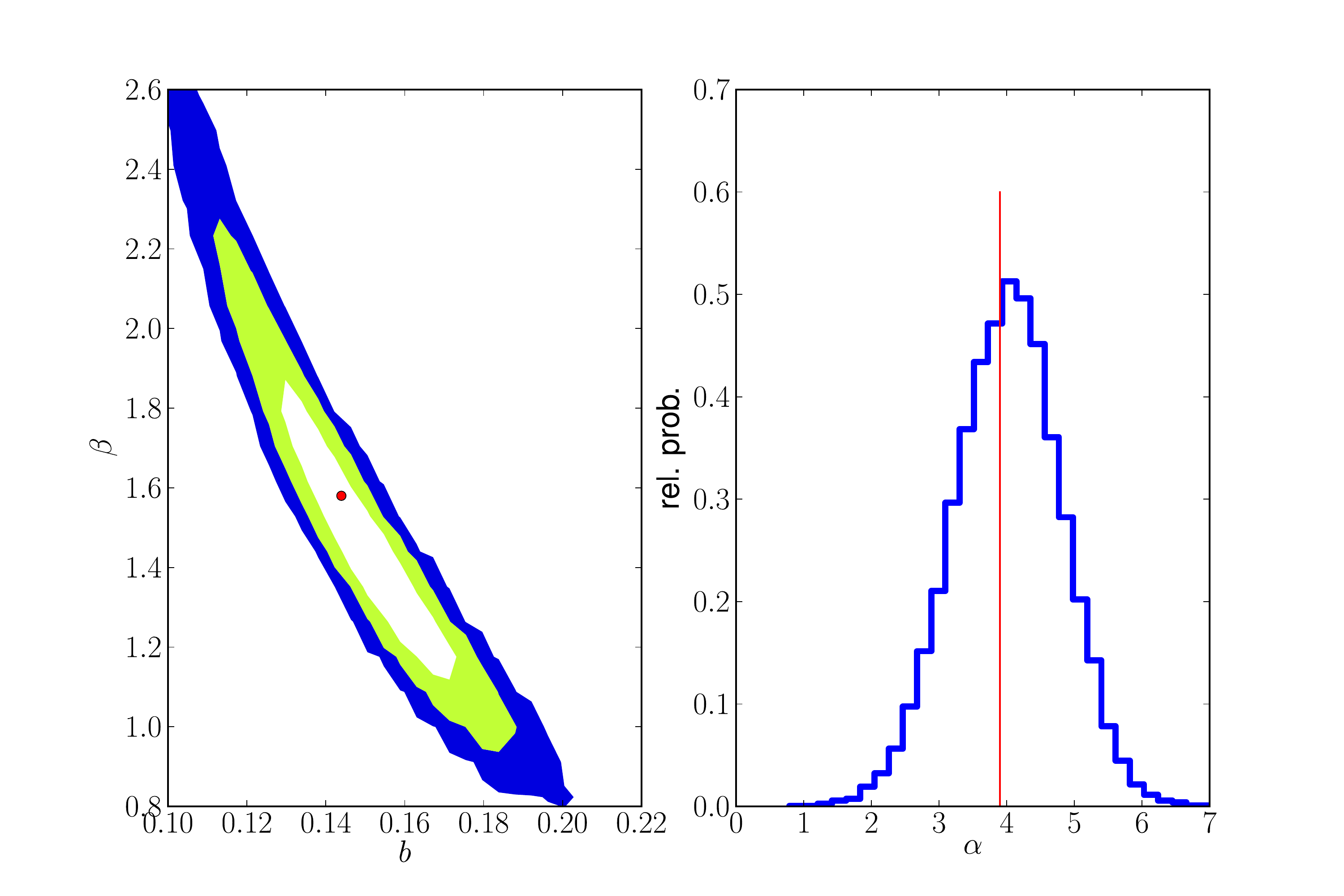} \\
  \end{tabular}
  \end{center}
  \caption{\label{fig:plusnoise} Results of
    fitting the data averaged over 30 mock datasets together with
    noise covariance for a single noisy realization and using only
    datapoints with $r>20\mpch$ in the fit. We show constraints on the
    $b-\beta$ plane and the probability histogram of $\alpha$ (which
    has negligible degeneracy with the other parameters).
    The input points are denoted by the red dot and the
    red line. The upper left plot is for the pure synthetic noiseless $\delta_F$
    values. The upper right plot is for synthetic data that have PCA
    continua and noise. The lower left plot is for the data that in
    addition to PCA continua are additionally painted with
    high column-density systems.  The bottom right panel is for
    synthetic data to which metals have been added as described in 
    \S \ref{subsec:formet} (with noise and continua but no DLA/LSS).
  }
\end{figure}

\begin{figure}
  \begin{center}
    \begin{tabular}{cc}
    \includegraphics[width=0.5\linewidth]{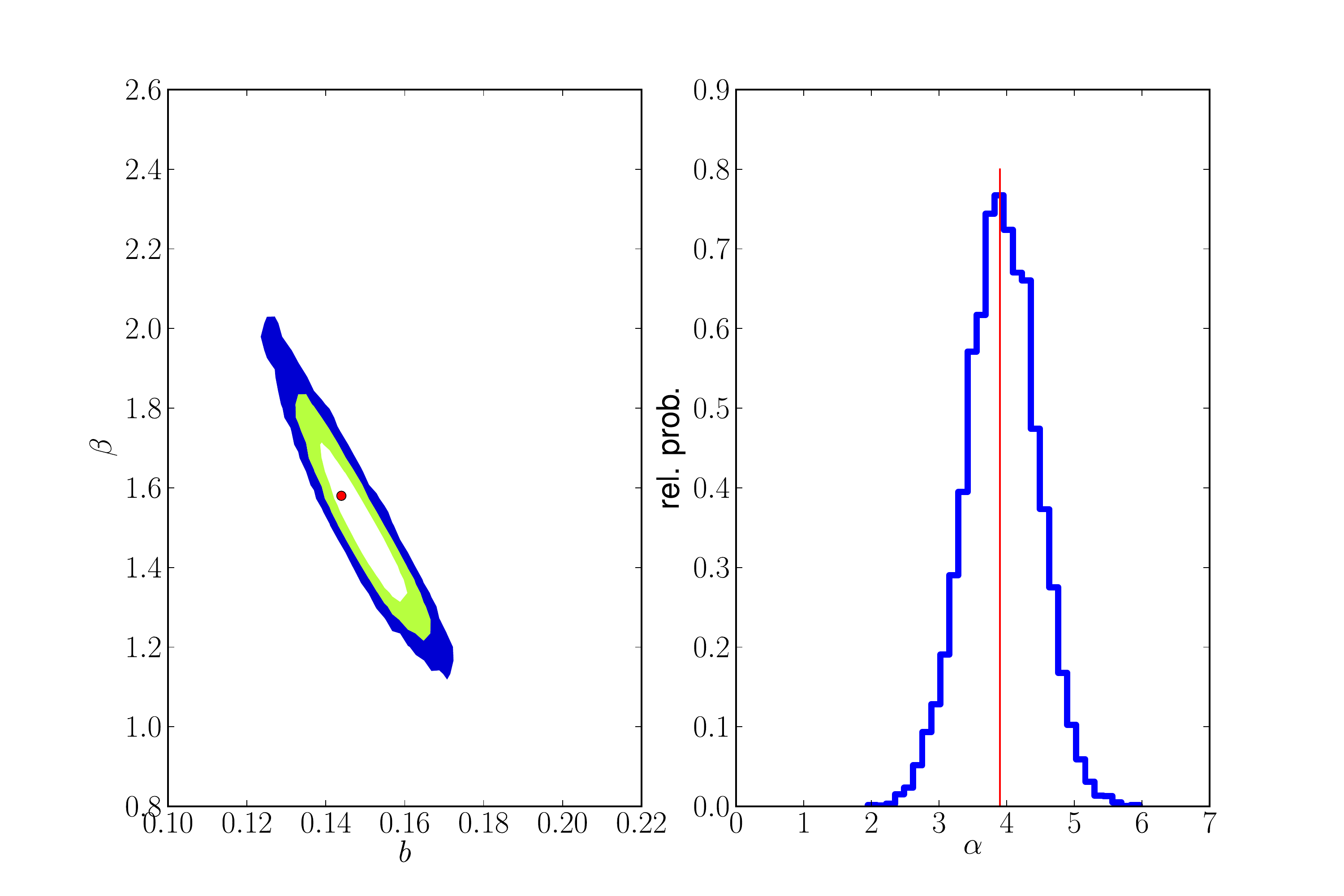}  & 
    \includegraphics[width=0.5\linewidth]{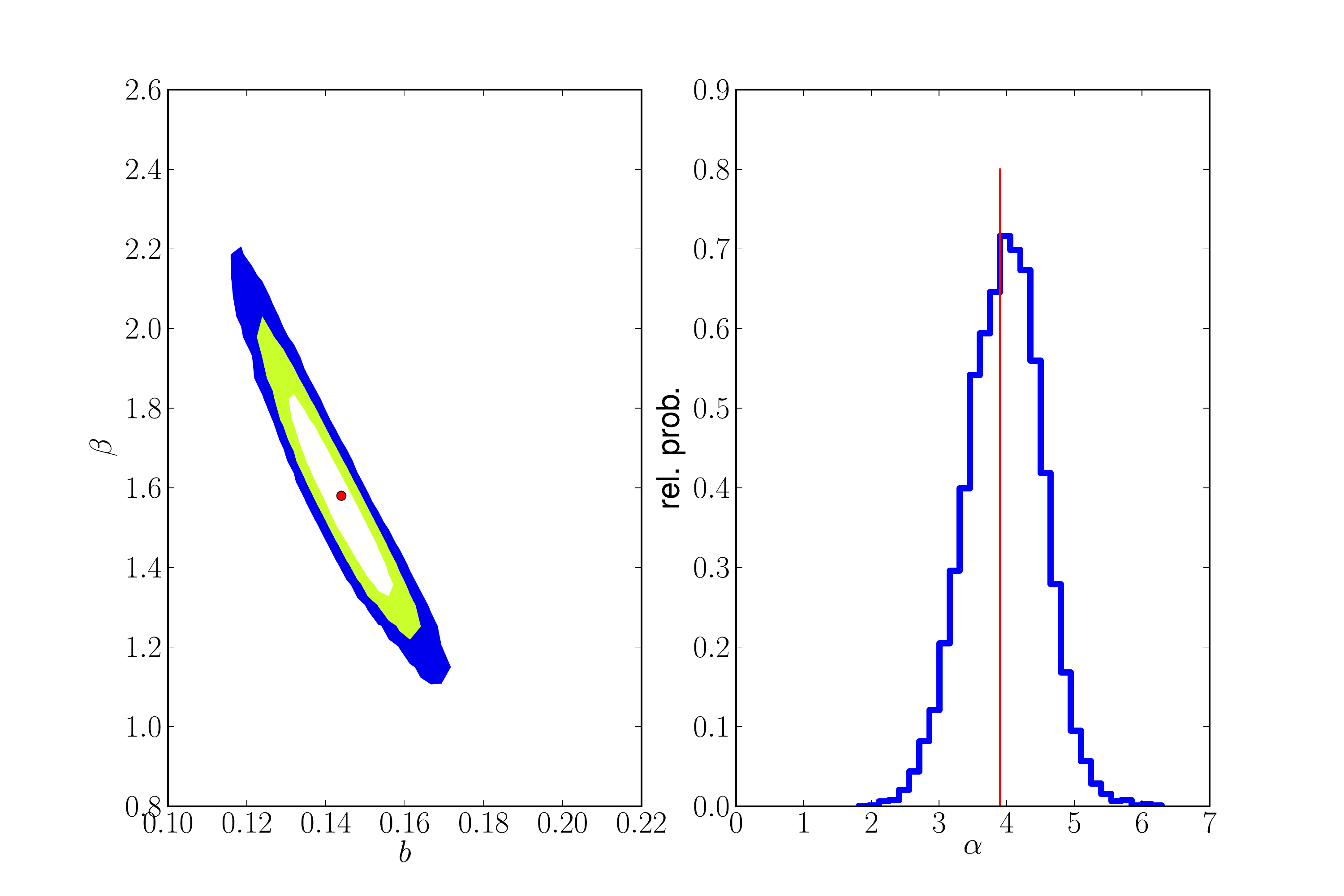} \\
    \includegraphics[width=0.5\linewidth]{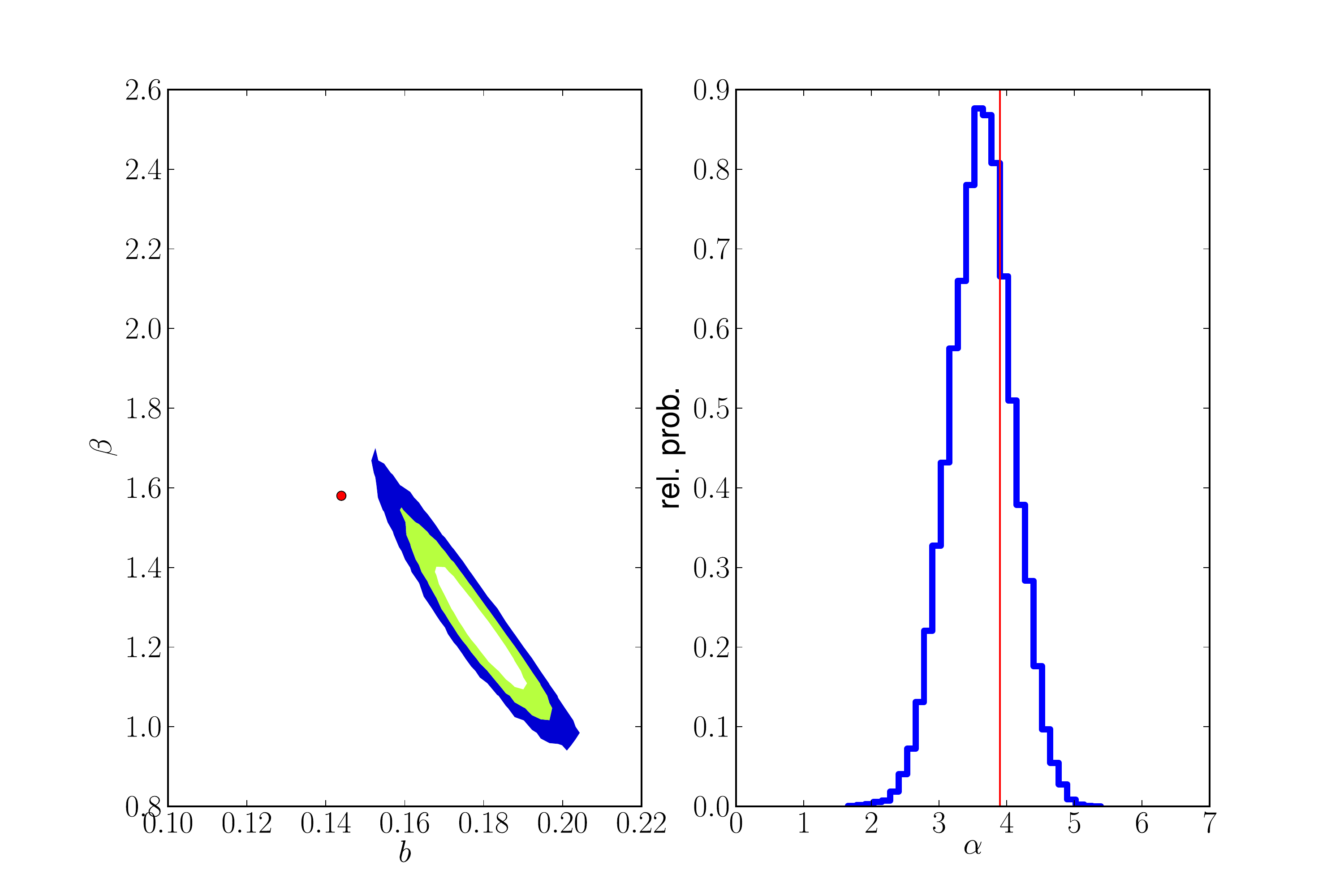} &
    \includegraphics[width=0.5\linewidth]{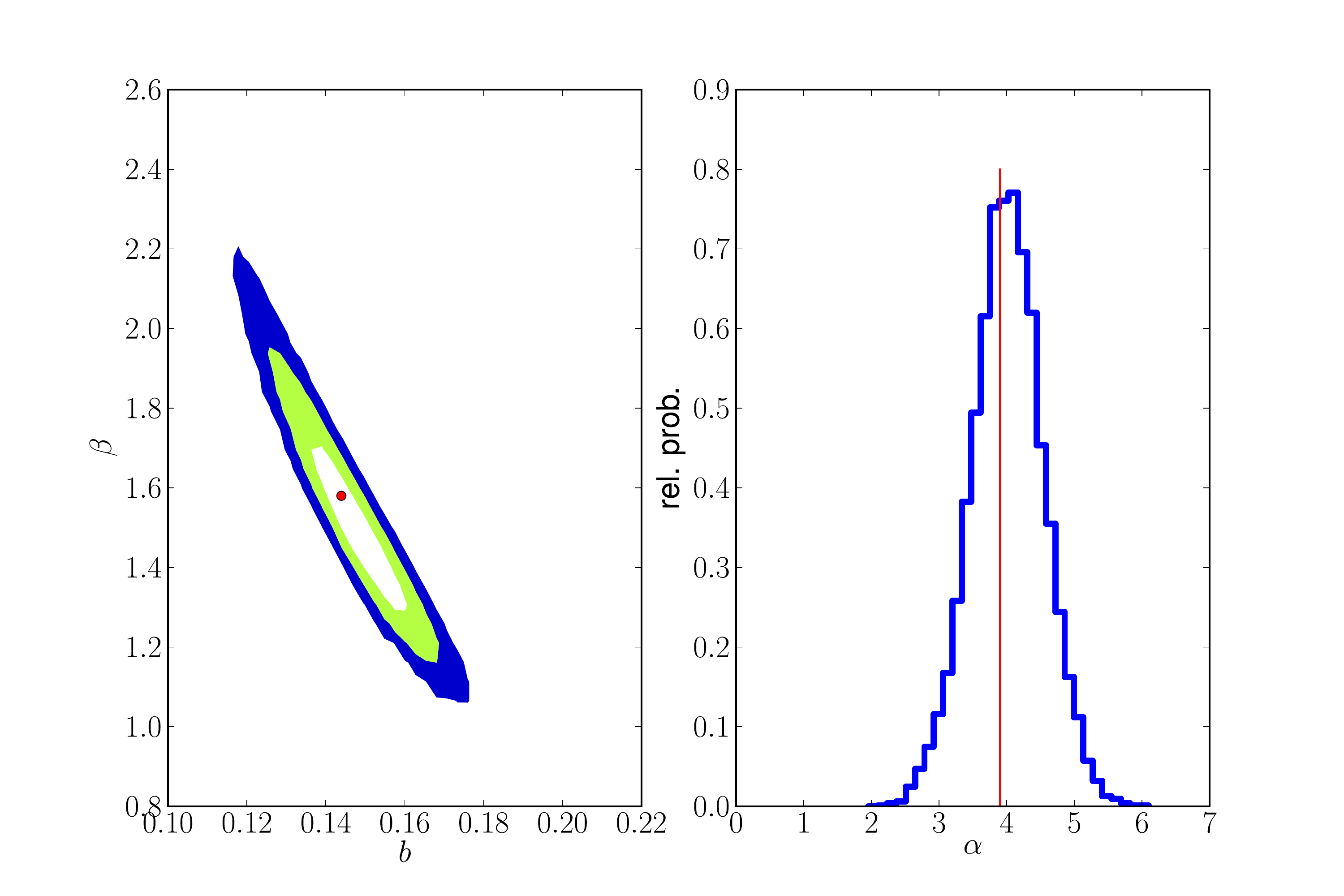} \\
  \end{tabular}
  \end{center}
  \caption{\label{fig:plusnoise2}  Same as figure \ref{fig:plusnoise}
    but for fitting all data with $r>10\mpch$. The axis limits were
    kept the same for easier comparison.}
\end{figure}

\begin{table}
  \centering
  \hspace*{-0.8cm}
  \begin{tabular}{c|cccc}
    Synthetic Data & bias & $\beta$ & $b(1+\beta)$ & $\alpha$ \\
    \hline
    Noiseless\\
    $r>20\mpch$ & $0.145\pm{0.013}$ & $1.58\pm^{0.24\ 0.54\ 1.03}_{0.19\ 0.35\ 0.49}$ & $0.374\pm{0.008}$ & $3.93\pm{0.75}$ \\
    Best fit    & $0.148$ & $1.53$ & $0.375$ & $3.95$ \\
    $r>10\mpch$ & $0.150\pm{0.007}$ & $1.50\pm^{0.13\ 0.28\ 0.43}_{0.11\ 0.22\ 0.32}$ & $0.373\pm{0.006}$ & $3.93\pm{0.54}$ \\
    Best fit   & $0.151$ & $1.47$ & $0.374$ & $3.93$ \\
    \hline 
    + Continuum/noise \\
    $r>20\mpch$ & $0.133\pm{0.017}$ & $1.75\pm^{0.39\ 0.93\ 1.83}_{0.29\ 0.52\ 0.72}$ & $0.366\pm{0.008}$ & $3.88\pm{0.84}$ \\
    Best fit & $0.140$ & $1.59$ & $0.364$ & $4.07$ \\
    $r>10\mpch$ & $0.143\pm{0.008}$ & $1.57\pm^{0.18\ 0.39\ 0.60}_{0.16\ 0.29\ 0.42}$ & $0.368\pm{0.006}$ & $3.98\pm{0.57}$ \\
    Best fit &    $0.145$ & $1.55$ & $0.369$ & $3.88$ \\
    \hline
    Continuum/noise/LLS/DLA \\
    $r>20\mpch$  & $0.172\pm{0.014}$ & $1.38\pm^{0.22\ 0.47\ 0.80}_{0.18\ 0.32\ 0.45}$ & $0.409\pm{0.007}$ & $3.50\pm{0.65}$ \\
    Best fit & $0.176$ & $1.33$ & $0.410$ & $3.50$\\
    $r>10\mpch$  & $0.179\pm{0.008}$ & $1.25\pm^{0.11\ 0.23\ 0.38}_{0.10\ 0.18\ 0.26}$ & $0.402\pm{0.005}$ & $3.63\pm{0.45}$ \\
    Best fit & $0.182$ & $1.20$ & $0.401$ & $3.71$ \\
    \hline
    Continuum/noise/forest metals\\
    $r>20\mpch$ & $0.149\pm{0.016}$ & $1.45\pm^{0.32\ 0.70\ 1.26}_{0.24\ 0.42\ 0.58}$ & $0.367\pm{0.008}$ & $4.04\pm{0.80}$   \\
    Best fit & $0.156$ & $1.35$ & $0.366$ & $4.12$ \\
    $r>10\mpch$ & $0.148\pm{0.009}$ & $1.50\pm^{0.17\ 0.38\ 0.63}_{0.15\ 0.28\ 0.39}$ & $0.369\pm{0.006}$ & $4.01\pm{0.52}$ \\
    Best fit & $0.149$ & $1.47$ & $0.369$ & $4.07$\\
    \hline
    Continuum/noise/LLS/\\DLA/forest metals\\
    $r>20\mpch$ & $0.193\pm{0.014}$ & $1.13\pm^{0.17\ 0.36\ 0.61}_{0.14\ 0.26\ 0.37}$ & $0.412\pm{0.007}$ & $3.66\pm{0.62}$ \\
    Best fit &  $0.197$ & $1.09$ & $0.412$ & $3.58$\\
    $r>10\mpch$ & $0.187\pm{0.007}$ & $1.15\pm^{0.10\ 0.20\ 0.31}_{0.09\ 0.17\ 0.25}$ & $0.404\pm{0.005}$ & $3.62\pm{0.43}$ \\
    Best fit &  $0.190$ & $1.12$ & $0.403$ & $3.60$\\
  \end{tabular}
  \caption{Results of parameter fittings for
    the average of 30 synthetic datasets and noise matrix for a single
    noisy measurement. The input values are $b=0.145$, $\beta=1.58$,
    $b(1+\beta)=0.375$ and $\alpha=3.8$. Errorbars are $1-\sigma$ confidence 
    limits
    except for  $\beta$ where we give the 1, 2 and 3 $\sigma$ error bars.
  }
  \label{tab:resmock}
\end{table}

In Figure \ref{fig:plusnoise} we show the results of the test
described above, when fitting with the data corresponding to
$r>20\mpch$. Figure \ref{fig:plusnoise2} contains the same material
but for the sample for $r>10\mpch$.  We show the same information also
in the tabular form in table \ref{tab:resmock}. It is important to
note that the inferred parameter constraints come with the usual
caveats about Bayesian inference (projections and priors) and therefore we also quote the
best-fit model, which should be at the position of the true theory, if
we had infinite number of synthetic datasets. 

The noise and continuum fitting do not strongly affect the inferred
bias parameters. It is important to stress that this is a non-trivial
conclusion: the continua in the synthetic data were simulated using
PCA eigen-components, while the fitting procedure assumed a much
simpler model for the continuum variations.  This test also confirms
the validity of the results of the Appendix \ref{sec:appDC} for the
present purpose. There is some evidence that the continuum fitting
may systematically decrease the bias,
but the effect is smaller than the
errorbars for one set of mocks. However, we find that the presence of the high
column-density systems increases the expected bias and lowers the
inferred value of $\beta$. This is not an artifact of the pipeline,
but simply the result of change in the theory. Forest metals have the
same effect, but to a somewhat lesser extent.  High column-density
systems and forest metals definitely warrant further investigation,
but this is beyond the scope of this article, and here we just note
that these two effects can affect the inferred values of bias,
$\beta$, and, to a lesser extent, $\alpha$. The inferred parameters
are skewed (at a $0.5-\sigma$ level) when fitting with linear models
using $r>10\mpch$ data due to incorrect assumption of fully linear
redshift-space distortions, especially at large values of $\mu$.

Finally, we note the surprisingly high degree of degeneracy in the
$b$-$\beta$ plane. This degeneracy is present in all measurements of
$\beta$ (see e.g. \cite{2009JCAP...10..007M}), but the high values of
$\beta$ of the \lyaf compared to typical galaxy populations makes it
particularly acute. At $\beta=1.5$, the power spectrum of purely
radial modes has $(1+1.5)^2=6.25$ times more power than purely
transverse modes and hence is measured much more precisely. If one
measures just radial modes, the data would exhibit a perfect
$b(1+\beta)$ degeneracy, which is only relatively weakly broken by the
measurements of the low $\mu$ modes.

\section{Results with the observed data}
\label{sec:results-with-real}

\begin{figure}
  \begin{center}
      \includegraphics[width=1.0\linewidth]{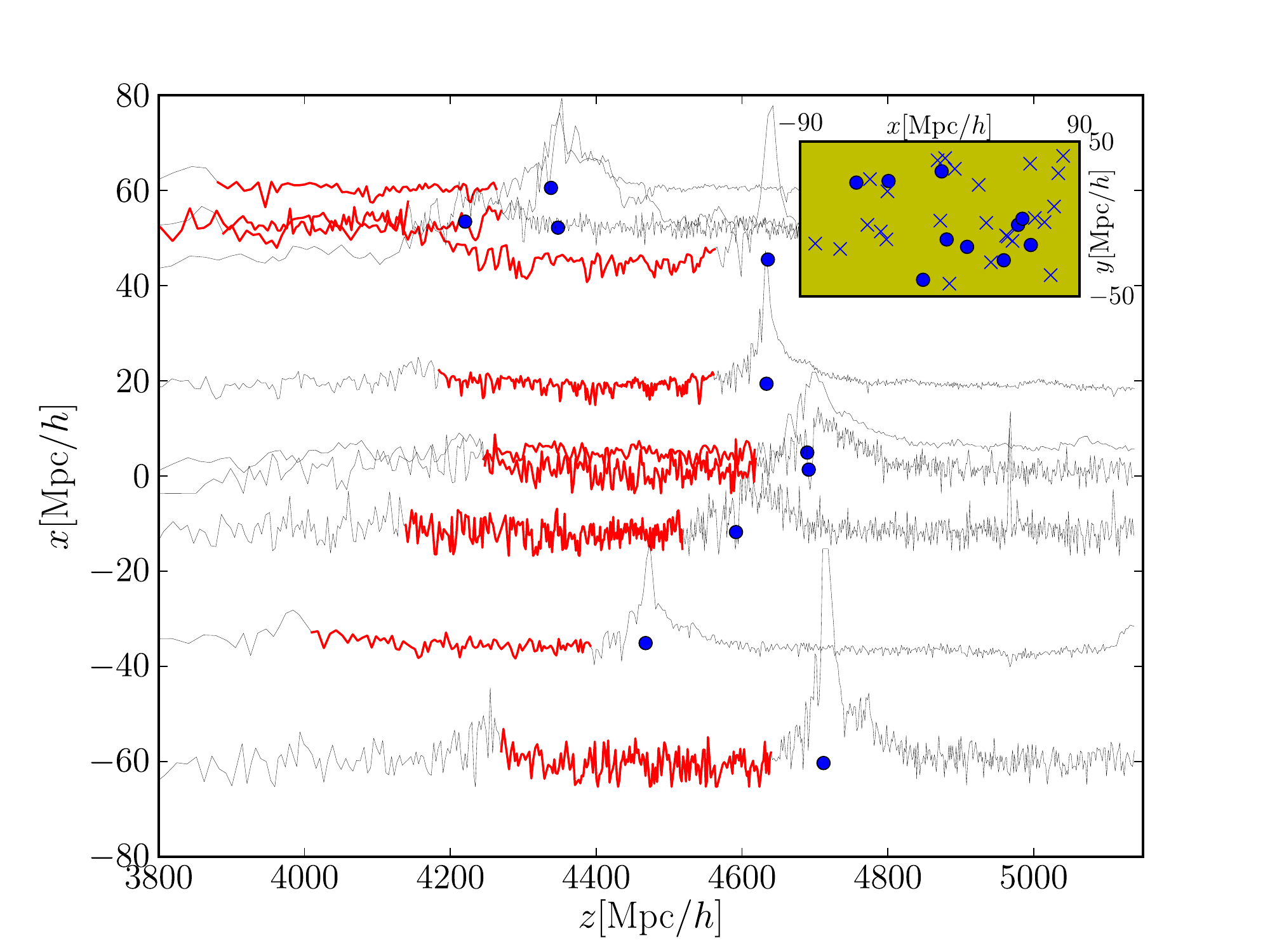}
  \end{center}

  \caption{\small \label{fig:wpl} Illustration of 3-d Lyman-$\alpha$
  forest measurement.  Inset panel shows the distribution of BOSS
  quasars in a patch of sky $170\hmpc \times 85\hmpc$ (comoving
  size at $z=2.5$).  The main panel plots spectra of the ten
  quasars marked by filled circles in the inset.  Other quasars are
  included in the analysis but omitted here for clarity.
  Wavelengths have been converted to equivalent comoving distance
  for Lyman-$\alpha$, and the Lyman-$\alpha$ forest regions are
  highlighted in red. Note that the scales for the vertical and horizontal 
axes are much different -- in fact the lines of sight are much closer together,
relative to their length, than they appear here. }
\end{figure}

In this section we discuss results derived from the observed data.  
Figure~\ref{fig:wpl} illustrates, qualitatively, our 3-d measurement
of structure traced by the Lyman-$\alpha$ forest, and it gives an
idea of the relevant scales.  The inset panel shows the
distribution of BOSS quasars in a patch of sky $140' \times 70'$,
corresponding to $170\hmpc \times 85\hmpc$ (comoving) at $z=2.5$
for our adopted cosmological model. 
The main panel plots the spectra of those quasars marked in
the inset panel by circles; we have omitted other quasars in
the field (shown by $\times$'s in the inset) to preserve clarity.
In the main panel, observed wavelengths have been converted to
redshifts for the Lyman-$\alpha$ transition and from redshift
to comoving distance using the standard cosmology.
Note the radical difference in scale of the two axes in the main panel:
the Lyman-$\alpha$ forest of a typical spectrum (highlighted
in red) spans $\sim 400 \hmpc$, while typical transverse separations
of neighbouring quasars are a few $\hmpc$ up to $\sim 20\hmpc$.

\begin{figure}

  \begin{center}
    \hspace*{-1cm}\begin{tabular}{cc}
      \includegraphics[width=0.5\linewidth]{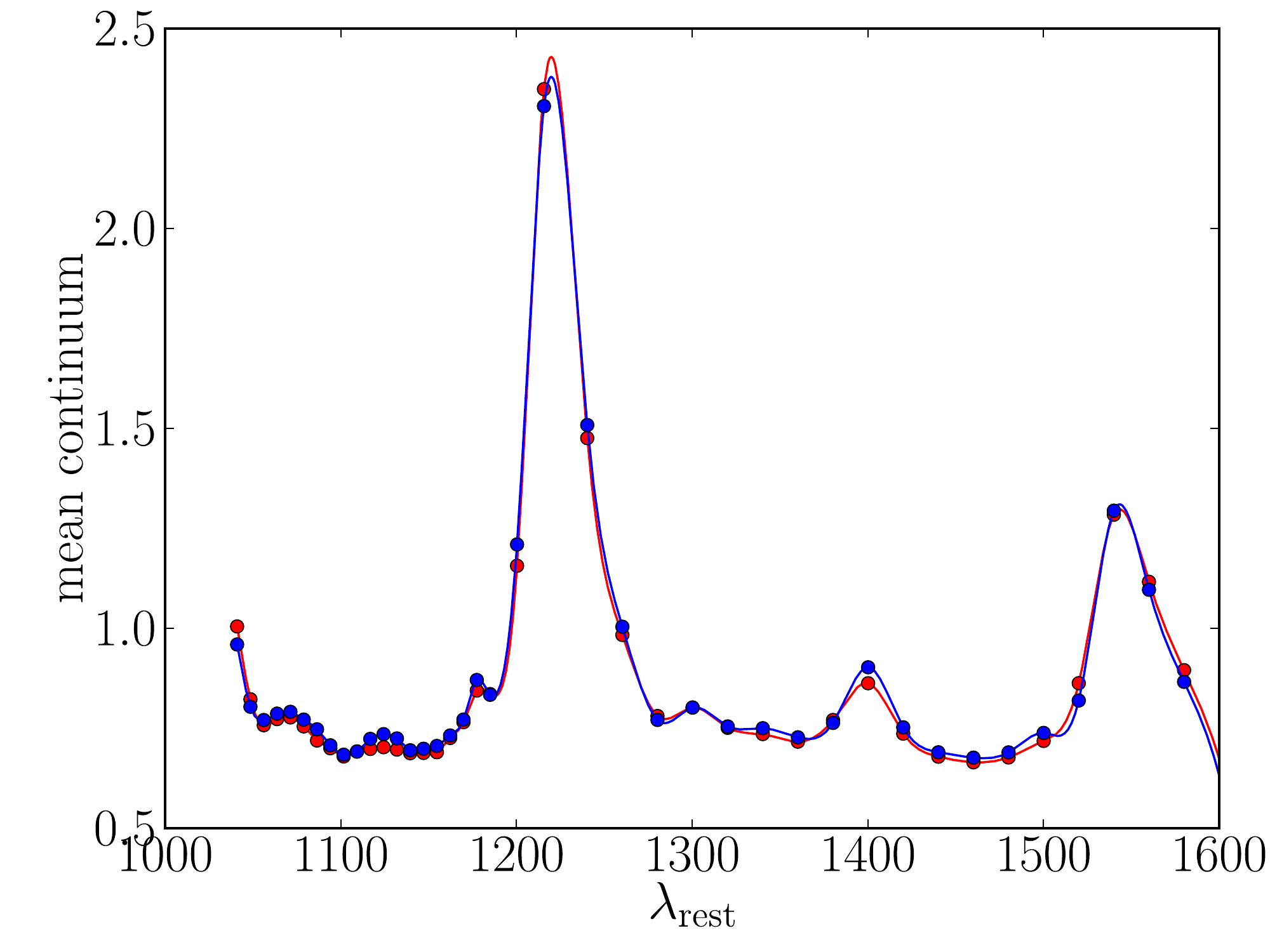}
      \includegraphics[width=0.5\linewidth]{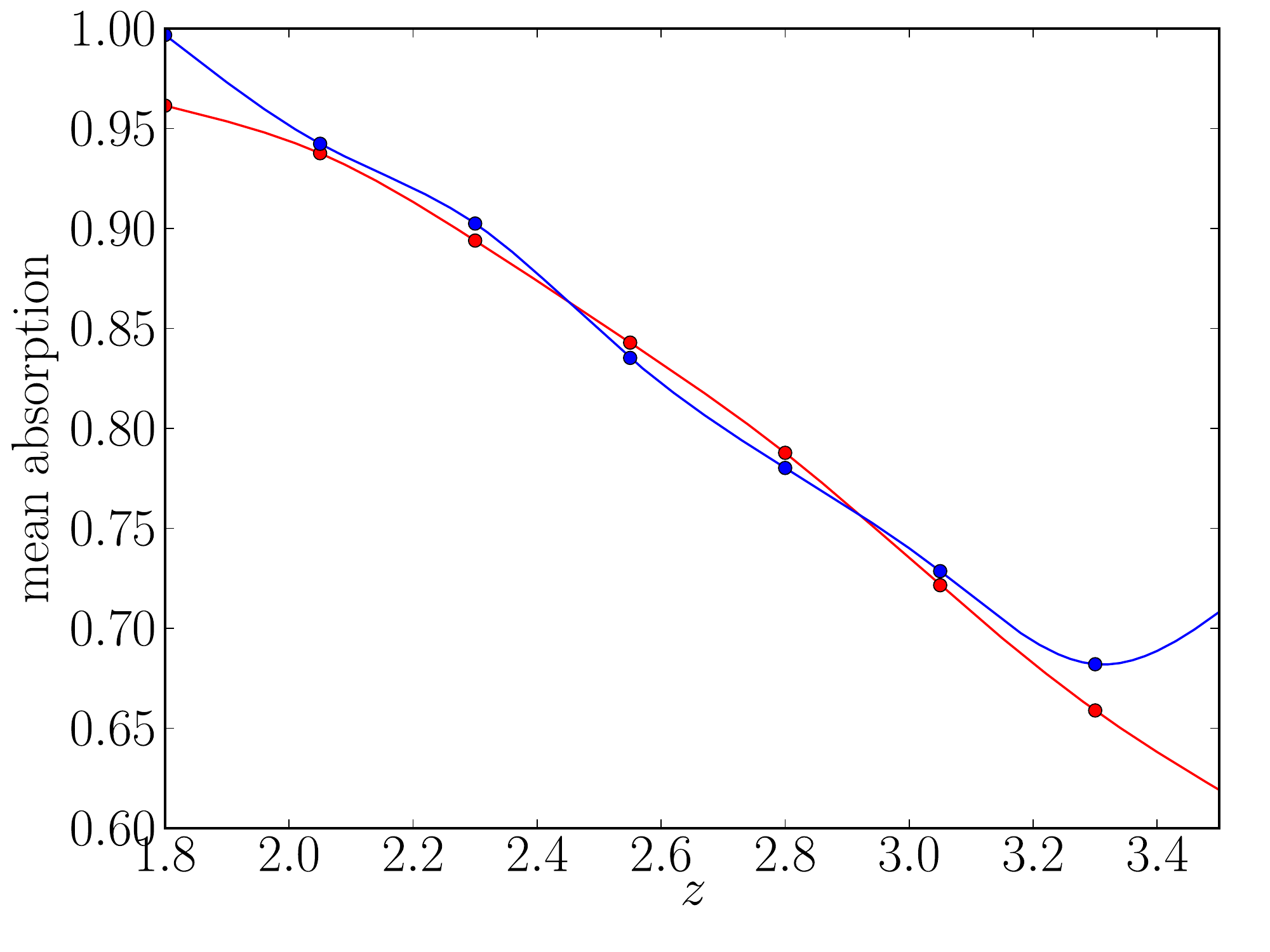}
    \end{tabular}
  \end{center}

  \caption{\small \label{fig:cfit}The mean continuum
    fits (left) and mean absorption fits (right), in arbitrary units,
    after thirty iterations of the
    continuum fitting pipeline for both the observed data (blue) and
    synthetic data (red). The
  curves diverge for $z<2$ and $z>3$ as there are virtually no data in
  those regions (see right panel of Figure \ref{fig:Nofz}).}
\end{figure}

Next we illustrate our continuum fitting process in Figure
\ref{fig:cfit} for the observed data and the synthetic data. The
left-hand panel shows the mean continuum, while the right hand panel
shows the mean absorption \emph{up to an arbitrary scaling factor}
(since it is completely degenerate with unabsorbed continuum
level). It is important to stress that no information from the actual
measured data went into this first generation of synthetic data. The
continua in the synthetic data were created from the
\emph{low-redshift} Hubble Space Telescope observations and the
agreement between real and synthetic data attests to the universality
of the quasar mean continuum. We do not plot the measurement errors
here, since the errors on the final correlation function are derived
using the measured correlations in the data, and these include all the
extra variance due to continuum errors.

\begin{figure}

  \begin{center}
    \includegraphics[width=0.5\linewidth]{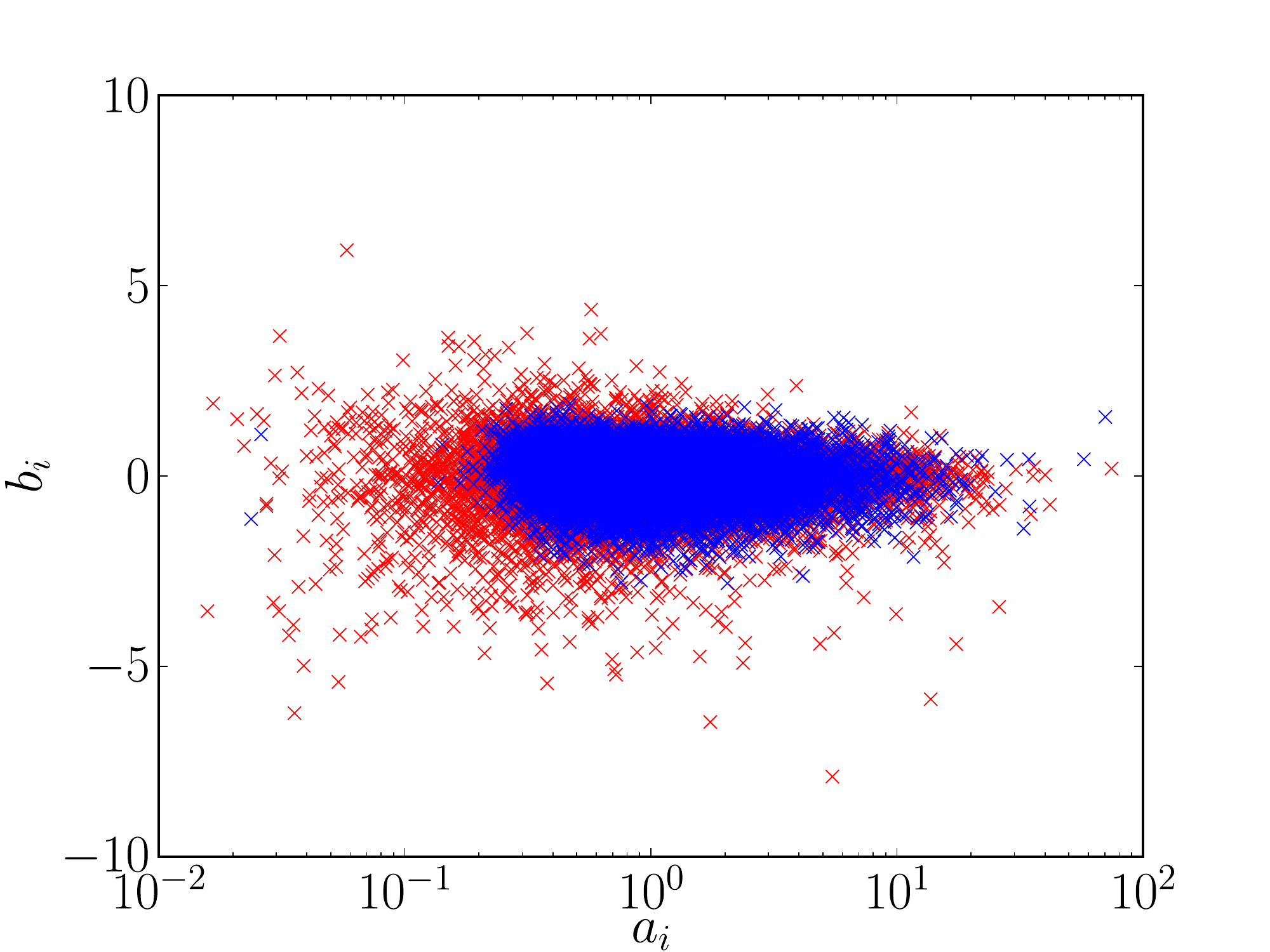}
  \end{center}

  \caption{\small \label{fig:parab} Distribution
    of parameters $a_i$ and $b_i$ for the data (red) and synthetic
    data (blue).}
\end{figure}

We illustrate the distribution of parameters $a_i$ and $b_i$ for real
and synthetic data in Figure \ref{fig:parab}. The distribution for the
observed data is considerably wider than that for the synthetic
data. This is most likely due to the presence of the
spectro-photometric errors in the data, but one should not exclude, in
principle, a wider intrinsic variation in the shape of the quasar
continua.  
However, we have shown that fitting out these parameters
does not affect the derived correlation function.

\begin{figure}

  \begin{center}
    \includegraphics[width=0.5\linewidth]{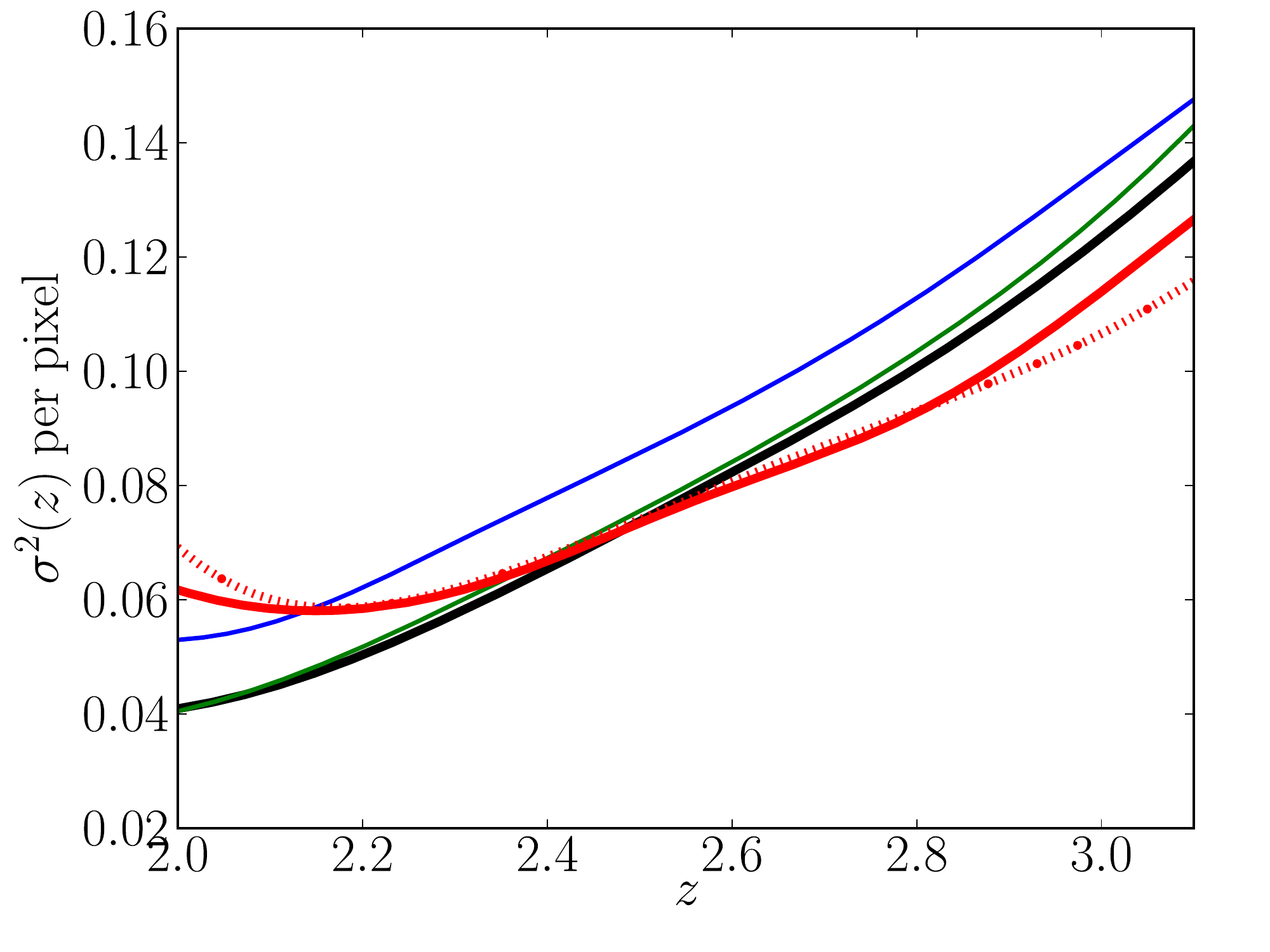}
  \end{center}

  \caption{\small \label{fig:psig} The remaining
    variance per pixel in the forest as a function of redshift after
    the observational noise has been subtracted. The thick
    black line is the value of the default synthetic data, while the
    blue line is the synthetic data with LLS/DLA added and the green line
    is with the forest metals added. The thick red line is for the data,
    and the dotted red line is for an alternative continuum fit to the data
    \cite{kgprep}.  
}
\end{figure}

In Figure \ref{fig:psig} we plot the variance per pixel of $\delta_F$,
after the observational noise has been subtracted. To
translate this quantity into a theory observable, one must take into
account the effects of the pixel size, the point spread function of the
telescope, and the variance introduced by the continuum fitting. We see
that at relatively high redshift the theoretical expectation obtained from the
synthetic data sets agrees within $\sim$ 10\%, although
there is an unexpected upward trend in the residual variance at
redshifts below 2.4. This will be further discussed in \S 7.
The solid red curve describes the data with our standard continuum
determination, while the dotted curve shows the results of using an
alternative continuum fitting method that will be described
elsewhere \cite{kgprep}.

\begin{figure}[h!]

  \begin{center}
    \includegraphics[width=1.0\linewidth]{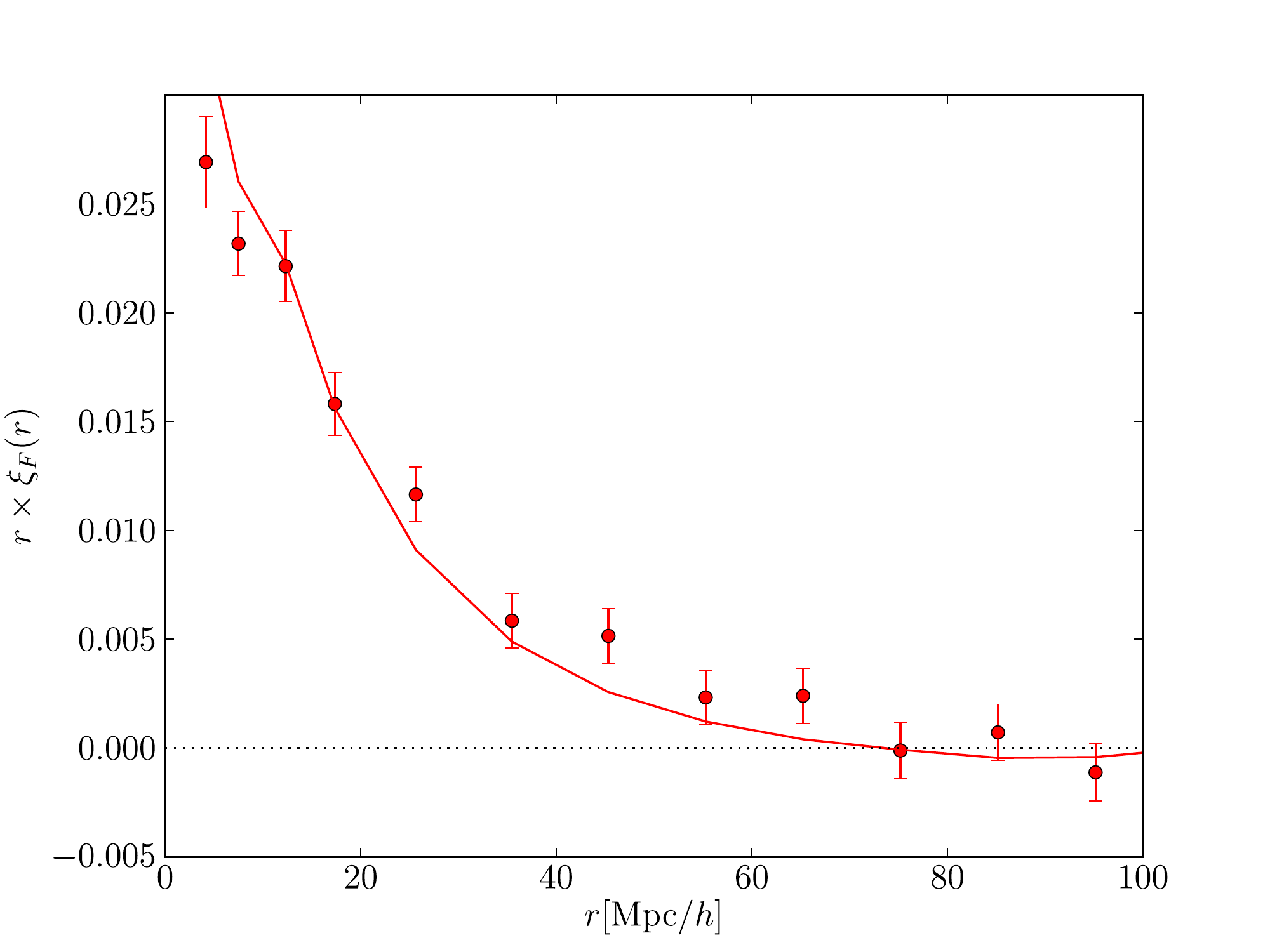}
\hspace*{-2cm}
    \begin{tabular}{cc}
    \includegraphics[width=0.6\linewidth]{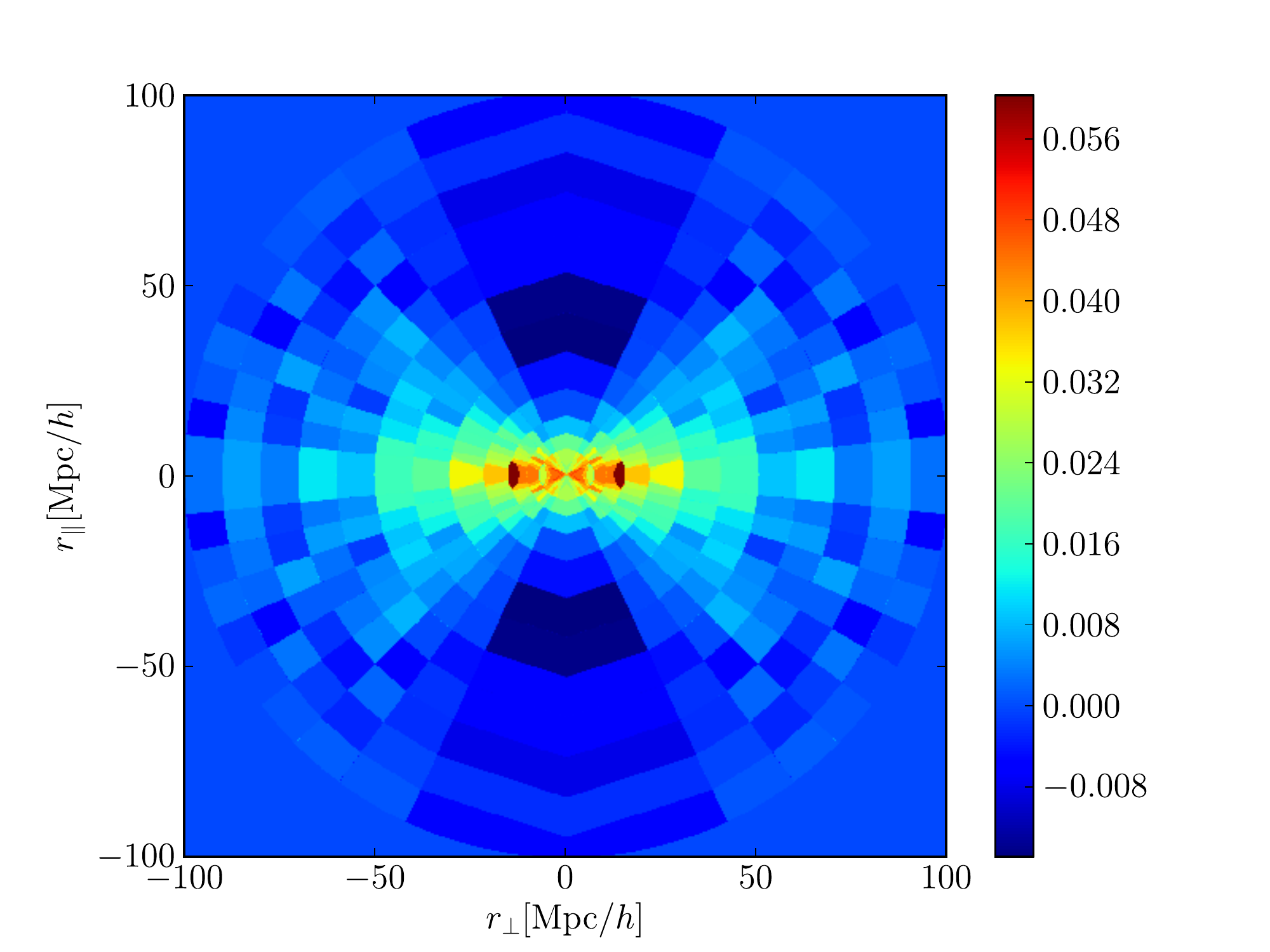} &
    \includegraphics[width=0.6\linewidth]{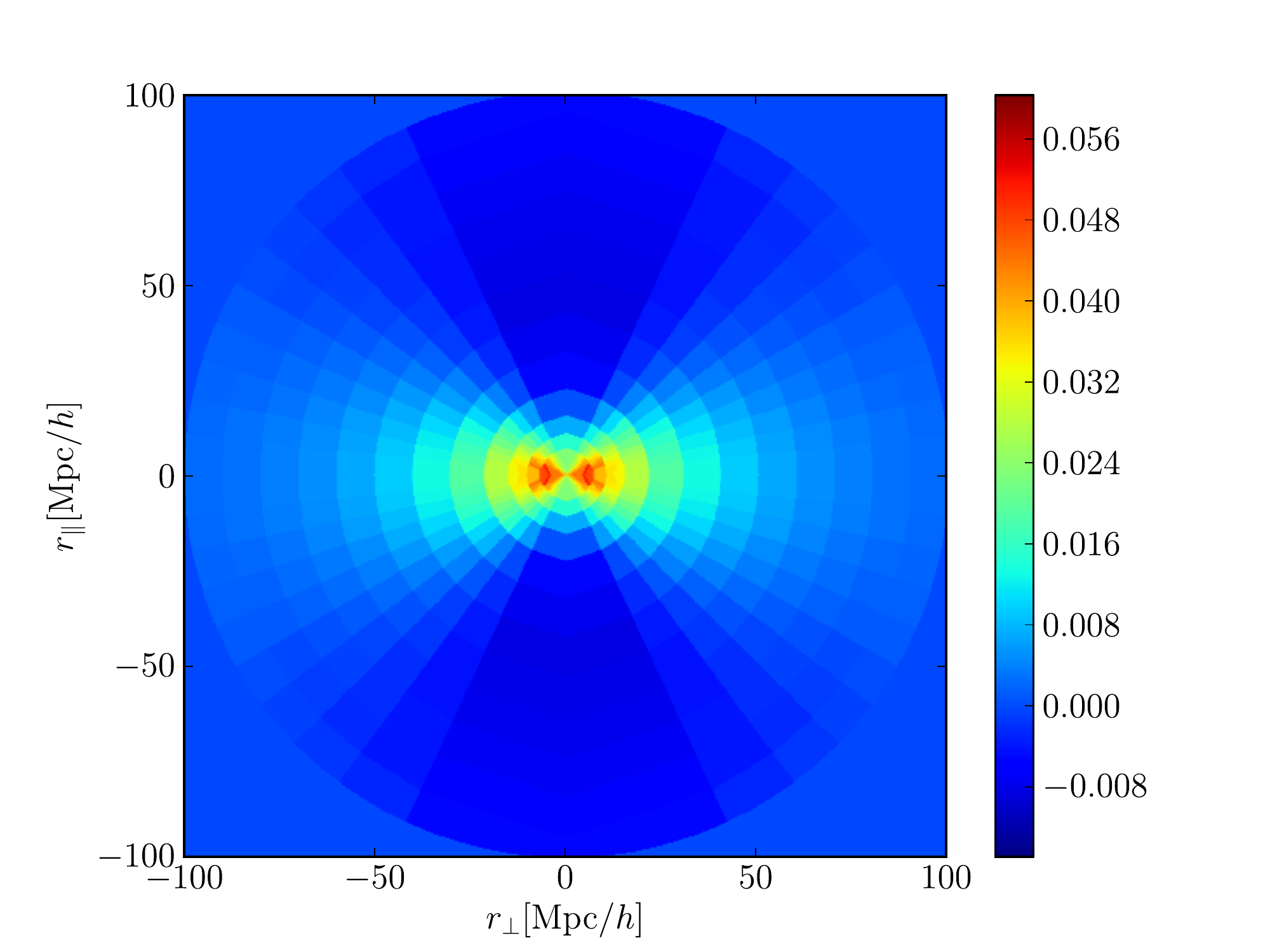}\\
  \end{tabular}
  \end{center}

  \caption{\small \label{fig:view} Primary measurement results in visual form. 
    Top
    plot shows the monopole of the correlation function,
    together with a best-fit two-parameter ($b$, $\beta$) linear
    model. The bottom two plots are redshift averaged data plotted
    in the plane $r_\perp-r_\parallel$, with each pixel plotted with
    the value corresponding to the nearest neighbor. The left panel
    corresponds to data, and the right panel to the corresponding
    best-fit theory.
  }
\end{figure}

Finally we proceed to the actual results on the correlation function.
Figure~\ref{fig:view} displays the measured correlation function as a
function of radius up to $100 \mpch$, together with the best fit
theory, after averaging over angle (upper plot), and also on a
two-dimensional grid of transverse ($r_\perp$) and radial
($r_\parallel$) separation (two lower plots).  These plots demonstrate
the main two results of this paper: we have detected correlations in
Lyman-$\alpha$ absorption in our 3-dimensional measurement out to
comoving separations of $60\hmpc$, and we have detected strong
redshift-space anisotropy in the flux correlation function.  Both
results are established at high statistical significance, and our
measurements are consistent with the predictions of a standard
$\Lambda$CDM cosmological large-scale structure model augmented by a
well motivated 3-parameter description of the relation between the
Lyman-$\alpha$ forest and the underlying dark matter distribution.
The parameter values --- describing linear bias, linear redshift-space
distortion, and redshift evolution --- are in accord with
{\it a priori} theoretical expectations, once we account for the impact of
high column density absorbers and metal lines.

We note also that our parameter errors are substantially larger for 
measurements on the real data relative to the mocks (not counting $\beta$, 
where the errors are highly dependent on the central value). Because the 
errors depend entirely on the
measured correlation function in each case, this implies a substantially 
different correlation function between the two cases. We have not investigated
carefully, but we suspect the difference is related to the large variance at
low $z$ in the real data (Figure \ref{fig:psig}), which is apparently not 
entirely compensated by boosted large-scale signal.

\begin{figure*}[h!]

\hspace*{-1.2cm}    \includegraphics[width=1.2\linewidth]{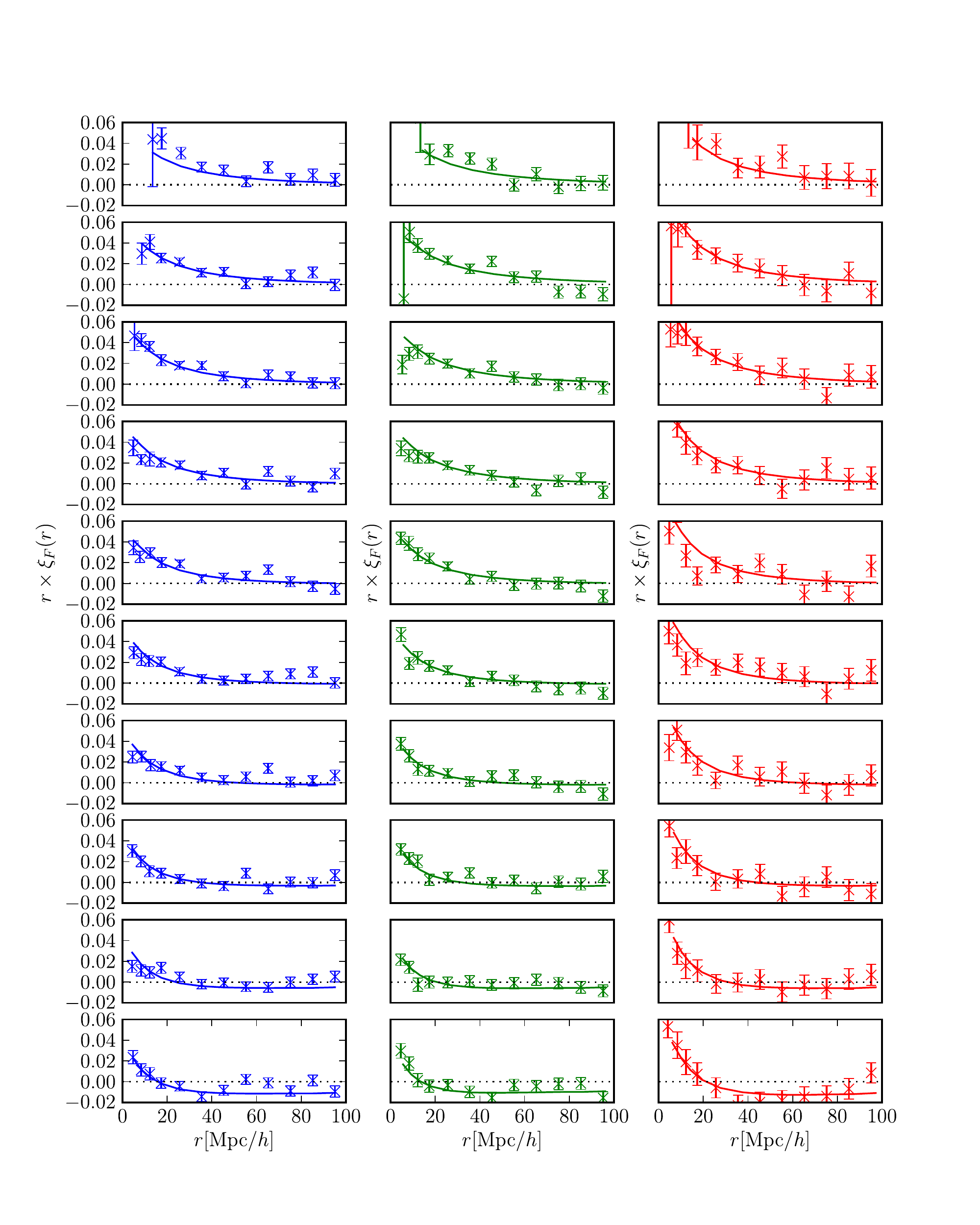}

\caption{\small \label{fig:allz} Measurements of the observed
  data. Columns correspond to the three redshift bins we use, with
  increasing redshift from left to right. The ten rows correspond to the ten
  bins of $\mu$, increasing from top to bottom. In each plot we show
  the measured $\xi_F$ as a function of separation
  for that particular redshift and $\mu$
  bin. The best-fit linear theory is over-plotted to guide the
  eye. The measurements are correlated and hence one should not
  evaluate ``$\chi^2$ by eye''.
}
\end{figure*}

\begin{figure*}[h!]

\hspace*{-1.2cm}    \includegraphics[width=1.2\linewidth]{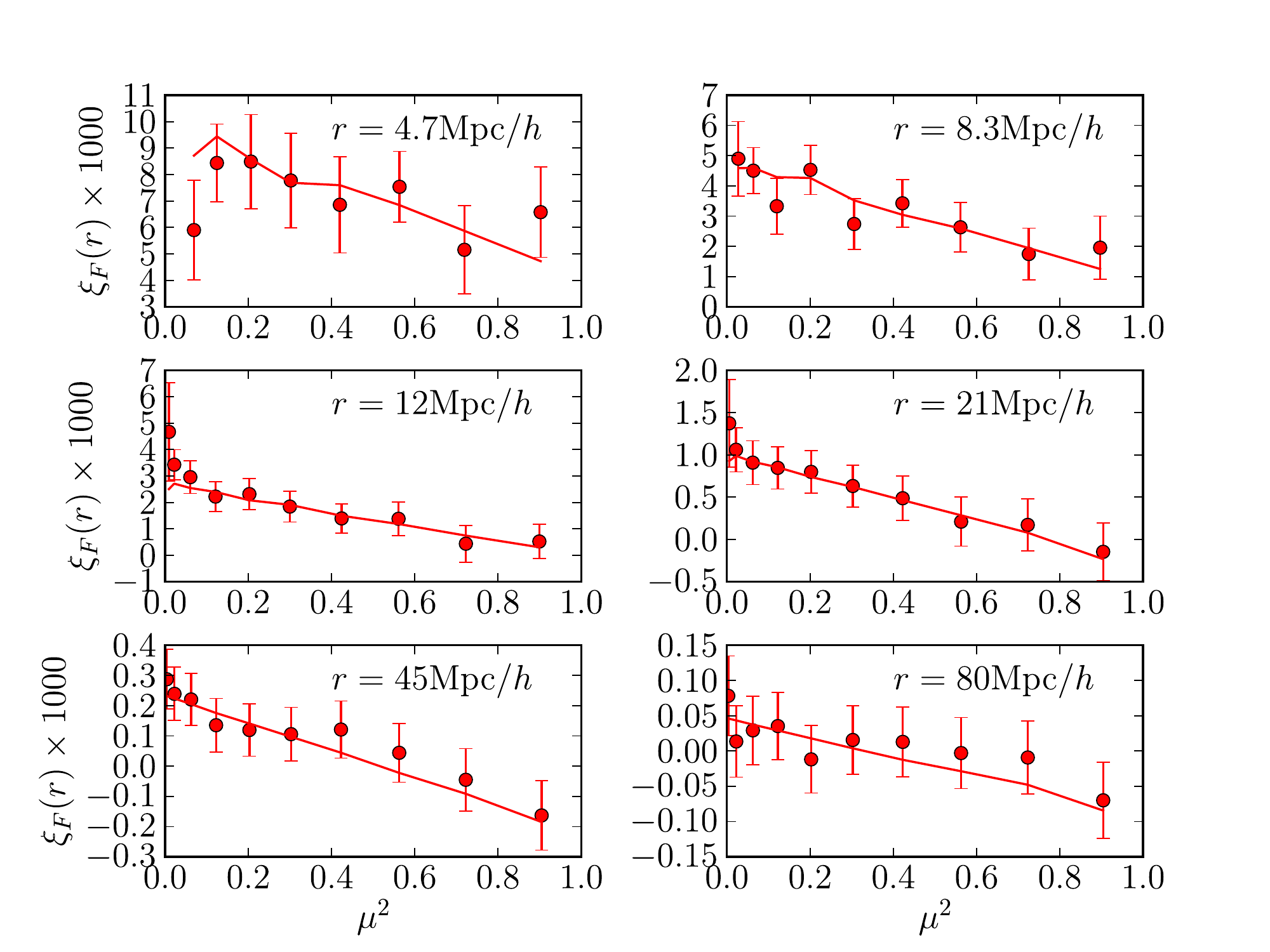}

\caption{\small \label{fig:mudep} Measurements of
the observed data. Each panel corresponds to redshift-averaged data at a
certain radius as a function of $\mu$. We also plot the best-fit
linear model to guide the eye. Measurements are correlated and
hence one should not evaluate ``$\chi^2$ by eye''. }
\end{figure*}

In Figure \ref{fig:allz}
we show the actual measured data-points of $\xi_F(r,\mu,z)$ in 30 panels
for each bin in $\mu$ and $z$, together with the best-fit
theory to guide the eye. We plot the square root of the diagonal
elements of the covariance matrix as error bars, but measurements are
correlated and therefore one should not attempt to evaluate $\chi^2$ by
eye. The points in this figure are our actual measurements
used in the fitting of the bias parameters. For easier
visualization of the results, we also convert them to a few
alternative forms.  In Figure \ref{fig:mudep} we average over redshift
bins and radial bins in some cases and plot the same datapoints as a
function of $\mu$.

\begin{figure}[h!]

  \begin{center}
    \includegraphics[width=1.0\linewidth]{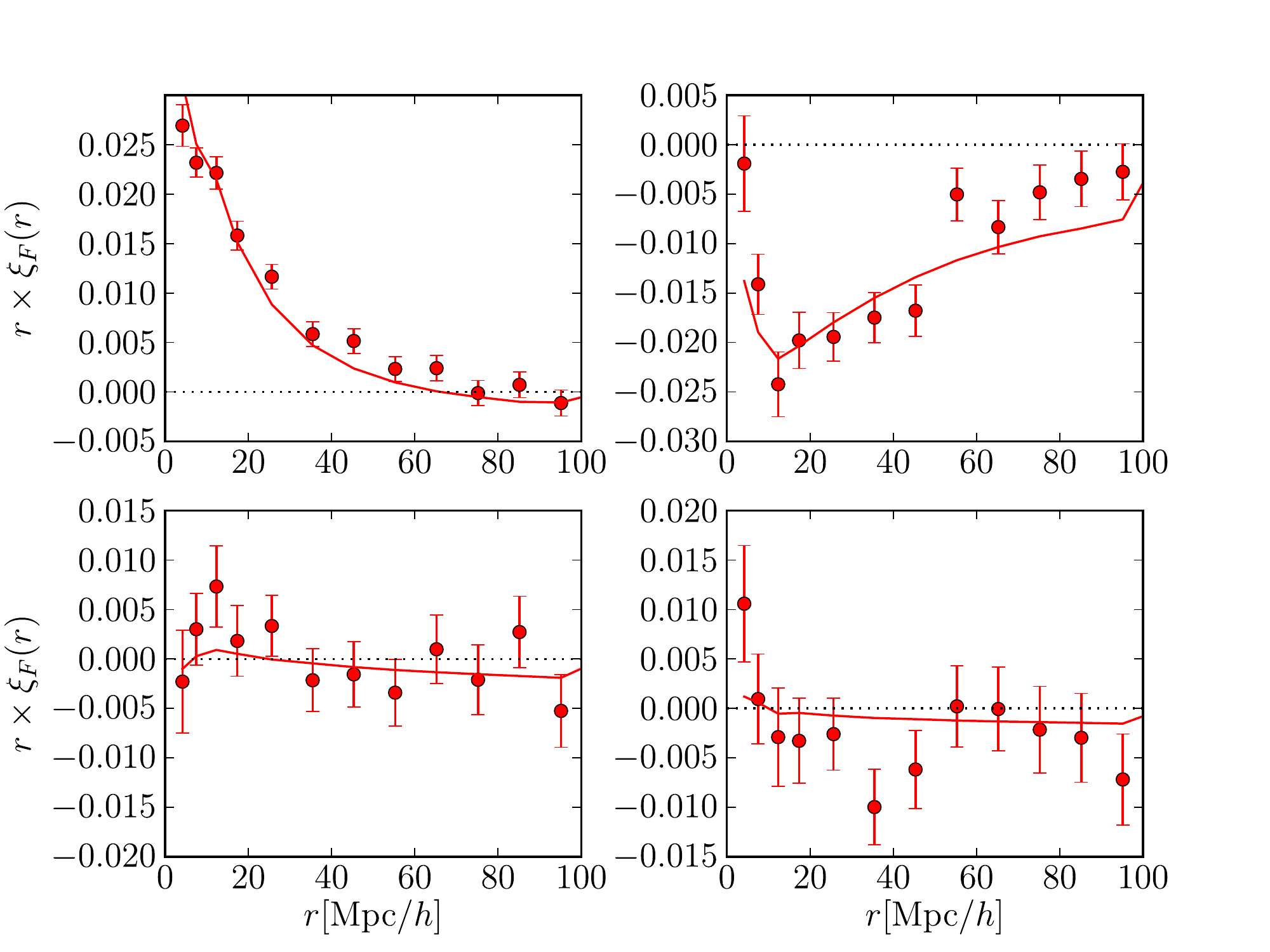}
  \end{center}

  \caption{\small \label{fig:multi} Results of Figure \ref{fig:allz}
    converted to multipoles. The four panels correspond to the
    redshift-averaged monopole, quadrupole (top row), hexadecapole and $\ell=6$
    moment (bottom row). Lines are best-fit theory. 
  }
\end{figure}

In Figure \ref{fig:multi} we convert the ten $\mu$ measurements,
averaged over redshift, into measurements of the multipoles of the
correlation function. We perform the same operation for the
best-fit theory. 
Our results are in striking agreement with the predictions of the
standard linear (i.e., extended Kaiser) theory: we see a strong 
monopole correlation
function that is proportional to the linear theory matter correlation
function and a strong
detection of a negative quadrupole, which is the signature of linear theory
large-scale redshift-space distortions.  
(Note that the mean removal effect and the uneven redshift of
individual points create a small $\ell=6$ moment for the theoretical
predictions, even though the $\ell=6$ moment is exactly zero in pure linear 
theory.) 

These results confirm that we have detected correlations in the \lya
forest flux in three dimensions out to a much larger scale than in
previous measurements, and that we have detected the linear redshift
distortions for the first time in the \lya forest. This demonstrates that,
on the large scales in which these linear correlations are measured,
the dominant source of the \lya forest transmission variations arise
from gravitational instability of primordial mass fluctuations.

\begin{figure}[h]
  \begin{center}
    \begin{tabular}{cc}
    \includegraphics[width=0.5\linewidth]{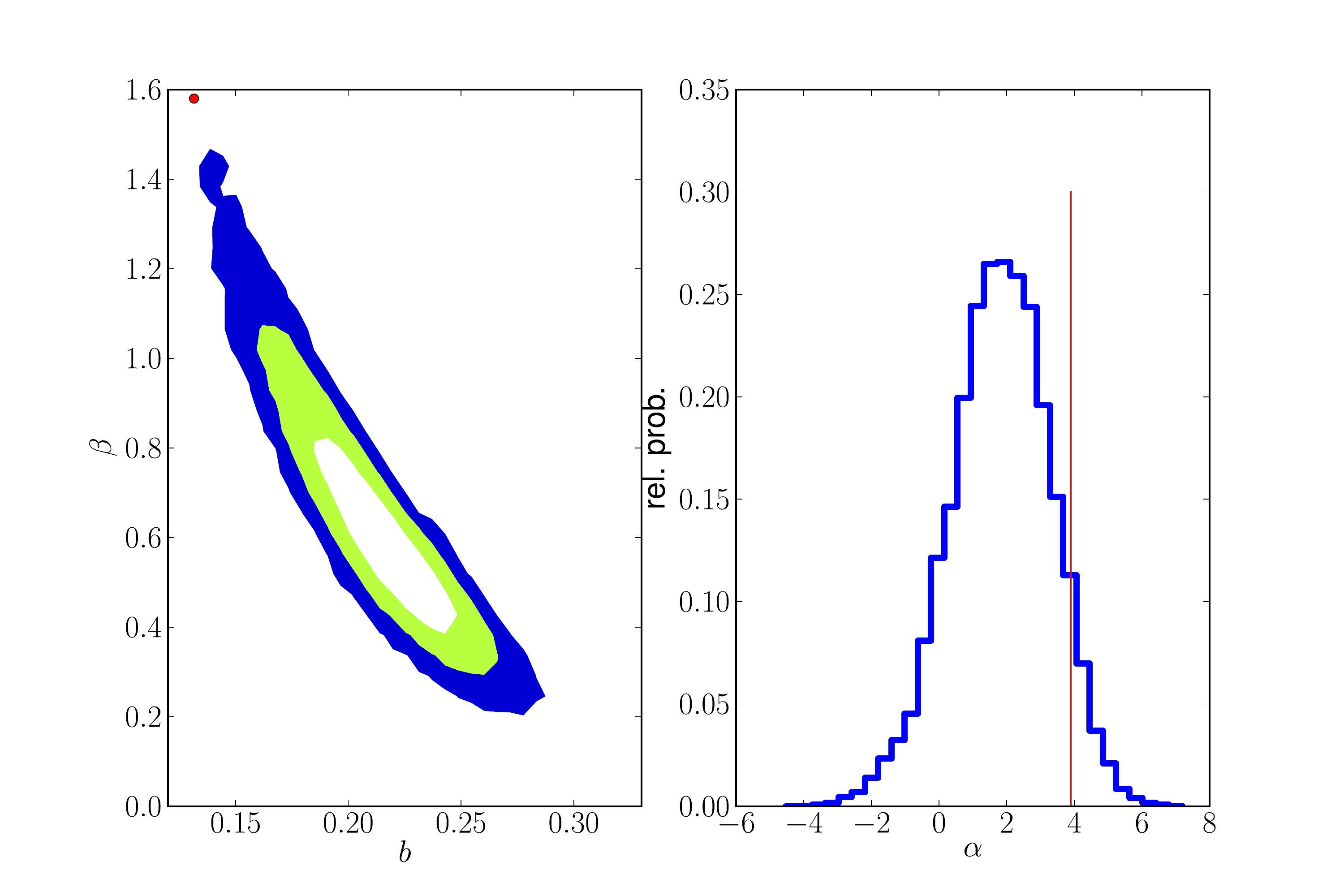}  & 
    \includegraphics[width=0.5\linewidth]{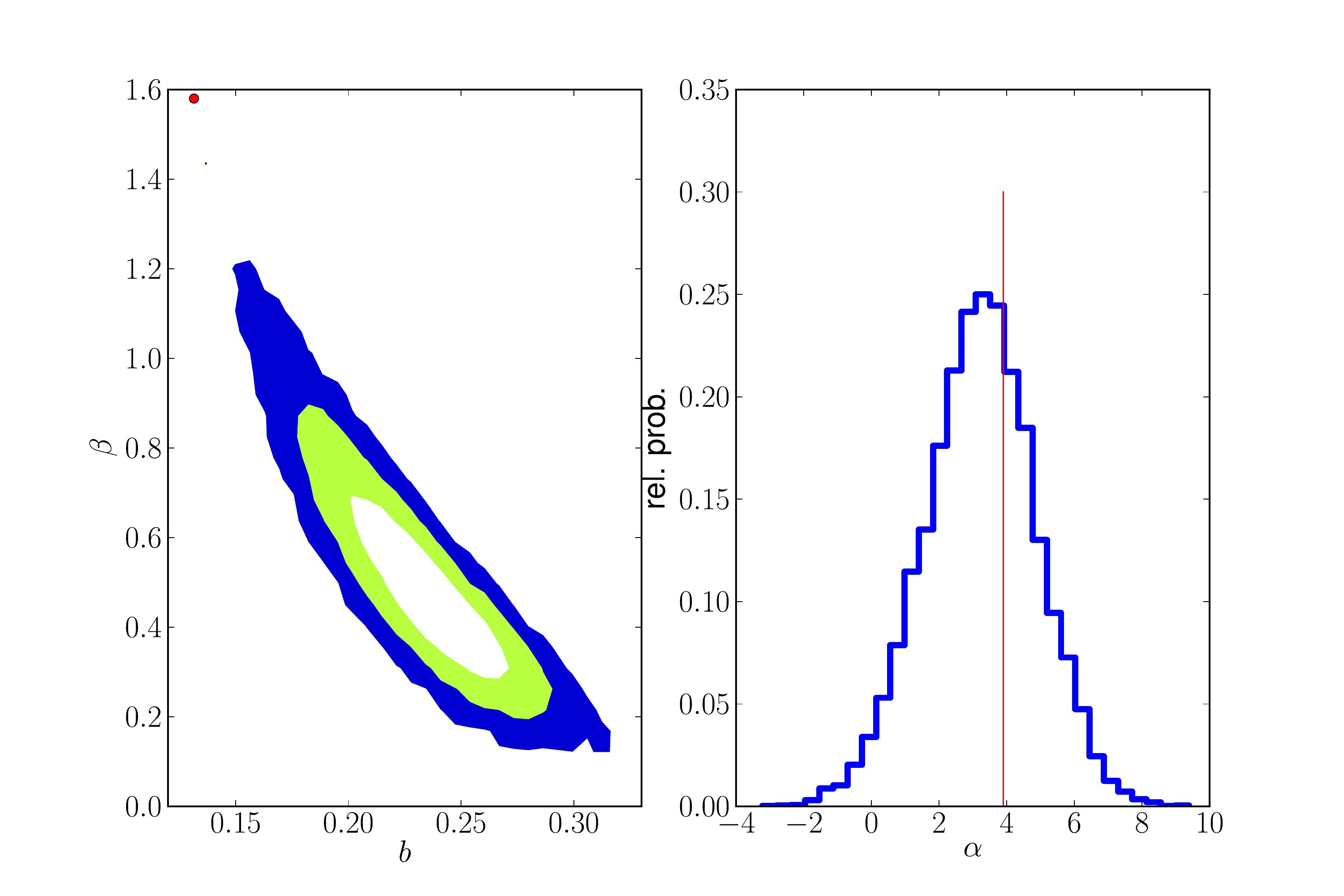} \\
    \includegraphics[width=0.5\linewidth]{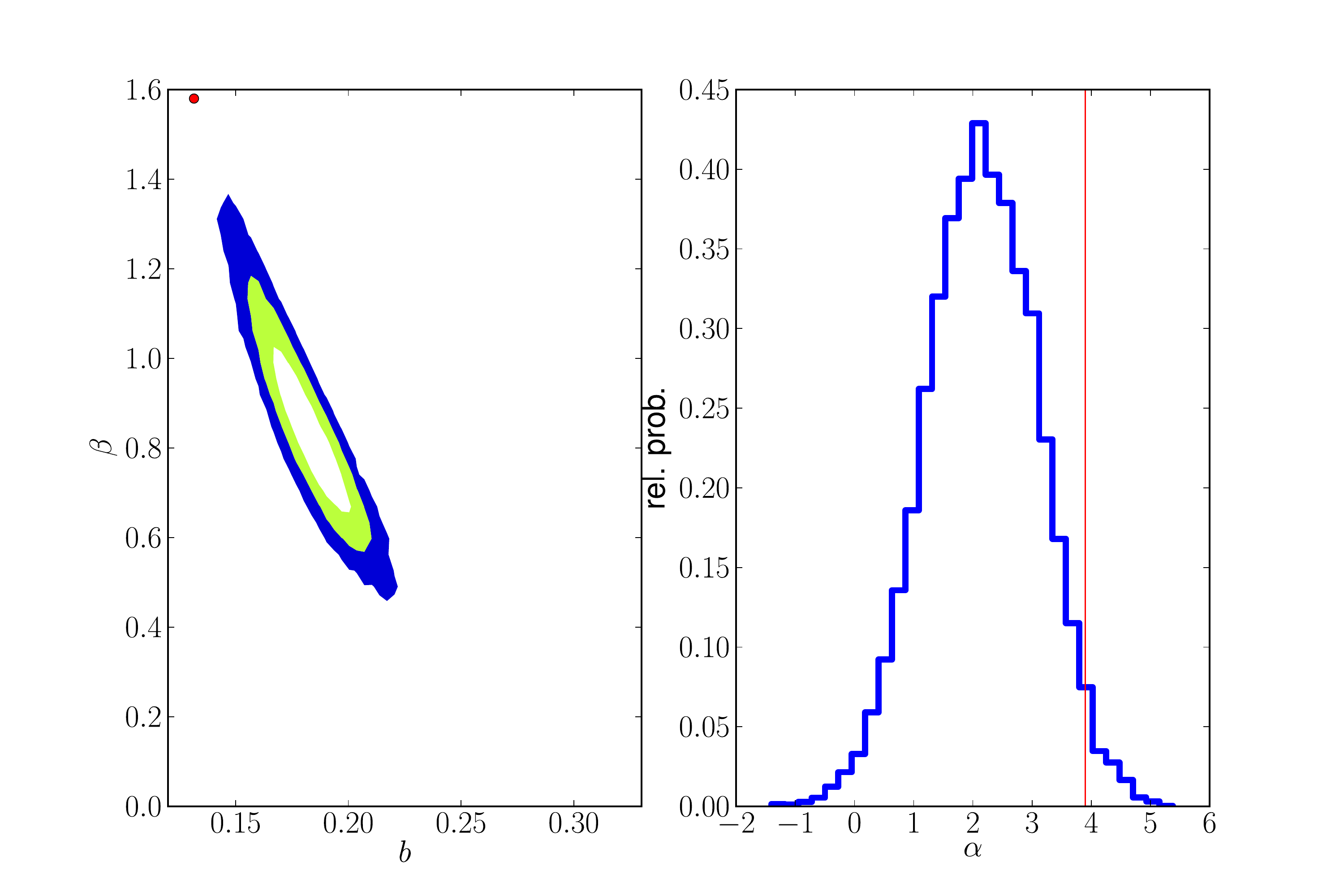} &
    \includegraphics[width=0.5\linewidth]{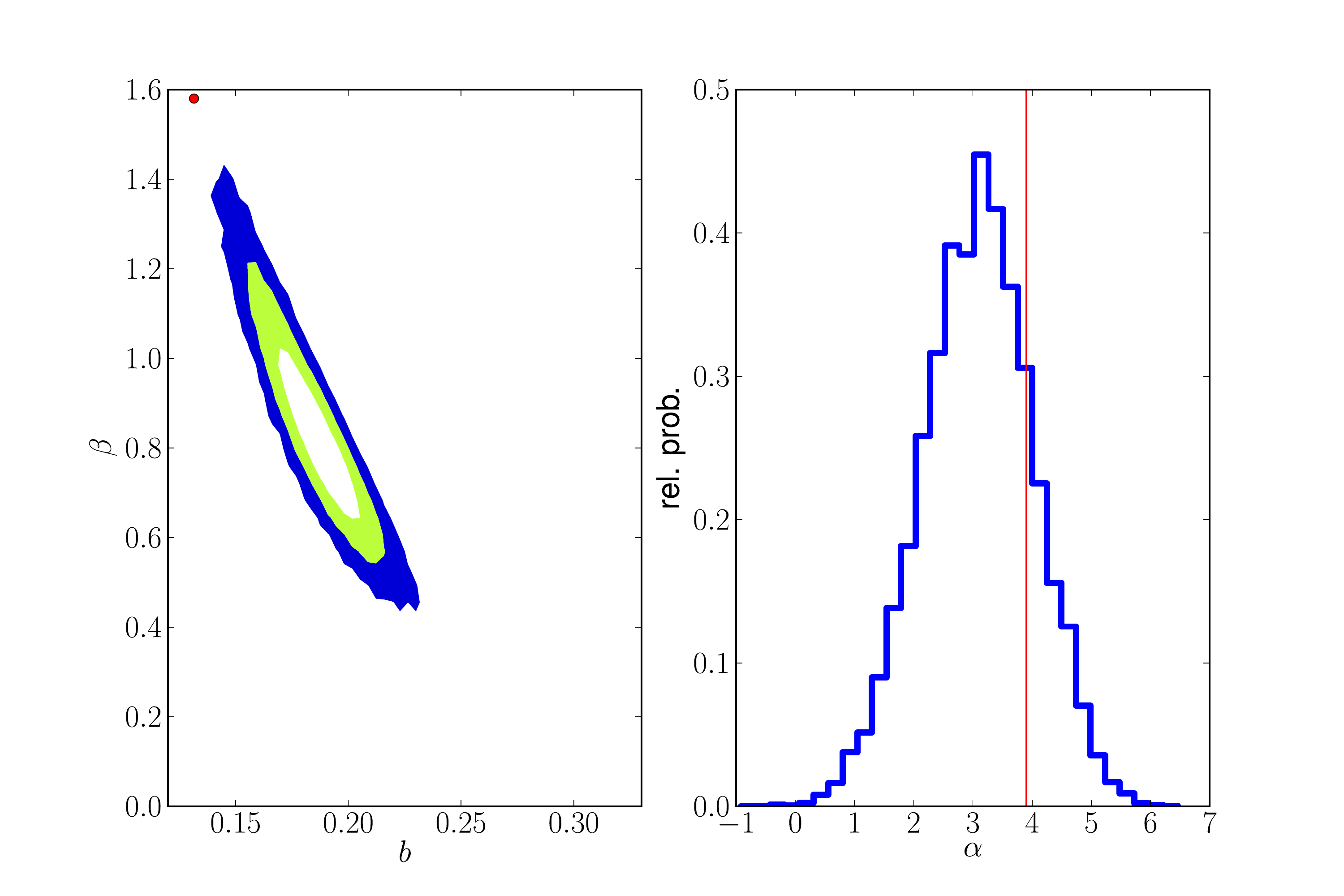} \\
  \end{tabular}
  \end{center}
  \caption{\label{fig:fitreal} Fits to the real data. The upper panels
    correspond to the data fitted using points with separations
    $r>20\mpch$ while the lower panel is for fits using points with
    $r>10\mpch$. The left-hand-side plots is for the default dataset,
    while the right hand side plot is for data that include quasars
    flagged as harboring DLAs by the FPG.  The red-point corresponds
    to the value that was used in the creation of the synthetic
    datasets. }
\end{figure}

To put this claim on a more quantitative basis, we fit 
the bias parameters as described in
section \ref{sec:fitt-bias-param}.
The results are plotted in Figure \ref{fig:fitreal} and given in 
Table \ref{tab:res}. The best-fit $\chi^2$ values are 233 with 237
degrees of freedom when fitting with points $r>20\mpch$ and 281 with 297
degrees of freedom when using points with $r>10\mpch$.

Compared to the simulations of reference \cite{2003ApJ...585...34M},
the data prefer lower values of $\beta$ and higher values of bias,
although the product $b(1+\beta)$ is of the right magnitude. We also
see a somewhat lower evolution with redshift.  The formal probability
that $\beta>1.5$ is $\sim 3\times 10^{-3}$ ($\sim 1\times 10^{-3}$ for
$r>10\mpch$ fitting; both cases probably dominated by the MCMC
sampling noise), and $\beta>1.0$ about $10\%$ (24\% for $r>10\mpch$
fitting). For fitting each redshift bin individually, the
probabilities are 11\%, 24\% and 48\% for $\beta>1.5$ and 33\%, 48\%
and 65\% for $\beta>1.0$ for the lowest, medium and highest redshift
bin respectively. We see that the low value $\beta$ for a single
$\beta$ fit is driven a lot by the lowest redshift bin. Our synthetic
data sets show that metals and high column density systems (LLS/DLAs)
can considerably lower the observed value of $\beta$.  Clearly, more
work will be required to explain in detail the values of bias
parameters that we find.

To determine the distance to which we have formally detected
correlations, we calculate the value of $\chi^2$ for a model with no
correlations and compare it to the $\chi^2$ of the three parameter
model.  Using this criterion, we have detected correlations up to a
distances $60\mpch<r<100\mpch$ at 3-$\sigma$ ($\Delta \chi^2>9$), and
up to $70\mpch<r<100\mpch$ at 2-$\sigma$ ($\Delta
\chi^2>4$). Similarly, we have detected redshift-space distortions
with high significance. For $r>20\mpch$, setting $\beta=0$ while
varying other parameters results in $\Delta \chi^2 \sim 48$, while
using distance $r>10\mpch$ one gets $\Delta \chi^2 \sim 120$. These
large $\Delta \chi^2$ values imply very high statistical
confidence. We quote this numbers as ``over $5\sigma$'' to conservatively take
into account potential imperfections in the error matrix.

\begin{table}
  \centering
  \hspace*{-0.8cm}
  \begin{tabular}{c|cccc}
    Data & bias & $\beta$ & $b(1+\beta)$ & $\alpha$ \\
    \hline 
    $r>20\mpch$  & $0.197\pm{0.021}$ & $0.71\pm^{0.21\ 0.49\ 0.87}_{0.16\ 0.27\ 0.39}$ & $0.336\pm{0.012}$ & $1.59\pm{1.55}$ \\
    $r>10\mpch$  & $0.175\pm{0.012}$ & $0.90\pm^{0.15\ 0.33\ 0.56}_{0.13\ 0.23\ 0.33}$ & $0.333\pm{0.008}$ & $2.09\pm{0.94}$ \\
    With LLS/DLA, $r>20\mpch$   & $0.217\pm{0.025}$ & $0.55\pm^{0.19\ 0.48\ 0.97}_{0.14\ 0.25\ 0.35}$ & $0.337\pm{0.014}$ & $2.99\pm{1.74}$ \\
    With LLS/DLA, $r>10\mpch$ & $0.180\pm{0.013}$ & $0.87\pm^{0.16\ 0.35\ 0.56}_{0.13\ 0.25\ 0.35}$ & $0.337\pm{0.009}$ & $3.11\pm{0.93}$ \\
    \hline
    $r>20\mpch$, \\
     1$^{\rm st}$  $z$-bin $z=2.0-2.2$ & $0.168\pm{0.033}$ &  $0.81\pm^{0.54\ 1.64\ 3.33}_{0.30\ 0.49\ 0.65}$ & $0.305\pm{0.017}$ & / \\
      2$^{\rm nd}$  $z$-bin $z=2.2-2.4$ & $0.167\pm{0.037}$ & $0.97\pm^{0.82\ 2.55\ 3.86}_{0.39\ 0.62\ 0.78}$ & $0.330\pm{0.019}$ & / \\
     3$^{\rm rd}$ $z$-bin $z=2.4-3.0$ & $0.164\pm{0.047}$ & $1.33\pm^{1.56\ 3.21\ 3.65}_{0.66\ 0.98\ 1.19}$ & $0.383\pm{0.033}$ & / \\
    \hline
    $r>20\mpch$, \\
    $\mu>0.1$  & $0.200\pm{0.021}$ & $0.63\pm^{0.19\ 0.45\ 0.78}_{0.15\ 0.27\ 0.38}$ & $0.325\pm{0.013}$ & $1.57\pm{1.77}$ \\
    $\mu<0.9$  & $0.206\pm{0.023}$ & $0.65\pm^{0.21\ 0.48\ 0.80}_{0.16\ 0.28\ 0.38}$ & $0.341\pm{0.013}$ & $1.33\pm{1.60}$ \\
    \hline
    $r>20\mpch$, \\
    1041\AA$<\lambda_{\rm rest} <$ 1120\AA & $0.207\pm{0.050}$ & $0.36\pm^{0.30\ 0.96\ 2.66}_{0.18\ 0.30\ 0.35}$ & $0.285\pm{0.054}$ & $-9.31\pm{13.87}$ \\
    1120\AA$<\lambda_{\rm rest} <$ 1185\AA & $0.118\pm{0.032}$ & $1.62\pm^{1.54\ 2.98\ 3.35}_{0.74\ 1.08\ 1.31}$ & $0.311\pm{0.024}$ & $3.40\pm{2.91}$ \\
    \hline
    $r>20\mpch$, \\
     $g<20.5$ & $0.137\pm{0.035}$ & $1.10\pm^{1.01\ 2.94\ 3.82}_{0.45\ 0.71\ 0.90}$ & $0.295\pm{0.042}$ & $-9.10\pm{7.55}$ \\
     $g>20.5$ & $0.162\pm{0.035}$ & $1.20\pm^{0.99\ 2.78\ 3.72}_{0.45\ 0.71\ 0.89}$ & $0.357\pm{0.022}$ & $2.64\pm{2.36}$ \\
    \hline
  \end{tabular}
  \caption{This table shows the results of the parameter fittings for
    the data and various systematics checks. All error bars are $1-\sigma$
    error bars, except for the $\beta$ parameter in which case we give
    $1$,$2$ and $3-\sigma$ confidence limits.
  }
  \label{tab:res}
\end{table}

\section{Systematic effects, cross-checks and data-splits}
\label{sec:syst-effects-cross}

In the previous section we have shown results of fitting the bias
parameters to the measured correlation function. In this section we
investigate how stable these results are against various divisions of
the data.  We always perform two basic cuts on the data: we never use
pairs of pixels from the same quasar, and we never use pairs of pixels
that are separated by less than $1.5\, $\AA. In addition,
we have performed several tests by splitting the data in a variety of
ways. Generally, for each split we measure
the bias parameters and compare these to the full dataset. If the
combined data were inconsistent with a subset of the data, this
might indicate some unknown systematic error.
 For example, if what we are measuring are truly cosmological
fluctuations, our results should be the same regardless of using
bright or faint quasars. We perform the following data splits:

\begin{figure}[h]
  \begin{center}
    \includegraphics[width=1.0\linewidth]{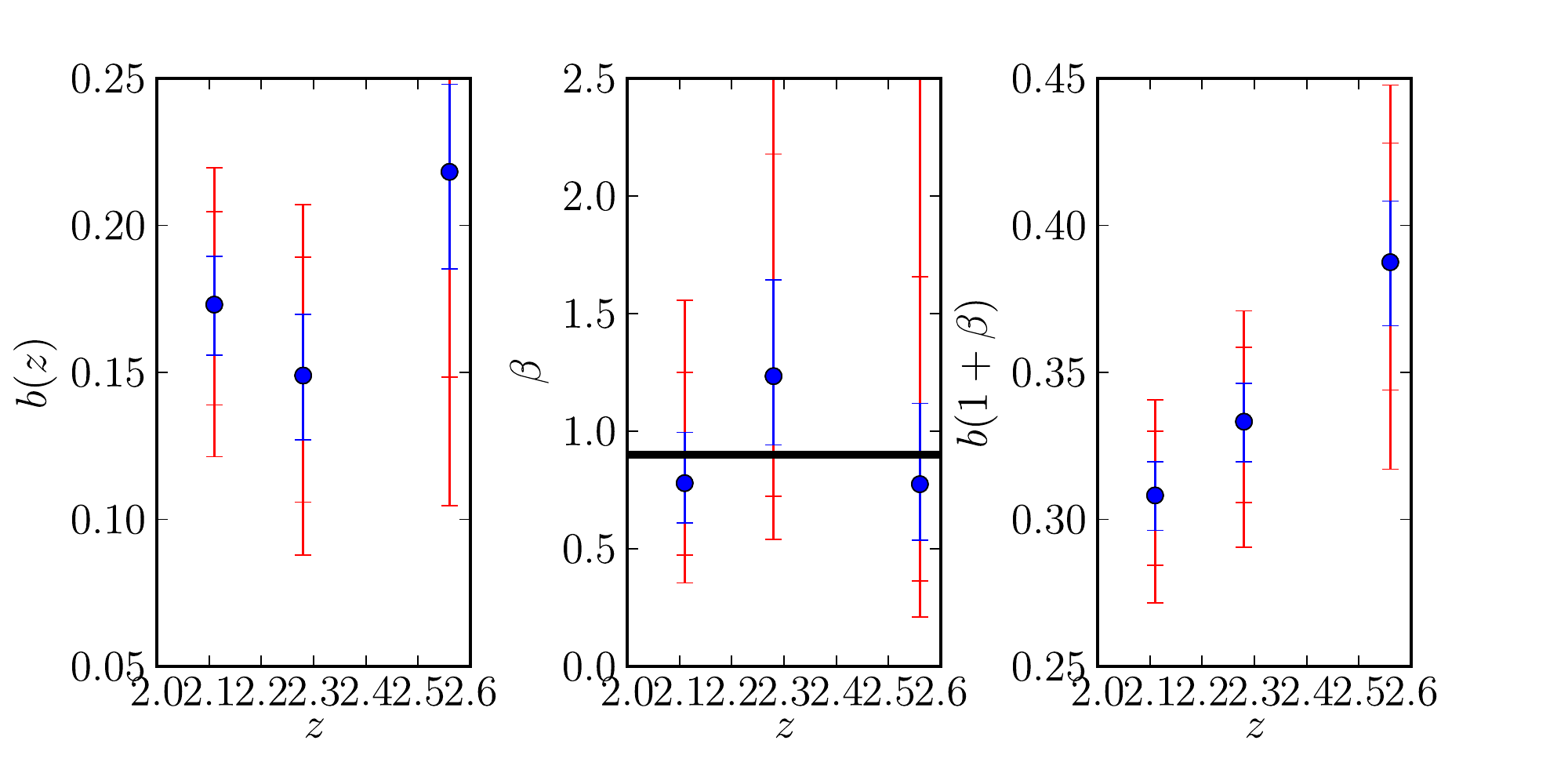}
  \end{center}
  \caption{\label{fig:koot} In this plot we show fits to the real
    data when the $b$ and $\beta$ parameters are allowed to be a
    function of redshift. We plot the 1,2 and 3 $\sigma$ error
    bars. The bias in this plot is with respect to the fiducial
    cosmological model with $\sigma_8=0.8$ at the redshift of interest,
    therefore the numbers cannot be directly compared with the fitted
    $\alpha$,$b$ parameters. We plot the value of $\beta$ determined
    from the overall fit as a black solid line. 
    All fits in this figure are limited to $r>10 \mpch$.
}
\end{figure}

\begin{figure}[h]
  \begin{center}
    \includegraphics[width=1.0\linewidth]{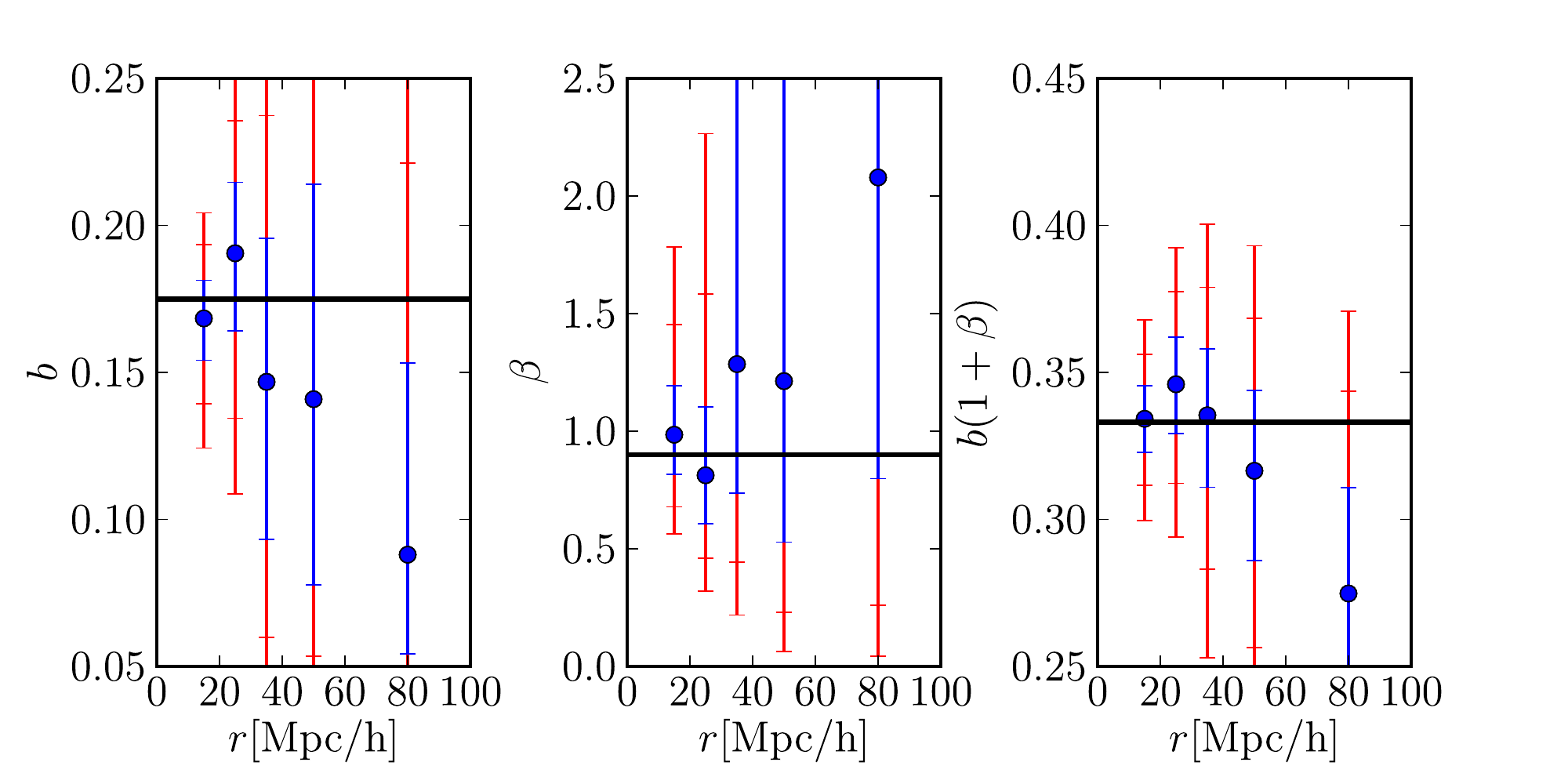}
  \end{center}
  \caption{\label{fig:shoot} Results of fits of the real data to the
    $b$ and $\beta$ parameters when they are allowed to be a function
    of scale. We plot constraints on $b$, $\beta$ and $b(1+\beta)$,
    and their 1,2 and 3 $\sigma$ error bars measured from the
    corresponding percentiles of MCMC chains. Note that $\beta$ and
    $b(1+\beta)$ have flat priors on them and that $b$ is a derived
    parameters.  The solid black thick lines correspond to the best
    fit parameters determined from fitting $r>10\mpch$ points.  }
\end{figure}

\begin{itemize}

\item \emph{Perform the fitting excluding the lowest $\mu$ or the
    highest $\mu$ bin.} In both cases, the central values shift and
  error bars increase, but the overall fit is not driven by either low
  and high $\mu$ values alone. 

\item \emph{Perform the fitting as a function of separation. We
    fit for bias-$\beta$ parameters in 5 radial bins.} Results of this
  test are plotted in Figure \ref{fig:shoot}. The value of
  $\alpha$ was  fixed at the fiducial value of $\alpha=3.9$, but
  $\alpha$ is not degenerate with other parameters. 

\item \emph{Perform the fitting as a function of redshift.}  We fit
  for the bias and $\beta$ parameters independently in each redshift
  bin. Results of this test are plotted in Figure \ref{fig:koot}.

\item\emph{ Split by \lyaf range.} We divided the total data into two 
  segments $1041{\rm\AA}-1120{\rm\AA}$ and $1120{\rm\AA}-1185{\rm\AA}$. Note
  that each of these sets contain only a quarter of the original pairs
  (the remaining half being in the cross-correlations between the two
  halves). The measurement of $\alpha$ is very poor using the short wavelength
end because there is a dearth of pixels at relatively high redshift in this 
case.

\item \emph{Split by quasar brightness}. We divided by $g$-band
  magnitude into quasars brighter or fainter than 20.5. Note that each
  of these sets contain only a quarter of the original pairs.
The measurement of $\alpha$ is very poor using the bright quasars
because there is a dearth of pixels at relatively high redshift in this case.

\end{itemize}

None of these tests resulted in a detection of a significant
systematic effect. Results of this exercise are also presented in
Table \ref{tab:res}.
Note, however, that these splits are not entirely
satisfactory, as the differences are hard to test at precision approaching the
overall precision. E.g., in the case where $\mu$ is restricted to be $>0.1$,
$b (1+\beta)$ changes from $0.336\pm 0.012$ to $0.325\pm 0.013$. This seems 
like a small change, but on the other hand it is almost $1~\sigma$ when only
a small fraction of the data is removed. If we think of $\mu>0.1$ and $\mu<0.1$
as two independent measurements of $b (1+\beta)$, the implied difference 
between the values is $0.074\pm 0.034$, i.e., a $2.2~\sigma$ difference. There 
is no
compelling reason to believe this is evidence for a systematic error (this was
chosen out of several possibilities as an example of a big difference), but it
highlights the fact that a split can differ by $\sim 6~\sigma$ by the scale of
the full-sample errors (i.e., 0.074/0.012) yet still not be decisive evidence 
for a systematic problem, i.e., big effects could hide in this kind of test.
In the cases where the
sample of quasars is split, the errors often expand catastrophically due to
a reduction in cross-correlation pairs, so in the end the tests are
not actually very powerful.
In the future,
we may try to squeeze out more powerful tests by, e.g., fitting for
some simple parameterized dependence of $b$ and $\beta$ on quasar
magnitude (or other properties), continuing to use the full data set
in the fit.  

\section{Discussion}
\label{sec:discussion}

We measure the redshift-space distortion parameter $\beta$ to be
between 0.44 and 1.2 at the 95\% confidence level.  This value is
lower than the theoretical prediction from numerical
simulations of the \lya forest in \cite{2003ApJ...585...34M}, $\beta
\simeq 1.47$, with small error bars for the particular model that was
analyzed there.
We have shown here that forest metal contamination and LLS/DLAs may
help explain this discrepancy: in the simple model we have adopted to
include LLS/DLAs in our synthetic data, these systems alone lower the
fitted value of $\beta$ from 1.58 to $\beta\simeq1.33$, while forest
metals alone lower it to $\beta\simeq1.35$. The two effects together
can lower it to $\beta\simeq 1.09$ (we used $\beta=1.58$ to start here,
instead of 1.47, because we had not yet adjusted the prediction of
\cite{2003ApJ...585...34M} to reflect a modern cosmological model).
Note that we have removed by-eye-identified DLAs from the data set, so any 
effect must be coming from ones that are missed, due to low column density 
and/or noise in the data. 

When we include identified DLAs in the analysis (Table
\ref{tab:res}), $b$ and $\beta$ change in the direction predicted by the mocks,
but it is actually quite difficult to estimate whether this change is 
consistent with a small or large fraction of the systems relevant to $b$ and
$\beta$ being included in this identified sample. Since optical depth is 
additive, the observed flux fluctuations  are 
$\delta_f(x)=(1+\delta_F(x))(1+\delta_D(x))$ where $\delta_F(x)$ is low density
forest and $\delta_D(x)$ high density. At this point one might be tempted to 
approximate this as $\delta_f(x)\simeq\delta_F(x)+\delta_D(x)$, which makes it 
straightforward to estimate $b$ and $\beta$ for one of the fields given the 
other two. This would make it possible to interpret the mocks as predicting 
$b$ and $\beta$ for the DLA/LLSs,
and then estimate exactly how, given that model, the observations with DLA/LLSs
should be related to ones without them. Unfortunately, as shown by 
\cite{2006PhRvD..74j3512M,2009JCAP...08..020M}, this linearized calculation 
is mathematically nonsensical. A careful study of a term like 
$\left<\delta_F(x)\delta_D(x) \delta_F(y)\right>$ shows that this is {\it not}
generically smaller than $\left<\delta_D(x)\delta_F(y)\right>$ -- a 
naive Taylor 
series approach is not valid because the product $\delta_F(x)\delta_D(x)$ 
applies locally, i.e., at a small-scale point where fluctuations are not small.
Pursuing this calculation to next-to-leading order shows that gravitational
evolution generically leads to a modulation of this local product by large
scale density modes, so the composite field $\delta_F(x)\delta_D(x)$ appears as
a standard linearly biased field on large scales. 

Somewhat puzzlingly, adding real DLAs increased our $\alpha$ result 
substantially
(Table \ref{tab:res}), while the mocks actually predict a
reduction. Clearly there are some imperfections in our treatment,
although sub-DLAs which are included in the mocks but not identified
in real data could in {\it principle} account for these oddities.  We
emphasize here that the degree to which LLS/DLA and metal lines may
lower the value of $\beta$ ought to depend on the way in which these
systems are inserted in our synthetic data sets, and on the way they
are cross-correlated with the \lya forest. The model we have adopted
to insert the DLA/LLS systems in our mock spectra should only be
considered as an illustrative example of their plausible effect.  An
observationally calibrated physical model of the distribution of these
systems will be required before reliable predictions can be made of
their impact on the value of $\beta$.

As discussed in the introduction, simulations with lower resolution/smoothing
scale than 
\cite{2003ApJ...585...34M} (including \cite{2003ApJ...585...34M}'s own low
resolution simulations) find considerably lower value of $\beta\sim 1.0$
(\cite{Slosar:2009iv}, Martin White private communication), so it is possible
that a low $\beta$ that survives other explanations is an 
indication of smoother-than-expected small-scale gas, or a flaw in the HPM 
modeling of pressure used in \cite{2003ApJ...585...34M}.

The second parameter that we measure is the bias. Bias is, of course,
completely degenerate with the assumed value of $\sigma_8$, which at
the moment is known to about 3\% in the simplest LCDM model
\cite{Komatsu:2010fb}. The parameter that we are really measuring from
the \lya forest observations is the product $b \sigma_8(z\sim 2.25)$.
We assumed a value of $\sigma_8(z=0)=0.8$ and our inferred bias varies
as the inverse of $\sigma_8(z=2.25)$ (where we mean $\sigma_8$ loosely
-- really, the linear power on the scale that we are measuring).  Our
bias is constrained to be between 0.16 and 0.24, a value which is
considerably higher than the $b\sim0.13$ obtained in
\cite{2003ApJ...585...34M} and \cite{2010ApJ...713..383W} (the latter
verified by later simulations of Martin White with higher resolution).
\cite{Slosar:2009iv} did obtain a higher value $b\sim 0.18$ from very
low resolution simulations, but the numerical smoothing in these
simulations is almost certainly much larger than any physical smoothing
of the IGM. These theoretical numbers are
for the uncontaminated forest; metal contamination associated with
forest absorption negligibly affects bias, but LLS/DLAs can raise the
bias considerably, by some 20\% in synthetic data.  The effect of
including or excluding quasars marked as DLAs is about 10\% on the
bias in our data. This is consistent with the fact that our non-DLA
flagged sample is still likely to be contaminated with high column
density systems at some level (see \S
\ref{sec:data-sample}). Regardless of the cause, the
higher-than-expected bias may improve BAO detection by creating a
higher-than-forecast signal.  

There is some evidence that our lowest redshift bin is the most
problematic. From Figure \ref{fig:koot} it is clear that it
contributes a lot to the overall signal on $\beta$ and that it drives
the low value of $\beta$ observed here. Moreover, Figure
\ref{fig:psig} indicates a greater variance than expected at the
lowest values of $z$ (this expectation is an extrapolation of the trend in
the SDSS observational measurement of \cite{2006ApJS..163...80M}, not 
necessarily a simulation prediction). 
The cause of this enhanced variance is not yet
understood. The bias evolution parameter $\alpha$ is similarly smaller than 
expected, indicating extra power at low $z$ on
large scales, although this is only significant enough to be suggestive.
At current sensitivity, the data are consistent with a
constant $\beta$ and bias, which suggests that any contaminant that is
affecting the large-scale correlation is itself a tracer of the large scale
structure. 

  Irrespective of these uncertainties on the values of $b$ and $\beta$
and their implications for the physics of the \lya forest, which will need to
be further investigated in the future, one main conclusion stands out
from this work: the correlation function of the \lya forest on the
scales of $10$ to $60 \mpch$ bears the signature of redshift distortions
consistently with the growth of linear density perturbations by
gravitational instability. The detailed physics of the \lya forest may
still be influenced by other processes on smaller scales, such
as galactic winds and outflows from quasar jets, but on the large scales
examined here, linear gravitational evolution must be the principal
process at work.

What are the main improvements that are desirable for the next
iteration of the analysis of the BOSS \lya forest data? We have
reasons to believe that our analysis is sufficient for the goals in
this paper, mainly because of the good fit we obtain to the standard
theoretical model and the tests performed in \S
\ref{sec:syst-effects-cross}. However, the next iteration may require
a more sophisticated analysis, especially to better understand the
lowest redshift bin. First, we need to remove the potential sources of
systematics by subtracting the 'signal' measured red-ward of the \lya
emission line. This should eliminate potential contamination from
low-redshift metal lines as well as any systematics associated with,
e.g., imperfect sky subtraction. With the present quasar sample we
have performed this test; however the sample used here did not include
low-$z$ quasars that would be required to do this subtraction in the
lowest redshift bin. The two higher redshift bins did not show any
measurable deviation from zero. Note, however, that such methods will
not solve the problem of metals lines that arise exclusively (or
almost exclusively) in the forest, as described in \S~\ref{subsec:formet}.
Second, the continuum
fitting could perhaps be improved by going beyond a fixed
continuum model 
(e.g.\ \cite{kgprep})
and more thoroughly investigating the spectro-photometric errors in the data. 
Third, our
results have shown that there is a real need to better understand and
filter out the high column density and metal-line systems systems
present in the data. These should be identified, possibly using the
absorption features red-ward of the \lya emission line, and either
corrected or removed from the data. 
Only the approximate impact of forest metals has been
explored here. As the survey grows, measurements of metal clustering and 
scatter in absorption strength will be included in the analysis. 
This will also provide greater precision and, perhaps, sensitivity to 
weaker metal lines.
We have not explored in this paper the impact of metals
associated with LLS/DLAs, and this is something we intend to address.
Fourth, additional 
physical effects might complicate the biasing of the observed
field with respect to the dark matter, such as temperature and
ionization fluctuations in the intergalactic medium
\cite{2010arXiv1010.5250M} (although see \cite{2010arXiv1007.3734L}
for a possible way to constrain these from the 1-dimensional 
\lya forest). These do not affect the measurement {\it per se},
but they do affect its interpretation.
Finally, the method for calculating the
correlation function (or power spectrum) should be made more
optimal. This should include a better treatment of the evolution of
the flux across the forest and an appropriate form of inverse
variance weighting.  The full problem is computationally intractable,
but one could apply the inverse covariance weighting on a per quasar
basis, an approximately optimal weighting suggested in
\cite{2007PhRvD..76f3009M,2011arXiv1102.1752M}, or do the full problem on 
coarse pixels.

One of the main ultimate goals of the measurement of the \lya forest
correlation function is to infer the angular diameter distance
$D_A(z)$ and the Hubble constant $H(z)$ at the observed redshifts,
using the position of the BAO peak. However, even at the smaller
scales at which we have made our measurements here, cosmological
constraints might be obtainable on the product $D_A(z) H(z)$ through
the Alcock-Paczy\'nski method
\cite{1979Natur.281..358A,1999ApJ...518...24M,1999ApJ...511L...5H}.
To illustrate the statistical potential of this test, we have
attempted to fit the observed data assuming an Einstein-de-Sitter
cosmology. If we simply rescale the radial and transverse distances,
keeping a constant form for the linear theory power spectrum, spurious
higher multipoles appear in the redshift-space correlation
function. This results in the best fit $\chi^2$ degrading by
$\Delta\chi^2 \sim 10$ when using only $r>20\mpch$, and by
$\Delta\chi^2 \sim 21$ when using $r>10\mpch$.  This procedure is not
a fully geometrical form of the Alcock-Paczy\'nski test, since we have
assumed that the real space correlation function has the shape
predicted by $\Lambda$CDM, but it shows that our data already have
enough power to detect any large deviations from the spacetime metric
of a flat, $\Lambda$-dominated universe.

A much stronger change appears if, in addition to changing distances,
we also change the underlying theory. The CDM correlation function for
$\Omega_m=1$ passes through zero in the radial direction at $\sim 28
\mpch$, clearly at odds with our data (see e.g. Figure
\ref{fig:view}).  The best-fit $\chi^2$ in this case increases by
$\Delta\chi^2 \sim 48$ when fitting $r>20\mpch$ and by $\Delta\chi^2
\sim 88$ when fitting with $r>10\mpch$.  We caution that this
procedure has not been tested with synthetic data; fitting
cosmological parameters goes beyond the scope of this paper. This
shows, however, that the shape of the correlation function may contain
substantial cosmological information in addition to the BAO feature if
systematic errors can be well controlled.

\section{Conclusions and Prospects}
\label{sec:conclusions}

For more than a decade, 1-dimensional analyses of the Lyman-$\alpha$
forest have provided a powerful quantitative tool for probing
structure in the high-redshift universe.  The BOSS quasar sample makes
it possible, for the first time, to treat large-scale structure in the
Lyman-$\alpha$ forest as a truly 3-dimensional phenomenon.  Although
this first-year BOSS quasar sample is only 10\% of the anticipated
final sample, it is already several times larger than the largest
previous sample used for cosmological analysis of the Lyman-$\alpha$
forest \cite{2006ApJS..163...80M}.  It is similar in size to the
entire sample of $z>2.1$ quasars from SDSS-I and SDSS-II
\cite{2010AJ....139.2360S}, and the order-of-magnitude higher surface
density of BOSS quasars makes it a much more powerful sample for
3-dimensional measurements.  We have achieved high-significance
detection of the angle-averaged flux correlation function out to
comoving separation of $60\hmpc$, and the shape of this correlation
function agrees well with the predictions of a standard $\Lambda$CDM
cosmological model.

Our measurements show the clear signature of redshift-space
anisotropy induced by large-scale peculiar velocities.
The agreement of the observed anisotropy
with the linear theory prediction of the extended Kaiser model
confirms the standard model of the Lyman-$\alpha$ forest as
structure that originates in the gravitational instability
of primordial density fluctuations \cite{1994ApJ...437L...9C,
1995ApJ...453L..57Z,1996ApJ...457L..51H,1996ApJ...471..582M,
1997ApJ...479..523B}.

We have fit our measurements with a 3-parameter model that describes
the linear bias of the forest ($b$), the redshift-space distortion
($\beta$), and the redshift-evolution of the correlation 
amplitude ($\alpha$).  Our estimated parameter values are within
the range of theoretical predictions, though the value of $\beta$
appears somewhat low (see \S\ref{sec:discussion}).  Our synthetic
data tests suggest that this low $\beta$ may be a consequence of
high column density systems (LLS/DLA) and metal-line absorption
within the forest.  Statistical errors estimated internally
from the data agree well with external estimates based on the
synthetic data sets, which suggests that we have identified any
observational or physical effects that have a large impact on
our measurements.

The tests in \S\ref{sec:discussion} show that assuming either an
$\Omega_m=1$ spacetime metric or an $\Omega_m=1$ CDM matter power
spectrum leads to substantially worse agreement with our measurements.
However, we have not attempted to derive cosmological parameter
constraints, instead fitting values of $b$, $\beta$, and $\alpha$
assuming an underlying $\Lambda$CDM cosmology.  Previous studies using
the 1-dimensional flux power spectrum have inferred the slope and
amplitude of the matter power spectrum by using cosmological
simulations to predict the bias of the flux power spectrum (including
its scale dependence) from first principles.  Even after marginalizing
over uncertainties in the IGM equation of state, these studies yield
valuable cosmological constraints (e.g.,
\cite{1999ApJ...520....1C,2000ApJ...543....1M,2002ApJ...581...20C,2004MNRAS.354..684V,2005PhRvD..71j3515S,2006JCAP...10..014S}

The BOSS Lyman-$\alpha$ forest measurements will allow these
tests to become much more powerful.  The measurement of the
1-dimensional power spectrum will itself become much more precise
with the large BOSS data sample, and division of the data set
into many subsamples of redshift, data quality, and quasar properties
will allow careful cross-checks for systematic errors.
Much stronger constraints can be obtained from three-dimensional
measurements, because of the additional information contained in the
cross-correlation of parallel lines of sight 
and because they allow for strong
tests based on redshift space distortions and the cosmological
dependence of the angular diameter distance and expansion rate. Fully
exploiting these data will require considerable analysis of the
systematic effects of the DLA/LLS and metal-line absorption and of
additional physical effects on the correlation function, such as those
due to variations in the ionizing background or in the
temperature-density relation induced by helium reionization
\cite{2010arXiv1010.5250M}.
However, BOSS data will provide many measurements with which
to constrain these models and test for observational or 
theoretical systematics.

The design goal of the BOSS quasar survey is to measure the angular
diameter distance $D_A(z)$ and Hubble parameter $H(z)$ at $z\approx
2.5$ from the BAO feature in Lyman-$\alpha$ forest clustering
\cite{2011arXiv1101.1529E}.
Forecasts using the formalism of \cite{2007PhRvD..76f3009M} imply
$1\sigma$ constraints of 7.7\% and 3.0\% on these two quantities,
respectively, from the full survey. These
errors are strongly correlated (similar to the $b$-$\beta$
degeneracy found in this paper), so it is more meaningful to quote
the forecast error on an overall distance scale dilation factor,
which is 1.9\%.  
Our present measurement of
clustering on sub-BAO scales is based on 10\% of the full BOSS data
sample and on first-pass versions of the spectroscopic reduction
pipeline and Lyman-$\alpha$ forest analysis procedures.  The good
agreement that we find with theoretical expectations reinforces the
promise of the Lyman-$\alpha$ forest as a tool to map the
high-redshift universe, to measure its expansion via BAO, and to
thereby constrain the origin of cosmic acceleration.

\section*{Acknowledgements}

We thank Martin White, David Kirkby and many others for useful
discussions.

Funding for SDSS-III has been provided by the Alfred P. Sloan
Foundation, the Participating Institutions, the National Science
Foundation, and the U.S. Department of Energy. The SDSS-III web site
is \texttt{http://www.sdss3.org/}.

SDSS-III is managed by the Astrophysical Research Consortium for the
Participating Institutions of the SDSS-III Collaboration including the
University of Arizona, the Brazilian Participation Group, Brookhaven
National Laboratory, University of Cambridge, University of Florida,
the French Participation Group, the German Participation Group, the
Instituto de Astrofisica de Canarias, the Michigan State/Notre
Dame/JINA Participation Group, Johns Hopkins University, Lawrence
Berkeley National Laboratory, Max Planck Institute for Astrophysics,
New Mexico State University, New York University, Ohio State
University, Pennsylvania State University, University of Portsmouth,
Princeton University, the Spanish Participation Group, University of
Tokyo, University of Utah, Vanderbilt University, University of
Virginia, University of Washington, and Yale University.

This work was supported in part by the U.S. Department of Energy under
Contract No. DE-AC02-98CH10886. AF and JM are supported by Spanish
grants AYA2009-09745 and CSD2007-00060.  Some computations were
performed on CITA's Sunnyvale clusters which are funded by the Canada
Foundation for Innovation, the Ontario Innovation Trust, and the
Ontario Research Fund.

\bibliographystyle{JHEP}
\bibliography{xip,cosmo,cosmo_preprints}

\newpage
\appendix

\section{Appendix: Removing the mean of the forest}
\label{sec:appDC}

Consider a toy model in which a quasar has a constant continuum and we
measure flux in pixels $i=1\ldots N$:
\begin{equation}
  f_i = \bar{f} (1+\delta_i+\epsilon_i),
\end{equation}
where $\delta_i$ is the underlying fluctuation field and $\epsilon_i$
our measurement error (we can rescale it by $\bar{f}$ without loss of
generality). By fitting a continuum to the set of points and
estimating the flux contrast, we actually estimate $\delta_i$ as:

\begin{multline}
  \delta_i'  = \frac{f_i-N^{-1} \sum_k f_i}{N^{-1}\sum_k f_i} =
\frac{\delta_i + \epsilon_i - N^{-1} \sum_k (\delta_k+\epsilon_k)}{1+
  N^{-1} \sum_k (\delta_k+\epsilon_k)}\\ \approx \left(\delta_i + \epsilon_i - N^{-1} \sum_k (\delta_k+\epsilon_k)\right)\left(1-N^{-1}\sum_k (\delta_k+\epsilon_k)\right).
\end{multline}

In the approximation, we have used the fact that the mean fluctuation
across the forest is much less than unity. Taking expectation value
over noise, one gets

\begin{equation}
  \left <\delta_i' \right> = \delta_i - N^{-1} \sum_k \delta_k
- \delta_i N^{-1} \sum_k \delta_k  + N^{-2} \sum_{kl}
\delta_k\delta_l  - N^{-1} \sigma_i^2 + N^{-2} \sum_k \sigma_k^2
\end{equation}
where we assumed diagonal noise vector $ \left < \epsilon_k \epsilon_l
  \right> = \delta^{K}_{kl} \sigma^2_k$
(neglecting the terms from the denominator above).

This means that, after fitting for mean continuum, the estimator is
not unbiased anymore and can, in principle, lead to change in
large-scale bias\footnote{Note that this calculation is incomplete as
  we are missing the third-order terms which contribute at the same
  order when computing a correlation function.}.This effect is the
1-dimensional analogue of the integral constraint for the correlation
function that is a direct consequence of $k=0$ mode being forced to be
zero. Our pipeline forces not only the overall mean of fluctuations to
be zero, but also mean in each individual quasars to be zero,
effectively imposing an integral constraint on the per-quasar as well
as per-survey basis.

We proceed to look at cross-correlation between two adjacent quasars
with respective flux contrast measurements $\delta'^A$ and $\delta'^B$
which we, for simplicity, assume to have forests of equal
length. Then, the trivial correlation function estimator gives

\begin{equation}
  \left< \delta'^A_i \delta'^B_j\right>  =   \left< \delta^A_i \delta^B_j\right> - 
\frac{1}{N} \sum_k (\left< \delta^A_i \delta^B_k \right> + \left< \delta^A_k \delta^B_j\right>)
+\frac{1}{N^2}\sum_k\sum_l   \left< \delta^A_k \delta^B_l \right>
+\mbox{h. o. correlators}
\label{eq:meansub}
\end{equation}
(neglecting the terms from the denominator above).

Therefore for a given pair of pixels, the process of removing the mean
component from the quasar results in measuring the true correlation
function minus the appropriately averaged correlation function averaged over
pixel pairs in the quasar. 

In practice, one does not need to simulate the full geometry of the
survey to calculate this effect; it is sufficient (as proven by tests on 
synthetic data) to assume that a
typical correlation function is averaged over some distance $\Delta r$
in the positions of both quasars:

\begin{equation}
  \xi_F'(r_\perp, r_\parallel) = \xi_F(r_\perp, r_{\parallel}) -
2 \frac{1}{\Delta r} \int _{-\Delta r/2}^{\Delta r/2} dr_1 \xi_F(r_\perp, r_{\parallel}+r_1)
+\frac{1}{\Delta r^2}  \iint_{-\Delta r/2}^{\Delta r/2} dr_1 dr_2 \xi_F(r_\perp, r_{\parallel}+r_1-r_2)
\end{equation}
where note that the $r_\parallel$ inside the integrals is not derived from
Eq. \ref{eq:meansub} -- we are approximating the distribution of relative quasar
redshifts by assuming that all quasars are at the same redshift in the 
$r_\parallel=0$ case, and then assuming that the slightly different weightings
of alignments for non-zero $r_\parallel$ lead effectively to a shift in 
alignment by exactly $r_\parallel$. 

We use this simple formula to account for the overall affect of
removing means from all spectra, with a single fitted $\Delta r$ in
each redshift bin, even though generally we could do a more careful
spectrum-by-spectrum calculation using the pixel-pair weights, because
mocks show that this approximation is good enough.

\section{Appendix: Errors of the trivial estimator}
\label{sec:appErr}

In this work we use the trivial correlation function estimator
(we drop the subindex $F$ in the flux correlation function in this
Appendix to reduce clutter):

\begin{equation}
  \bar{\xi} (r,\mu)  = \frac{\sum_{{\rm pairs}\ i,j} w_i w_j\delta_{f_i} 
\delta_{f_j}}{\sum_{{\rm pairs}\ i,j} w_i w_j},
\end{equation}

It is clear that the expectation value of this estimator is the true
correlation function, regardless of which weights $w_i$ are used:

\begin{equation}
\left<  \bar{\xi} (r,\mu)\right>  = \frac{\sum_{{\rm pairs}\ i,j} w_i
  w_j \left< \delta_{f i} \delta_{fj}\right>}{\sum_{{\rm pairs}\ i,j} w_i
  w_j} = \left< \delta_{fi} \delta_{fj}\right> \frac{\sum_{{\rm pairs}\ i,j} w_i
  w_j}{\sum_{{\rm pairs}\ i,j} w_i
  w_j} = \xi(r,\mu).
  \end{equation}
  (where the correlation function is assumed to be constant within a
  bin).  In other words, the estimator is un-biased.  The co-variance
  of this estimator is given by (we use $A$ and $B$ to denote two
  $(r,\mu)$ pairs):

\begin{equation}
\left<  \bar{\xi} (A) \bar{\xi} (B)\right>  = 
\frac{\sum_{{\rm pairs}\ i,j \in A, {\rm pairs}\ k,l\ \in B} w_i
  w_j w_k w_l\left< \delta_{fi} \delta_{fj} \delta_{fk}
    \delta_{fl}\right>}{\sum_{{\rm pairs}\ i,j \in A} w_i
  w_j \sum_{{\rm pairs}\ i,j \in B} w_i w_j}
\end{equation}
The 4-point term in brackets can be expanded using Wick's theorem,
which yields three terms, the first of which is separable, giving
\begin{equation}
\left<  \bar{\xi} (A) \bar{\xi} (B)\right>  = 
\xi(A)\xi(B) + \frac{\sum_{{\rm pairs}\ i,j \in A, {\rm pairs}\ k,l \in B} w_i
  w_j w_k w_l (\xi_{ik}\xi_{jl}+ \xi_{il}\xi_{jk})}{\sum_{{\rm pairs}\ i,j \in A} w_i
  w_j \sum_{{\rm pairs}\ i,j \in B} w_i w_j}
\end{equation}
and hence the covariance matrix of errors is given by
\begin{equation}
C_{AB}  \left<  \bar{\xi} (A) \bar{\xi} (B)\right>  - \xi(A)\xi(B) =
 \frac{\sum_{{\rm pairs}\ i,j \in A, {\rm pairs}\ k,l \in B} w_i
  w_j w_k w_l (\xi_{ik}\xi_{jl}+ \xi_{il}\xi_{jk})}{\sum_{{\rm pairs}\ i,j \in A} w_i
  w_j \sum_{{\rm pairs}\ i,j \in B} w_i w_j}.
\end{equation}
We use the raw
measurement of the correlation function from the data for $\xi$ here,
including the noise contribution to the diagonal (in an averaged sense, not 
pixel-by-pixel).
Strictly speaking this estimator is true only for Gaussian fields and
the corrections to it are of the order of bispectrum.

\section{Appendix: Measuring forest metal absorption}
\label{sec:appMet}

A modified version of the approach described by 
 \cite{2010ApJ...724L..69P} is used
 to measure metal absorption associated with the \lya forest. We 
 eliminate the requirement that stacked pixels be a local
flux minimum, which was intended to limit the stacking of wings from 
stronger lines. This is preferable because our goal is not to measure 
metallicity, but to measure a signal in order to reproduce it.
 We combine the individual
spectra using the arithmetic mean with a 3\% outlier clipping and
continuum fit to correct for uncorrelated absorption giving us a composite
transmitted flux, $F_{c}$, of stacked systems.

We select pixels to stack by virtue of their normalized flux 
$F_n \equiv F/\bar {F}$, where the mean transmitted flux, $\bar F$, is 
determined using the method set out in \S~\ref{sec:cont-fit}. Seven 
composite spectra were produced with
$F_n<0.4$ (above which the metal signal was negligible). We retain the
requirement from \cite{2010ApJ...724L..69P} that pixels be $0.5\sigma$
from saturation (a standard choice in pixel optical depth techniques)
in order to obtain a clean measure and minimize the contribution of
LLS/DLAs. In each of our composite spectra, we measure $F_{c}$ at 
line center for 7 metal transitions:
\ion{Si}{2} ($1193$\AA), \ion{Si}{3}
($1207$\AA), \ion{N}{5} ($1239, 1242$\AA), \ion{Si}{2} ($1260$\AA),
\ion{O}{1} ($1302$\AA), \ion{Si}{2} ($1304$\AA). Since we do not set a
requirement that all stacked absorption be \lya, we allow that some 
lines in the composite arise from stacking metal
lines. 
The only resolved contamination of
this sort arises from stacking \ion{Si}{3} lines.
We have
measured $F_{c}$ at line center for 7 wavelengths where this
`shadow' signal would be present (see \cite{2010ApJ...724L..69P} for
details). We include this shadow signal in the construction of
mocks as if it
came from, e.g., an additional metal line at $\sim 1225$\AA, even
though this implies that the original underlying mock field represents not
just a hydrogen field, but also an identical \ion{Si}{3} field unphysically
offset by $\sim 20 \hmpc$ in the radial direction. We don't think this 
affects our results.

This process provides a look-up table of 7 \lya line strengths and
normalized flux decrements ($D_c \equiv 1- F_{c}$) at 14 fixed
spectral locations. This is shown in Table~\ref{tab:metals}.  It
should be noted that we do not limit ourselves to high-significance
lines (as in dedicated metal line studies). Instead we include all
metal lines that may introduce contamination, and are seen in some
composite spectra. We constrain the metal signal to be positive, which
biases our results upward (e.g., if a line did not really exist, we
would on average add one), but we do not think this affects our
results, because generally the non-detections have fairly tight upper
limits. We ignore the fact that \ion{N}{5} is a doublet and the two
lines are treated independently. The very low absorption seen in the
weaker \ion{N}{5} renders it negligible and is only included for
completeness.
 
No significant evolution in the metal line
strengths is seen, but as the size of the BOSS survey grows we will revisit 
the analysis. Here we use the full redshift range used in this flux correlation
analysis.

\begin{sidewaystable}\footnotesize
  \centering
  \begin{tabular}{ccccccccccccccccccc}
   & & &  &
    \multicolumn{7}{c}{$(10^3 D_{c})$ from correlation with \lya} &
    \multicolumn{7}{c}{$(10^3 D_{c})$ from correlation with Si\,{\scriptsize III} }\\
     \cline{5-11}    \cline{13-19} 
      \multicolumn{4}{c}{Species}&
    \ionalt{Si}{2}  & 
     \ionalt{Si}{3}  &
     \ionalt{N}{5}   &  
      \ionalt{N}{5}  &
      \ionalt{Si}{2}  &       
\ionalt{O}{1}  & 
\ionalt{Si}{2}  &
& 
   \ionalt{Si}{2}  &
      \lya  &
     \ionalt{N}{5}  &  
      \ionalt{N}{5} &
      \ionalt{Si}{2}  &
\ionalt{O}{1} &
 \ionalt{Si}{2}  \\
\hline
 \multicolumn{4}{c}{$\lambda$ }  & 
$1193$\AA& 
$1207$\AA& 
$1239$\AA &
$1242$\AA &
$1260$\AA &
$1302$\AA&
$1304$\AA & 

&
$1202$\AA& 
$1225$\AA&
$1248$\AA &
$1252$\AA &
$1270$\AA & 
$1312$\AA& 
$1314$\AA 
\\

\hline
 \multicolumn{3}{c}{Stacked $F_n$} &&&&&&&&&&&&&&&\\
 \cline{1-3}
Min & Max & Mean &&&&&&&&&&&&&&& \\
\cline{1-3} \cline{5-19}

0.02 & 0.1  & 0.074 & &
40 $\pm$ 10 & 
123  $\pm$ 8  & 
8  $\pm$ 7  & 
5  $\pm$ 8  & 
50  $\pm$ 8  & 
40   $\pm$ 10   & 
30   $\pm$ 10   & 
& 
0  $\pm$ 6  & 
13  $\pm$ 8  & 
1  $\pm$ 9  & 
0  $\pm$ 7  & 
1  $\pm$ 9  & 
10   $\pm$ 10   & 
0  $\pm$ 10 \\

 0.1 & 0.15  & 0.128  & &
 8  $\pm$ 9  & 
 53  $\pm$ 7  & 
 11  $\pm$ 6  & 
 0  $\pm$ 6  &
 10  $\pm$ 6  & 
 0  $\pm$ 8  & 
 12  $\pm$ 9  & 
 &
 2  $\pm$ 6  & 
 17  $\pm$ 6  & 
 0  $\pm$ 6  & 
 0  $\pm$ 7  & 
 12  $\pm$ 7  & 
 21  $\pm$ 9  & 
 0  $\pm$ 8  \\
 
 0.15 & 0.2  & 0.177  & &
 5  $\pm$ 6  & 
 20  $\pm$ 7  &
  5  $\pm$ 5  & 
  0  $\pm$ 4  & 
  6  $\pm$ 5  & 
  13  $\pm$ 6  & 
  12  $\pm$ 5  & 
   &
 1 $\pm$ 6  & 
  14  $\pm$ 4  & 
   2  $\pm$ 5  & 
  0  $\pm$ 4  & 
  9  $\pm$ 6  & 
  7  $\pm$ 6  & 
  0  $\pm$ 6   \\
  
 0.2 & 0.25  & 0.227  & &
 0  $\pm$ 6  & 
 8  $\pm$ 4  & 
 3  $\pm$ 4  & 
 3  $\pm$ 4  & 
 3  $\pm$ 4  & 
 1  $\pm$ 4  & 
 0  $\pm$ 5  & 
 &
 4  $\pm$ 5  & 
 10  $\pm$ 3  & 
 3  $\pm$ 4  & 
 0  $\pm$ 4  & 
 4  $\pm$ 3  & 
 4  $\pm$ 6  & 
 10  $\pm$ 5 \\
 
0.25 & 0.3  & 0.276  & &
0  $\pm$ 5  & 
0  $\pm$ 4  & 
5  $\pm$ 3  & 
3  $\pm$ 3  & 
0  $\pm$ 3  & 
1  $\pm$ 4  & 
4  $\pm$ 4  & 
&
0  $\pm$ 5  & 
10  $\pm$ 3  & 
3  $\pm$ 3  & 
0  $\pm$ 3  & 
3  $\pm$ 4  & 
0  $\pm$ 5  & 
6  $\pm$ 5 \\

0.3 & 0.35  & 0.326  & &
0  $\pm$ 5  & 
1  $\pm$ 4  & 
2  $\pm$ 3  & 
1  $\pm$ 3  & 
0  $\pm$ 3  & 
0  $\pm$ 3  & 
0  $\pm$ 4  & 
&
2  $\pm$ 4  & 
5  $\pm$ 3  & 
2  $\pm$ 3  & 
0  $\pm$3  & 
0  $\pm$ 3  & 
0  $\pm$ 4  & 
0  $\pm$ 4 \\

 0.35 & 0.4  & 0.376  & &
 0  $\pm$ 4  & 
 0  $\pm$ 4  & 
 0  $\pm$ 3  & 
 1  $\pm$ 2  & 
 1  $\pm$ 3  & 
 0  $\pm$ 3  & 
 3  $\pm$ 3  &  
 &
 0  $\pm$ 4  & 
 10  $\pm$ 2  & 
 0  $\pm$ 3  & 
 3  $\pm$ 3  & 
 2  $\pm$ 3  & 
 2  $\pm$ 4  & 
 5  $\pm$ 3  \\

   \end{tabular}
  \caption{The flux decrement at line center, $D_{c}$, measured 
  in the composite spectra of absorbers within the \lya forest as 
  described in Appendix~\ref{sec:appMet}. Signal measured as 
  a result of stacking both \lya and \ion{Si}{3} is shown. The 
  correlated signal due to 8 transitions are measured.
  }
  \label{tab:metals}
\end{sidewaystable} 

\end{document}